\documentclass[11pt, fleqn]{article}

\usepackage{geometry}
\usepackage{comment}

\usepackage{soul}
\usepackage{subcaption}

\usepackage{jheppub}
\usepackage{amsmath,amssymb,amsthm}
\usepackage{bm}
\usepackage{slashed}
\usepackage{graphicx}
\usepackage[T1]{fontenc}
\usepackage{fix-cm}

\usepackage{tabu}
\usepackage{physics}

\makeatletter

\newcommand*\bigcdot@[2]{\mathbin{\vcenter{\hbox{\scalebox{#2}{$\m@th#1\bullet$}}}}}
\makeatother

\newlength{\dummysp}
\newcommand{\mzb}{m_{z,\beta}}
\newcommand{\sqmzmi}{\sqrt{m_{z,\beta}^2+M_i^2}}

\usepackage[font=scriptsize]{caption}
\usepackage{mwe}

\def\R{{\mathbb R}}
\def\S{{\mathbb S}}
\def\Z{{\mathbb Z}}

\def\tr{\,{\rm tr}\,}

\title{Topological terms and anomaly matching in effective field theories on $\R^3 \times \S^1$: I. Abelian symmetries and intermediate scales}
\author{Erich Poppitz and F. David Wandler}
\affiliation{Department of Physics, University of Toronto, Toronto, ON M5S 1A7, Canada}
\emailAdd{poppitz@physics.utoronto.ca}  \emailAdd{f.wandler@mail.utoronto.ca}    

\abstract{We explicitly calculate the topological terms that arise in IR effective field theories for $SU(N)$ gauge theories  on \(\R^3 \times \S^1\) by integrating out all but the lightest modes. We then show how these terms match  all global-symmetry 't Hooft anomalies of the UV description. We limit our discussion to  theories with abelian 0-form symmetries, namely those with one flavour of adjoint Weyl fermion and one or zero flavours of Dirac fermions. While anomaly matching holds as required, it takes a different form than previously thought. For example, cubic- and mixed-$U(1)$ anomalies are matched by local background-field-dependent topological terms (background TQFTs) instead of chiral-lagrangian Wess-Zumino terms. We also describe the coupling of 0-form and 1-form symmetry backgrounds in the magnetic dual of super-Yang-Mills theory in a novel way, valid throughout the RG flow and consistent with the monopole-instanton  't Hooft vertices. We use it to discuss the matching of the mixed chiral-center anomaly in the magnetic dual.}

{  }

\begin{document}

\maketitle

\section{{}{Introduction}}
\label{sec:1}

This paper is about the connection between two recent developments in nonperturbative gauge theories which we find quite  exciting and worthwhile. 

The first, more than a decade old, is the realization that complicated and analytically untractable problems in four-dimensional (4d) asymptotically free gauge theories---such as confinement and chiral symmetry breaking---yield to theoretically controlled analysis once spacetime is compactified to $\R^3 \times \S^1$, where $\S^1$ is interpreted as a spatial circle and is not encoding finite temperature. This first development is due to the work of \" Unsal (and, later, collaborators). The study of gauge theories on $\R^3 \times \S^1$ yields new analytical insight into their nonperturbative dynamics, a   rare luxury in four dimensional gauge theories. Further, 
this insight is not restricted to supersymmetric theories and, in 
 many cases,  is known or expected to lead to results whose  validity (qualitatively) extends to the physically relevant limit of large circle size.
 By now, there is too much literature on circle-compactified gauge dynamics to refer to in the Introduction. We only mention a few early references  \cite{Unsal:2007vu,Unsal:2007jx,Unsal:2008ch,Shifman:2008ja} and the more recent review \cite{Dunne:2016nmc}. We also stress that there is no inherent magic in the circle-compactification setup. The two pillars of the approach are the (approximate) center stability and the existence of a small parameter  related to the circle size: semiclassical calculability holds for $\Lambda N L \ll 2 \pi$, where  $L$ is the $\S^1$ size and $\Lambda$ is the strong coupling scale of the $SU(N)$ gauge theory.

 The second exciting development is only a few years old. It was realized, after \cite{Gaiotto:2014kfa,Gaiotto:2017tne,Gaiotto:2017yup}, that there are previously missed anomaly matching conditions, generalizing those of 't Hooft  \cite{tHooft:1979rat}. These conditions severely constrain (but, usually, do not uniquely determine) the infrared (IR) phases of gauge theories. These, as we shall call them, ``new 't Hooft anomaly matching conditions'' involve discrete symmetries, including spacetime ones, as well as higher-form symmetries, notably the  ``center symmetry'' of pure $SU(N)$ Yang-Mills theory. The latter is well-understood on the lattice, see \cite{Greensite:2011zz} for  review, but appears here in new disguises and has new interesting applications.

In this paper, we do not aim to find new solutions or propose new phases of previously unsolved gauge theories. Instead we shall connect the above two developments in a  modest way, yet one that has not been fully fleshed out. We want to understand, in a detailed manner, how the new anomaly matching conditions manifest themselves within  the weakly-coupled nonperturbative dynamics of circle-compactified gauge theories. As will be reviewed below, solving for the IR  dynamics on $\R^3 \times \S^1$ involves a tower  of effective field theories (EFTs), each valid at consecutively lower energy scales. Some of these EFTs are obtained after an electric-magnetic duality
transformation. Every EFT captures the degrees of freedom and the perturbative and nonperturbative effects important within the range of its validity. The solution of the circle-compactified theory is found in a theoretically controlled way, hence there is no doubt that the appropriate new and old 't Hooft anomalies are matched. In fact, we shall see that our results  explicitly confirm this expectation; one might view  the matching as an argument in  favour  of small-circle calculability. 

Despite the expected matching of 't Hooft anomalies,  it is still interesting to  give a detailed description thereof within the calculable $\R^3 \times \S^1$ setup. Such a description entails\footnote{We assume familiarity with 't Hooft anomalies. A  (necessarily rough) reminder is that  't Hooft anomalies  arise when, upon introducing nondynamical background fields gauging the global symmetries, the theory can not retain invariance under the corresponding background gauge transformations. In other words, promoting these background fields to dynamical ones would be impossible without introducing 't Hooft's spectators \cite{tHooft:1979rat}, or, equivalently, anomaly inflow from a higher dimensional bulk; a recent review along these lines is in \cite{Cordova:2018cvg}.} determining how background fields gauging the global symmetries are coupled to the  EFTs that capture the dynamics at different scales. These background field couplings  allow one to see the  manifestation of  't Hooft anomalies  in each energy range.\footnote{A few papers \cite{Aitken:2018mbb, Aitken:2018kky,Cox:2019aji,Tanizaki:2019rbk,Anber:2020xfk} have dealt with related issues in the calculable small-circle setup, at various levels of explicit detail. Closest to our immediate subject, the coupling of backgrounds gauging the  global symmetries in deformed Yang-Mills theory (dYM) \cite{Unsal:2008ch} was studied in \cite{Tanizaki:2019rbk}; a  comment relevant to dYM is in footnote \ref{footnotedym} after eq.~(\ref{thooftSYM}). The results of Section \ref{sec:4.2.2} are also relevant.} 

We can think of at least two reasons to study this question.
One is that determining these background-gauging couplings is, by itself, an interesting exercise in constructing EFTs. We shall see that the details of the 't Hooft anomalies' matching can be  intricate.\footnote{There may be other applications of our results, such as studying topological phases on the circle, but we shall not discuss these further, see \cite{Kan:2019rsz}.}
We also expect that understanding the EFT's coupling to global symmetry backgrounds will shed light on how various topological quantum field theories (TQFTs) appear in the deep IR, in a setup where the dynamics is understood at all scales. Such IR TQFTs have been deemed necessary   to match the new 't Hooft anomalies, including anomaly inflow on the worldvolumes of domain walls. It would be interesting to see how they arise from the calculable dynamics;  related studies in different contexts are in \cite{Hansson:2004wca,Hidaka:2019mfm}.

This already rather lengthy paper is entirely devoted to the first item above. 

\subsection{The scope and main results of this paper}
\label{sec:1.1}

To be concrete, we consider a small subset (but, we think, a fairly representative one) of the theories whose dynamics can be analytically  studied on $\R^3 \times \S^1$. We now list the 4d theories we study in this paper, along with a brief account of their global symmetries, 't Hooft anomalies,  symmetry realization, and the results of our findings.
  
We believe that many of the remarks we make below in the context of the particular theories discussed are of more general validity.
  
 {\flushleft  {\bf A+F}}: An $SU(N)$ gauge theory with a single massless adjoint Weyl fermion and a single massless Dirac fermion in the fundamental representation $F$.\footnote{With center stability ensured by a double-trace deformation or by the addition of a few massive adjoint multiplets as in \cite{Unsal:2008ch,Shifman:2008ja}.} The global symmetries are: a chiral $U(1)_A$, acting on both the Dirac fermion and the adjoint, and a vector $U(1)_V$ acting on the Dirac fermion. 
In the massless $A+F$ case, there are only ``old''  't Hooft anomalies,  the mixed $U(1)_A U(1)_V^2$ and the cubic $U(1)_{A}$ anomalies (we do not study anomalies involving gravity in this paper). While the $A+F$ theory has no new 't Hooft anomalies, it is still instructive to discuss the fate of the ``old'' ones in the $\R^3 \times \S^1$ EFT in  some detail. We do so in Sections~\ref{sec:3.2} and \ref{sec:4.1}, where we study the matching of anomalies in the electric and magnetic EFTs, respectively. These EFTs describe physics below the scale $1/(NL)$.

The $\R^4$ expectation for the $A+F$ theory is that fermion-bilinear condensates break $U(1)_A$ and the 't Hooft anomalies are matched by coupling of the Goldstone field  to the various global backgrounds via a Wess-Zumino term in the chiral lagrangian.\footnote{Recent lattice studies on mixed representation theories are in \cite{Bergner:2020mwl}, but with Dirac adjoint fermions.}
On $\R^3\times \S^1$, this theory was studied in \cite{Poppitz:2009tw}, where it was argued that the small and large circle phases of the center-stabilized theories are  continuously connected, with $U(1)_A$ broken both at small and large radius. 
The physics of the Goldstone mode was described by a  magnetic dual  EFT, which incorporates nonperturbative magnetic bion effects. 

Our new results regarding this theory are as follows. First, the topological terms, induced by loop effects on $\R^3 \times \S^1$, are calculated after carefully regulating the theory in a manner preserving both gauge  and 4D Lorentz invariance. These calculations, whose details are presented in Appendices~\ref{appx:A}, \ref{appx:B}, constitute the main technical result of this paper.
Second, we show that the nature of the Goldstone field is determined by these topological terms.\footnote{This is similar to how Goldstone \cite{Affleck:1982as} identification has been argued to occur in 3D theories \cite{Aharony:1997bx}. The discussion of this paper using loop-induced topological terms in locally-4D theories is new.}
Finally, we argue that, while there is a Goldstone field associated with the $U(1)_A$-breaking on $\R^3 \times \S^1 $ \cite{Poppitz:2009tw},  similar to $\R^4$, the matching of the $(U(1)_A)^3$ and  $U(1)_A (U(1)_V)^2$ anomalies in the small-circle EFT  is not due to its Wess-Zumino coupling, but is instead due to the local, background-only dependent terms, arising from integrating out the Kaluza-Klein modes of the $A+F$ fermions in the center-stabilized theory. 

The matching of the cubic and mixed anomalies by local background-dependent terms, without  IR poles \cite{Dolgov:1971ri,Frishman:1980dq,Coleman:1982yg}, is only possible due to the breaking of Lorentz symmetry by the compactification, which also gives mass to the relevant charged fermions. This has, in fact, been appreciated in the string literature \cite{Corvilain:2017luj,Corvilain:2020tfb}. Our contribution here is to combine this realization with a discussion of the calculable nonperturbative dynamics on $\R^3 \times \S^1$, including electric-magnetic duality. Further, our calculations in Coulomb-branch backgrounds are more general and  significantly more involved than those previously considered. In particular, the use of gauge-invariant Majorana mass regulators for real-representation fermions  on $\R^3 \times\S^1$    is a new technique. 

It is sometimes stated that the IR EFT on $\R^3 \times \S^1$ should only match the ``new" anomalies involving higher-form symmetries and the matching of the 0-form anomalies is ignored. The calculations done here explicitly show that 0-form anomalies are also matched, but are due to local  terms depending on the background fields gauging the 0-form symmetries, both continuous or discrete. We expect that this continues to hold in the small-circle setup also for theories with nonabelian global symmetries.

{\flushleft  {\bf SYM}}: An $SU(N)$ gauge theory with a single massless adjoint Weyl fermion. This is super Yang-Mills theory (SYM). The global symmetries are $\Z_{2N}^{(0)}$ and $\Z_{N}^{(1)}$, the 0-form chiral and 1-form center symmetries,\footnote{We use the superscripts $(0)$ to denote 0-form symmetries and $(1)$ to denote 1-form ones.} which have a mixed 't Hooft anomaly of the ``new'' type. There is also a cubic $\Z_{2N}^{(0)}$ anomaly. On $\R^4$ (see e.g. \cite{Intriligator:1995au,Shifman:1999mv} for reviews), the theory is known to break $\Z_{2N}^{(0)} \rightarrow \Z_2^{(0)}$ and have $N$ vacua with domain walls (DW) between them.  The global symmetry realization and vacuum structure is the same on $\R^3 \times \S^1$ and the theory is gapped and confining at all $\S^1$ sizes \cite{Seiberg:1996nz}.

In Section~\ref{sec:3.1}, we determine the couplings of the Coulomb-branch EFT (valid at energy scales larger than mass gap but smaller than $1/(NL)$) to the background fields for the 0-form and 1-form global symmetries. We also discuss the matching of the various anomalies. The topological terms from Appendix \ref{appx:A} once again play an important role in this discussion. Similar to the cubic-$U(1)_A$ anomaly of the $A+F$ theory, the 0-form $\Z_{2N}^{(0)}$ cubic anomaly is also matched by local terms. The matching of the new mixed chiral-center symmetry anomaly in the electric theory is discussed  in Section~\ref{sec:3.1.2}. 

The introduction of chiral and center symmetry backgrounds in the electric and magnetic theories and the way the anomaly manifests itself in the magnetic dual EFT is discussed at great length in 
Section~\ref{sec:4.2}. We  gauge the 1-form symmetry in a manner valid at all scales,  including the UV $SU(N)$ theory, without using an embedding in the emergent center of the Coulomb branch theory \cite{Cordova:2019uob,Tanizaki:2019rbk}.
The results on background couplings and anomaly matching obtained here should form the basis of further studies of the interplay of  dynamics and anomaly matching at lower scales.

{\flushleft  {\bf A+R}}: An $SU(N)$ gauge theory with a single massless adjoint Weyl fermion and a single massless Dirac fermion in a complex representation $R$ (in other words, there is a left-handed Weyl fermion in $R$ and another one in the complex conjugate $R^*$). We do not discuss the dynamics of this theory in this paper. However, our calculations of topological terms on $\R^3 \times \S^1$ are done for Dirac fermions in arbitrary representations $R$ in arbitrary holonomy backgrounds and allow a discussion of anomaly matching.\footnote{For completeness, the continuous global symmetries and their 't Hooft anomalies are as for the $A+F$ theory, but now there is also a discrete chiral symmetry, $\Z_{2 T(R)}^{(0)}$, acting only on the Dirac fermion (this is an independent symmetry only for $R$ of $N$-ality $n_R$ greater than unity). 
It has a cubic  anomaly and mixed anomalies with the continuous symmetries. For gcd$(N, n_R) = 1$, there is no 1-form center symmetry, but one can turn on 't Hooft fluxes for the colour and $U(1)_V$ fields, leading to the so-called new ``BCF'' 't Hooft anomaly, see \cite{Anber:2019nze,Anber:2020xfk,Anber:2020gig}. The topological terms we calculated allow one to study how this anomaly is matched. }
 
\subsection{Outline}
\label{sec:1.2}

A brief outline of this paper is as follows. In Section~\ref{sec:2}, we describe the $\R^3 \times \S^1$ setup. We introduce notation appropriate to describe the long-distance dynamics on the $SU(N)\rightarrow U(1)^{N-1}$ Coulomb branch, introduce the relevant  Cartan subalgebra fields, discus the appropriate scales, and pay particular attention to the description of the global center symmetries, see the end of Section~\ref{sec:2.2}. The turning on of various background fields for the global symmetries is discussed in Section~\ref{sec:2.3}.

In Section~\ref{sec:3}, we describe the results of the sometimes lengthy calculations of Appendices~\ref{appx:A} and \ref{appx:B} of the topological terms induced by integrating out  the Kaluza-Klein modes of the fermions. We discuss in turn the matching of various anomalies in the electric EFT in SYM (Section~\ref{sec:3.1}) and the $A+F$ theory (Section~\ref{sec:3.2}).
In Section~\ref{sec:4}, we describe the coupling to background fields and anomaly matching in the corresponding  magnetic dual EFTs.

Appendix~\ref{appx:A} contains the details of our adjoint-fermion calculations. We calculate their contributions 
to various background-field dependent topological terms using a gauge and 4D Lorentz invariant Pauli-Villars regulator with Majorana mass terms. In Appendix~\ref{appx:B} we repeat this for a general Dirac fermion in a representation $R$. The calculations given in detail in these appendices are among our main results. 
A proof of a useful infinite sum is in Appendix~\ref{appx:D}. 

Finally, in Appendix~\ref{appx:C},
we make some comments on the loop-induced Chern-Simons term in the  unbroken-$SU(N)$ phase.  This remark may be relevant for the study of small-$L$ topological phases, but we do not pursue this further. We only note that 
our calculations  imply that there is a subtlety regarding the classical equivalence between a background Wilson line and the non-periodic boundary conditions for the fermions on $\S^1$, in cases where the relation between the two involves an anomalous-$U(1)$ field redefinition.
\subsection{Outlook}
\label{sec:1.3}

Many circle-compactified theories were left out of the above discussion. 
Notably, we did not consider theories with nonabelian global symmetries (e.g. multiflavor ones). There are many interesting examples in the literature, including ones where the IR physics is not gapped \cite{Unsal:2007vu,Unsal:2007jx,Unsal:2008ch,Shifman:2008ja,Anber:2017pak,Anber:2019nfu}.    Notably, a version of QCD on $\R^3 \times \S^1$ with an  abelian chiral lagrangian was studied in \cite{Cherman:2016hcd}, somewhat similar to our $A+F$ theory. The results here allow for a generalization to this and other theories. In addition, we did not study 
anomalies involving gravity, nor did we study global anomalies; to complete the understanding of anomaly matching, it would be desirable to investigate these as well.
Another avenue for future study is to use our results here for a further study of the TQFTs arising in the deep IR in the theories with calculable dynamics.

\section{{}{The $\R^3 \times \S^1$ setup: notation, relevant scales, and EFTs}}\label{sec:2}
As described in the Introduction, Section \ref{sec:1}, we study $SU(N)$ Yang-Mills theory with two kinds of massless fermions: massless adjoint Weyl fermions and  massless  Dirac fermions in an arbitrary representation $R$. In this paper, we restrict the discussion to at most a single flavor of each field.
In the nonsupersymmetric case, to achieve (approximate) center stability, one adds a number of massive extra adjoints, or a double-trace deformation (an explicit description of center stabilization will not be essential for us and can be found in \cite{Unsal:2008ch,Shifman:2008ja}).

The essential feature ensuring semiclassical calculability on $\R^3 \times \S^1$ is that due to the deformation (or in SYM, the nonperturbative dynamics) the theory dynamically abelianizes: $SU(N) \rightarrow U(1)^{N-1}$ at a scale of order $1\over NL$. The unbroken $U(1)^{N-1}$ is the maximal Abelian subgroup of $SU(N)$. Perturbatively, the $U(1)^{N-1}$ gauge bosons (we shall often call them ``Cartan photons'') remain massless, while the non-Cartan $W$-bosons have mass, the lightest of which is $2 \pi \over NL$ at the center symmetric point.
  We assume that $\Lambda N L \ll 2 \pi$ so that there is a  separation of scales between the mass of the lightest $W$-boson, ${2 \pi\over NL}$, and the strong coupling scale $\Lambda$. This guarantees that the gauge coupling remains weak  at all scales. 
  
   The 3d EFT describing  physics at distance scales $\gg NL$ is thus a theory of the Cartan subalgebra fields (in SYM, both bosonic and fermionic). The EFT also incorporates semiclassical nonperturbative effects. In SYM, these lead to the generation of a mass gap, confinement, and chiral symmetry breaking. As discussed many times, see \cite{Unsal:2007vu,Unsal:2007jx,Unsal:2008ch,Shifman:2008ja,Dunne:2016nmc}, despite being 3d, this EFT ``remembers'' many features of the 4d theory. Here we focus on  the various 't Hooft anomalies of the 4d theory.
To study these, in this paper we  integrate out all the massive non-Cartan fermions (sometimes, we also study the effect of the light Cartan fermions). Information about the anomalies is encoded in  the topological terms they induce in the long-distance $\R^3$ theory. To extract it, we shall include background fields for the global symmetries and study the topological terms involving these fields.  

\subsection{{}{Notations} }
\label{sec:2.1}

We denote the $\R^4$ coordinates by $x^M$, with $M=0,1,2,3$. The $\S^1$   coordinate is $x^3 \equiv x^3 + L$,  while $x^\mu$ with $\mu = {0,1,2}$ are coordinates in $\R^3$. The Levi-Civita tensors in $\R^4$ and $\R^3$ are $\epsilon^{0123} = +1$ and $\epsilon^{012}=+1$ and the metric has a $(+,-,-,-)$ signature.

 As the theory abelianizes
 at distances greater than $NL$, our EFT involves Cartan subalgebra fields. We shall use indices $a,b$ ($a,b = 1,..., N-1$) to denote the Cartan components of the various fields. 
Sometimes, we will replace Cartan subalgebra indices with arrows.
 Thus, the $a$-th Cartan component of the $SU(N)$ gauge field 
is $A^a = A^a_\mu dx^\mu$   (equivalently, the 1-form $\vec{A}$) and $F^a = dA^a = {1\over 2} F_{\mu\nu}^a dx^\mu \wedge dx^\nu$  is the corresponding Cartan-subalgebra field strength (equivalently, we use the 2-form $\vec{F}$).  The Cartan components of the gauge field in the compact direction are $A_3^a$. They play the role of a compact Higgs field responsible for the $SU(N) \rightarrow U(1)^{N-1}$ breakdown and a more appropriate notation for them will be introduced below. 
 
 Our Lie-algebraic conventions are as follows. The simple roots are $\vec\alpha_k$, taken to have squared length $2$; the fundamental weights are $\vec w_k$, with $\vec\alpha_k \cdot \vec w_p = \delta_{kp}$, $k,p=1,...N-1$. We denote by $\beta^+$ the set of all positive roots. More explicitly, these can be labelled as $\vec\beta^{AB}$, with $1 \le A < B \le N$ (the simple roots are a subset, $\vec\alpha_k = \vec\beta^{k, k+1}$). The positive roots are related to the weights of the fundamental representation, $\vec\nu^A$ ($A = 1,...,N$) as $\vec\beta^{AB} = \vec\nu^A -\vec\nu^B$. The weights of the fundamental representation obey $\vec\nu^A \cdot \vec\nu^B = \delta^{AB} - {1\over N}$. We shall also use the facts that  $\sum\limits_{\beta^+} \beta_a \beta_b \equiv \sum\limits_{1 \le A < B \le N} \beta^{AB}_a \beta^{AB}_b = N \delta_{ab} \equiv {C(A)\over 2} \delta_{ab}$ and $\sum\limits_A \nu^A_a \nu^A_b = \delta_{ab} \equiv C(F) \delta_{ab}$, where $C(R)$ denotes the quadratic Casimir of the representation $R$ (above we used $R= A$ (adjoint) or $R=F$ (fundamental)). The Weyl vector $\vec\rho$, which determines the center-symmetric value of the holonomy (\ref{defofphi}), equals the sum of all fundamental weights  $\vec w_k$ or half the sum of all positive roots $\vec\beta^{AB}$. When we consider a general representation $R$, we shall use $\vec\lambda_i$ to denote its weights, $i=1,...,{\rm dim} R$, with $\sum_j \lambda_j^a \lambda_j^b = C(R)\delta^{ab}$.

 \subsection{{}{Dynamical fields, scales, and center symmetries in the Coulomb branch EFT} }
\label{sec:2.2}

 It is convenient to introduce new notation to describe the fluctuations of $A_3^a$ in the EFT. The non-Cartan components of $A_3$, as well as the nonzero Kaluza-Klein (KK) modes of $A_3^a$ on $\S^1$ are gauged away in the ``unitary'' gauge (see e.g.~\cite{Aitken:2017ayq}).  
We shall denote by  $2 \pi \vec{\phi}$   the Cartan-valued $\S^1$ holonomy. We define the origin of $\vec{\phi}$ so that $\langle \vec\phi \rangle = 0$ corresponds to the center symmetric point. Explicitly, the definition of $\vec\phi$, relating it to $\vec A_3$, is\footnote{\label{footnoteconvention}Notice that, compared to most previous work on the subject, $2 \pi \vec\phi\big\vert_{{\rm this \; paper}} =  \vec\phi\big\vert_{{\rm older \; refs.}}$, e.g. \cite{Davies:2000nw,Argyres:2012ka,Anber:2015wha}.}
\begin{equation}\label{defofphi}
L A_3^a  \equiv {2 \pi \rho^a \over N} + {2 \pi \phi^a } .
\end{equation} Here $\vec\rho$ is the Weyl vector. Thus, $2 \pi \vec\phi$ denotes the $\S^1$ holonomy shifted by  the center symmetric value. Using (\ref{defofphi}), the $N$ 
 eigenvalues of the fundamental Wilson loop $\Omega ={\rm diag}(\omega_1, \ldots \omega_N)$ along the $\S^1$  are 
 \begin{equation}\label{holonomyevs}
 \omega_A = e^{i \vec\nu_A \cdot \vec{A}_3 L} = e^{ i {2 \pi \over N}({N-1\over 2} - A)} e^{i 2 \pi \vec\nu_A \cdot \vec\phi}~, ~ A = 1, \ldots, N.
 \end{equation} This makes it obvious that $\vec\phi=0$ corresponds to the center symmetric point, $\tr \Omega^{k} = 0$, $k \ne N \Z$.  
The periodicity of $\vec\phi$ in the $SU(N)$ theory is in the root lattice, i.e. under large gauge transformations $\vec\phi \rightarrow \vec\phi + \vec\alpha_k c^k$, $c^k \in \Z$ ($k=1,...N-1$). One way to see this is to note that Wilson line operators in the fundamental representations, whose (gauge invariant) eigenvalues are given in (\ref{holonomyevs}), are periodic under root-lattice shifts of $\vec\phi$.\footnote{For use below, note that in the $SU(N)/\Z_N$ theory, on the other hand, the line operators are in the adjoint representation and their eigenvalues, similar to (\ref{holonomyevs}), but with $\vec\nu_A$  replaced by roots $\vec\beta^{AB}$, are periodic under weight-lattice shifts of $\vec\phi$.}

If any of the $\R^3$ directions are compactified (e.g. on a large three-torus), the  holonomies of $\vec{A}$ around 1-cycles also have  periodicity in the root lattice. For noncontractible $2$-surfaces, we also have that $\oint \vec F =  2 \pi \vec\alpha_k \; c^k$, $c^k \in \Z$, for $U(1)^{N-1}$ fields descending from an $SU(N)$ bundle, as appropriate for backgrounds corresponding to dynamical gauge fields.\footnote{One can argue for this quantization by considering a probe fundamental particle in the $SU(N)$ theory. Its wavefunction changes by $e^{i \oint\limits_C\vec\nu_A \cdot  \vec A} = e^{i \int\limits_{S, \partial S=C} \vec\nu_A \cdot \vec F}$ upon parallel transport around a closed loop $C$. Imagining that $S$ lies, e.g. inside a two-torus, the latter expression is well defined, i.e. independent on the choice of $S$ only if the integral $\oint \vec F$ belongs to the root lattice (times $2 \pi$). Again, for future use we note that this quantization condition, like the periodicity of $\vec\phi$, changes when the $SU(N)/\Z_N$ theory is considered: the probe particle wavefunction now involves a root-lattice vector and the quantization of $\oint \vec F$ is now $2 \pi \times$ a weight-lattice vector.}
Similarly, over 1-cycles, the periodic scalars can have monodromies  $\oint d \vec\phi =    \vec\alpha_k c^k$, implied by their root-lattice periodicity.

The 1-form $\vec A$ and 0-form $\vec \phi$ introduced in this Section are the dynamical bosonic fields in the $\R^3$ EFT governing physics at scales $\gg NL$.
In both  SYM and $A+F$ theories, the Cartan subalgebra fermions $\vec\psi$  also remain light and participate in the EFT.\footnote{Sometimes, we shall also call this theory the ``Coulomb branch'' EFT.}
In terms of these Cartan subalgebra fields, the tree-level kinetic Lagrangian of our IR theory   on $\R^3$ is that of a  free theory of $U(1)^{N-1}$ gauge fields $\vec A$ with a decoupled set of $N-1$ scalar fields $\vec\phi$ and the Cartan components of the fermions $\vec\psi$ (two component Weyl fermions, see Appendix 
\ref{appx:A} for the UV fermion lagrangian and definition of   $\bar\sigma^M$):
\begin{eqnarray}
\label{LIRkinetic}
L^{IR, kinetic} = - {L \over 4 g^2}\; \vec F_{\mu\nu} \cdot \vec F^{\mu\nu} + {2 \pi^2 \over g^2 L}\; \partial_\mu \vec\phi' \cdot \partial^\mu \vec\phi'  +i \vec\psi^{\; \dag} \bar\sigma^\mu \partial_\mu \vec\psi,
\end{eqnarray}
where $\vec\phi' \equiv \vec\phi - \langle \vec\phi \rangle$ denotes the fluctuation of (\ref{defofphi}) around its expectation value.
Here, the coupling $g$ is the 4d coupling taken at the scale ${1 \over L}$. At one loop order, $W$-boson loops lead to mixing between the different Cartan fields and hence to different couplings of the different $U(1)$ factors; these mixings are suppressed in the $\Lambda N L \ll 1$ limit and we shall ignore them for simplicity (see \cite{Anber:2014lba} for their calculation). There are also $U(1)^{N-1}$-gauge invariant terms suppressed by inverse powers of $m_W = {2 \pi \over NL}$, which we also ignore.

In the nonsupersymmetric case of an $A+F$ ($A+R$) theory, there is  a potential term for the scalars, due to the center stabilization mechanism that ensures abelianization. The scalars $\vec\phi$ then obtain     mass of order $m_{\vec\phi} \sim {\sqrt{g^2 N}\over LN} \ll {1 \over LN} \sim m_W$. As $m_{\vec\phi}$ is smaller than the $W$-boson mass, one is still justified keeping the $\vec\phi$ fields  in the Cartan subalgebra EFT (\ref{LIRkinetic}) valid at energies $\ll {1 \over LN}$. However, at much lower scales $\ll {g \sqrt{N} \over L N}$, one should integrate out $\vec\phi$ and transition to an EFT only involving $\vec{A}$ and $\vec\psi$.

{\flushleft{Finally,}} we also comment on the global center symmetries of our $\R^3 \times \S^1$ Coulomb-branch EFT (\ref{LIRkinetic}). We note that the usual description of the global center symmetries' action in the $\R^3 \times \S^1$ setup uses an embedding into the emergent center symmetries of the Coulomb branch theory. This way to represent the action of the center symmetries  is described below. In Section \ref{sec:4.2.2}, we give a description that does not use such an embedding and discuss its relation to the present one. The description of Section \ref{sec:4.2.2} has the advantage of being valid at all scales, as it makes no use of emergent symmetries, and  is particularly useful when introducing background fields gauging the center symmetries.

{\flushleft{\bf 1-form electric center symmetry:}}   The IR theory has an $[U(1)_e]^{N-1}$ emergent 1-form electric center symmetry, but this is not a symmetry of the UV  $SU(N)$ theory, which only has a
$\Z_N^{(1)}$ symmetry. Its reduction to $\R^3$  acts on topologically nontrivial (or infinite) Wilson loops in $\R^3$ by multiplication by a $\Z_N$ phase.
The $\Z_N^{(1)}$ symmetry of the UV theory can be embedded in the emergent electric center. To see the $\Z_N^{(1)}$ action, consider  the EFT objects that are charged under the global center symmetry. In the electric description of the theory, the charged objects are electric Wilson lines, $e^{i \oint\limits_C \vec A \cdot \vec \nu_B}$, which are the eigenvalues of fundamental Wilson loops along noncontractible (or infinite, in $\R^3$) loops $C$, and $B=1,...,N$ labels the eigenvalues. Under center symmetry, noncontractible loops are multiplied by $\Z_N$ phases. This can be achieved by postulating the global symmetry transformation\footnote{\label{centerfootnote}The choice of $\vec w_1$ for the direction of embedding of $\Z_N^{(1)}$ in $[U(1)_e]^{N-1}$ is not unique. Another  choice was  made in \cite{Cordova:2019uob,Tanizaki:2019rbk}, related to ours by a Weyl transformation.   }
\begin{equation}
\label{globalcenter1}
\Z_N^{(1)}: ~  \vec A \rightarrow \vec A  -N \vec w_1 \epsilon^{(1)},
\end{equation}
where $\epsilon^{(1)}$ is a closed 1-form (so that $\vec{F} = d \vec{A}$ is invariant)
obeying, along noncontractible loops,
\begin{equation}
\label{globalcenter2}
\oint\limits_C \epsilon^{(1)} = {2 \pi k \over N}, ~ k \in \Z, ~ d \epsilon^{(1)} = 0, 
\end{equation}
and $\vec w_1$ is the highest weight of the fundamental representation.
  Then, using $\vec w_1 \cdot \vec \nu_B = - {1 \over N} + \delta_{B1}$, the Wilson loop transforms as\begin{equation}
\label{globalcenter3}
\Z_N^{(1)}: e^{i \oint\limits_C \vec A \cdot \vec \nu_B} \rightarrow e^{- i {(\delta_{B1} - {1 \over N}) \oint\limits_C N \epsilon^{(1)}}} \; e^{i \oint\limits_C \vec A \cdot \vec \nu_B} =  e^{ i {2 \pi k\over N}} \; e^{i \oint\limits_C \vec A \cdot \vec \nu_B}~,
\end{equation}
precisely as desired for the 1-form $\Z_N^{(1)}$ center symmetry.

{\flushleft{\bf 0-form emergent magnetic center symmetry:} }$[U(1)_m]^{N-1}$, due to the fact that the currents $j^a = * F^a$ are conserved in the long-distance theory (\ref{LIRkinetic}) due to the Bianchi identities $d F^a=0$. One can introduce background 1-form gauge fields $a^a$ for this emergent symmetry via CS couplings of the form $k_{ab}  a^a \wedge * j^b =k_{ab} a^a \wedge F^b$. This symmetry is also broken by the UV $SU(N)$ dynamics.

{\flushleft{\bf ``0-form'' electric center symmetry:}} $\Z_N^{\S^1}$ is inherited from the 1-form center symmetry of the 4d $SU(N)$ theory (this could be called  ``$\S^1$-component of the 1-form $\Z_N$ symmetry of the $\R^4$ $SU(N)$ theory'' but for brevity we shall refer it to as above). $\Z_N^{\S^1}$ acts on the $\S^1$ fundamental Wilson loop $\Omega = {\rm diag}( \omega_1, \ldots \omega_N)$, with eigenvalues $\omega_A = e^{i L \vec A_3 \cdot \vec\nu_A}$ given in (\ref{holonomyevs}), upon multiplication by a $\Z_N$ phase. It is clear from (\ref{holonomyevs})  that shifting the holonomy by the weight vector $\vec w_1$ multiplies $\omega_A$ by a $\Z_N$ phase, i.e.
\begin{eqnarray}
\label{zeroformcenter}
\Z_N^{\S^1}: ~ L \vec A_3 &\rightarrow& L \vec A_3 - N \vec w_1 \epsilon^{(0)},  ~ {\rm with  } \;  \epsilon^{(0)} = {2 \pi   \over N}, \\
\omega_A &\rightarrow& e^{i {2 \pi   \over N}} \omega_A~, \nonumber
\end{eqnarray}
owing to $\vec\nu_A \cdot \vec w_1 = \delta_{A1} - {1 \over N}$. We   stress the analogy of these transformations to (\ref{globalcenter1}, \ref{globalcenter2}) (we have set $k=1$) stemming from the fact that the 0-form center is the component of the $\R^4$ 1-form center along the compact direction. 
Under the action of the 0-form center (\ref{zeroformcenter}), the $\S^1$ Wilson loop transforms as \begin{eqnarray}
\label{zeroformcenter1}
\Z_N^{\S^1}: ~ \tr \Omega^p  &\rightarrow& e^{i {2 \pi  p  \over N}} \tr \Omega^p,
\end{eqnarray}
as appropriate for a 0-form center symmetry  $\Z_N^{\S^1}$. We can use (\ref{zeroformcenter}) to find the  action of the 0-form center on the EFT field $\vec\phi$ of (\ref{defofphi}), $\vec\phi \rightarrow \vec\phi - 2 \pi   \vec w_1$.

The  action of the global 0-form center  from (\ref{zeroformcenter}), however, is not convenient for use in the EFT, because the shift of $L \vec A_3$ by $2 \pi \vec w_1$ brings the field outside its fundamental domain, also known as the Weyl chamber. To elucidate, we note that the masses of the non-Cartan $W$-bosons, see e.g. \cite{Aitken:2017ayq}, labeled by the positive-root vectors $\vec\beta^{AB}$, with $A < B$, are given by\footnote{For brevity, we omit the expectation value signs in the discussion that follows.}  $L m^{AB} = |L \vec A_3 \cdot \vec \beta^{AB} + 2 \pi z|$, where $z \in \Z$ is the Kaluza-Klein number.
As discussed in  \cite{Argyres:2012ka,Anber:2015wha}, the Weyl chamber is a connected region in the $\vec A_3$-space containing the center symmetric point $L \vec A_3 = {2 \pi \vec\rho \over N}$ such that no $W$-bosons become massless inside this region. Furthermore, points outside the Weyl chamber are gauge equivalent to the points inside the Weyl chamber.

Explicitly, the Weyl chamber  consists  of points in $\vec A_3$ space such that $\vec\alpha_p \cdot \vec A_3 >0$ ($p=1,...,N-1$) and $- \vec\alpha_N \cdot \vec A_3 < {2 \pi \over L}$. This is a set in the $N-1$ dimensional vector space  bounded by  $N$ hyperplanes---a triangle for $SU(3)$, tetrahedron for $SU(4)$, etc.  It is clear from the mass formula that at the boundaries of the Weyl chamber, some $W$-bosons become massless, causing the Coulomb branch EFT to break down. The center symmetric point $\vec A_3  L = 2 \pi\vec\rho/N$ is in the geometric center (of the triangle, tetrahedron, etc.) of the Weyl chamber. 
It is easy to see that shifting $\vec A_3 L$ by $-2 \pi\vec w_1$, as per  (\ref{zeroformcenter}), takes the field outside 
 the Weyl chamber. The shift also rearranges the heavy $W$-boson spectrum by mixing KK modes,\footnote{Using (\ref{defofphi}), the $W$-boson masses can be rewritten as ${L m^{AB} \over 2 \pi}$ =$|{B-A \over N}  +  \vec\phi \cdot \vec\beta^{AB} + z|$. Noting that $\vec w_1 \cdot \beta^{AB} =\delta^{A1}$,  (\ref{zeroformcenter}), acting as $\vec\phi \rightarrow \vec\phi - 2 \pi k \vec w_1$, changes this to $|{B-A \over N} +  \vec\phi \cdot \vec\beta^{AB} + z - k \delta^{A1}|$, showing the KK shift.} but this change is not detected by the EFT, since $W$-bosons are integrated out.

 It is preferable to have the 0-form center symmetry defined so that the symmetry-transformed field does not go outside the Weyl chamber. This is so especially in the presence of nonperturbative center-stabilizing potentials  for $\vec A_3$, where the vev determines the action of various semiclassical objects.
   This can be accomplished by supplementing (\ref{zeroformcenter}) by a  cyclic Weyl transformation $\cal P$, a discrete $SU(N)$ gauge transformation.
  Explicitly, it is described as follows:\footnote{The transformation of (\ref{zeroformcenter2}) maps the Weyl chamber to itself. It represents a discrete $\Z_N \in SO(N-1)$ rotation around the center-symmetric point. See \cite{Anber:2015wha} for a pictorial illustration for $SU(3)$. In  the standard $N$-component basis of the weight vectors, $\cal P$ is simply a cyclic shift of its components. }
\begin{eqnarray}
\label{zeroformcenter2}
\Z_N^{\S^1}: L \vec A_3 &\rightarrow& {\cal P}  L \vec A_3 - N   \vec w_1   \epsilon^{(0)}, ~{\rm with } \; \epsilon^{(0)} = {2 \pi \over N}~,\\
 \vec A_\mu &\rightarrow&{\cal P} \vec A_\mu   ~,\nonumber \end{eqnarray}
where, for an arbitrary vector in the weight space $\vec{v}$, 
\begin{eqnarray}
\label{zeroformcenter3}
{\cal P} \vec{v} &=& s_1 s_2 ... \ldots s_{N-1} \vec{v}, ~{\rm where} ~ s_p \vec{v} \equiv \vec{v} -  \vec\alpha_p (\vec\alpha_p \cdot \vec v)~.
\end{eqnarray}
 Put in  words, ${\cal P}$ is the product of Weyl reflections w.r.t. all simple roots, generating a $\Z_N$ subgroup of the Weyl group (recall that $|\vec\alpha_p|^2 = 2$). 
 The second line in (\ref{zeroformcenter2}), the action of $Z_N^{\S^1}$ on the  $\R^3$ 1-form gauge field, is due to the  discrete gauge transformation $\cal{P}$ to bring $\vec \phi$ in the fundamental domain \cite{Anber:2015wha}.
 For future use, we list the action of ${\cal P}$ on some important group-lattice vectors:
\begin{eqnarray}
\label{zeroformcenter4}
{\cal P} \vec\alpha_k &=& \vec\alpha_{k+1 ({\rm mod}N)}~, ~k = 1, ..., N,\nonumber \\
{\cal P} \vec\nu_A &=& \vec\nu_{A+1 ({\rm mod}N)}~, ~A = 1, ..., N,\nonumber \\
{\cal P} \vec w_l &=& \vec w_{l} - (\alpha_1 + ... +\alpha_l) ~, ~l=1,..., N-1, \\
{\cal P} {2 \pi \vec\rho \over N} &=& {2\pi \vec\rho \over N} - 2 \pi \vec w_1~. \nonumber
\end{eqnarray}

 With (\ref{zeroformcenter2}) instead of (\ref{zeroformcenter}), we find, using (\ref{zeroformcenter4}) and the fact that $\cal P$ is an $N-1 \times N-1$ orthogonal matrix acting in Cartan space,
 \begin{eqnarray}
 \label{zeroformcenter5}
\Z_N^{\S^1}: \omega_A  \rightarrow e^{- i {2 \pi \vec\nu_A \cdot \vec w_1}} e^{i  \vec\nu_A \cdot {\cal P}  \vec A_3 L} = e^{i {2\pi \over N}} e^{i {\cal P}^{-1} \vec\nu_A \cdot \vec A_3 L} =  e^{i {2 \pi \over N}}  \omega_{A-1 ({\rm mod} N)} ~,
\end{eqnarray}
which implies that $\tr \Omega^q \rightarrow e^{i {2 \pi q \over N}} \tr \Omega^q$, as in (\ref{zeroformcenter1}) (clearly,  this is expected---since (\ref{zeroformcenter}) and (\ref{zeroformcenter2}) differ by an $SU(N)$ gauge transformation, their action on gauge invariant operators should be identical). 

The detailed remarks here are useful for understanding the action of the global 0-form and 1-form center symmetries, where the parameters $\epsilon^{(0)}$ and $\epsilon^{(1)}$ obey (\ref{zeroformcenter2}) and  (\ref{globalcenter2}). When turning on backgrounds for these symmetries, especially in the magnetic dual of our EFT (\ref{LIRkinetic}), we shall revisit the center symmetry action, see Section \ref{sec:4.2.2} and find that using an embedding of $SU(N)$ into $U(N)$, as in the UV $SU(N)$ theory, is required.

\subsection{{}{Background fields}}
\label{sec:2.3}

To study anomalies, we will introduce background fields gauging the various global symmetries. 
Consider first background gauging of the 1-form global $\Z_N^{(1)}$ symmetry. One way\footnote{The other way is by explicitly introducing 2-form gauge backgrounds for the $\Z_N^{(1)}$ symmetry, in the continuum formalism of \cite{Kapustin:2014gua}. This will be used later in the paper, see Section \ref{sec:4.2.2}.}  to achieve this is to 
 introduce gauge  backgrounds from   a $SU(N)/\Z_N$ bundle, instead of a $SU(N)$ bundle, e.g. nontrivial 't Hooft fluxes \cite{tHooft:1979rtg}. Thus, in  backgrounds gauging the $1$-form $\Z_N^{(1)}$ symmetry,  the fluxes over noncontractible 2-cycles are $\oint \vec F =  2 \pi \vec w_k c^k$, while the monodromies of the scalars  $\vec\phi$ over 1-cycles are $\oint d \vec \phi =  \vec w_k c^k$, i.e. take values in the weight lattice, as opposed to the  root lattice for dynamical $SU(N)$ backgrounds.

  We also introduce nondynamical gauge backgrounds for the classical global $U(1)_A$ symmetry of SYM (and the $U(1)_A$ symmetry of the $A+R$ theory). For SYM, the restriction of the $U(1)_A$ field to a $\Z_{2N}^{(0)}$ field, following  \cite{Kapustin:2014gua}, will be described later. Thus, we denote by 
$B = B_\mu dx^\mu$ the 1-form field gauging the classical global  $U(1)_A$ symmetry of the fermions. We also introduce a constant $B_3$ background, which we denote by $W \equiv B_3$ (the vev of the $\S^1$ holonomy of the background field for the global $U(1)_A$ is then $ {L W}$). Gauge transformations act as $\Delta B_M = \partial_M \omega$, $M=0,1,2,3$. A large gauge transformation with $\omega = {2 \pi x^3 \over L}$ shifts $W$ by $2 \pi \over L$.  We can write the $U(1)_A$ background  as a 4d 1-form $B_{4d}$ defined as:
\begin{equation}
\label{axialbackground}
B_{4d} = B_\lambda(x) dx^\lambda + \left(W + B'_3(x)  \right) dx^3~.
\end{equation}
We indicated that the $\R^3$ 1-form $B_\lambda dx^\lambda$ and scalar $B'_3$, describing the fluctuation of $B_3$ around its vev $W$, depend on $x^\mu$. These are the $\R^3$-only dependent backgrounds that can be coupled to the KK zero-mode EFT.

Similarly to $U(1)_A$, we introduce background fields for the $U(1)_V$ symmetry of the $A+F$ ($A+R$) theory. We denote these by $V = V_\lambda dx^\lambda$. The expectation value of the $\S^1$ holonomy of $U(1)_V$ is denoted by $\mu L = \oint\limits_{\S^1} \langle V_3 \rangle dx^3$.   Under a large $U(1)_V$ transformation, $\mu$ shifts by $2 \pi \over L$. Similar to (\ref{axialbackground}), we write this background as a 4d 1-form $V_{4d}$ as
\begin{equation}
\label{vectorbackground}
V_{4d} = V_\lambda(x) dx^\lambda + \left(\mu + V_3'(x)  \right) dx^3~,
\end{equation}
where we denoted by $V_3'$ the $\R^3$-dependent fluctuation of $V_3$ around its vev $\mu$.

In our calculations, we included nontrivial Wilson lines for the $U(1)_{A/V}$ backgrounds  for several reasons. Firstly, as our main goal here is the study of anomalies,  $W$ and $\mu$  can be used to probe the   anomalies for large $U(1)$ transformations on $\S^1$, as they shift by $2 \pi \over L$ under such transformations. Further, a nonzero $U(1)_V$ holonomy $\mu$ is (in some cases, see Appendix \ref{appx:B})  necessary 
to avoid the presence of massless charged states in the long-distance EFT and thus avoid strong coupling in our EFT. 
Finally,  in the presence of nonzero Wilson lines the theory may have  topological IR phases and understanding of the topological terms induced when $W \ne 0$ would be an important ingredient in their study. This goes beyond our scope here, and we only note that the infrared dynamics may be substantially altered when $W \ne 0$.\footnote{\label{footnotedynamicsw}
As this goes outside our main topic, we only briefly comment on the dynamical effect of   nonzero $W$, which manifests itself  at small $L$ in various ways. In SYM, turning on  even an infinitesimal $W$ breaks supersymmetry and induces a ``GPY'' \cite{Gross:1980br} potential on the Coulomb branch. However, the strength of the potential is controlled by $|WL|$ and can be made arbitrarily small, so that to all orders in perturbation theory, the Coulomb branch can be treated as approximately flat. In the non-supersymmetric case, using center stabilization with  massive adjoint flavors, the effect of an infinitesimal $W$ on the one-loop potential is small, producing an infinitesimal shift away from the center symmetric vev. However, despite  the approximate preservation of the flatness of the Coulomb branch, or of center symmetric holonomy, $W$ can induce another more important effect: a quantized Chern-Simons term, giving topological mass to the  (previously massless) Cartan gauge bosons.  This topological mass is of order ${g^2 \over L}$, much larger than the exponentially small mass due to nonperturbative effects,  altering the IR dynamics.}

\section{{}{Topological terms and anomaly matching in the  $\R^3 \times \S^1$ electric EFT}}
\label{sec:3}

Having set our notation in the preceding Sections, we can now write down the results for the topological terms in our long-distance EFT,  generated by integrating out the massive components of the various fermions. 
 These terms depend  on the dynamical light fields, $\vec A, \vec \phi$, as well as on the background fields gauging the global symmetries, $B, W, V, \mu$ from (\ref{axialbackground}, \ref{vectorbackground}) (as appropriate for the given theory).
 
For both SYM and $A+F$ (and more generally, $A+R$) theories, we integrate out the non-Cartan components of the adjoint fermions, which obtain mass from the  nonzero expectation value of the holonomy (\ref{defofphi}), taken to be (near) center-symmetric. We also  integrate out the nonzero KK modes of the Cartan components of the adjoint fermions. In addition, in the presence of nonzero $W$, the KK zero-modes of the Cartan components obtain mass and we consider integrating them out, too.

For the $A+F$ ($A+R$) theory, we, in addition, integrate out the   Dirac fermion in an arbitrary representation $R$. It obtains mass due to the expectation value of the holonomy  and possibly $\mu \ne 0$.  

We defer to the Appendix for the many details of the derivation. In what follows, we only present and discuss the results. 
Regarding the calculation, here we stress the most salient point: we   employ  a regulator preserving the   gauge-  and 4d Lorentz symmetries and regulating all fermion loops---a set of three gauge invariant Pauli-Villars (PV) fields. The adjoint PV sector consists of three copies of adjoint Weyl fields with  gauge invariant Majorana masses. The three Dirac PV fields have the usual Dirac mass terms.  The masses and statistics of the three PV fields obey the relation  (\ref{PVmassrelation}) (the calculation of the triangle diagram of Section \ref{appx:A.5} is an example of  its utility).

We stress that the use of $\zeta$-function regularization in the calculation of topological terms can sometimes lead to misleading results. First, it is not a regulator for all graphs (e.g. the triangle, witness the calculation in Section \ref{appx:A.5}). Second,  the gauge invariant adjoint PV regulator does not respect periodicity   w.r.t. ${2 \pi}$ shifts of the Wilson line $WL$, due to the Majorana mass term (recall that these shifts correspond to large gauge transformations). However,    this periodicity  is preserved in a $\zeta$-function regularization. Thus, as we shall see below, the PV regulator  correctly reproduces the anomalies in large gauge transformations, while the $\zeta$-function does not.\footnote{The string-motivated literature on anomalies in circle compactifications  has already noted that ``$\zeta$-function is not enough'' \cite{Corvilain:2017luj,Corvilain:2020tfb}, essentially for the reasons mentioned.} 

\subsection{{}{Topological terms from a   Weyl adjoint and anomalies in SYM}}

\label{sec:3.1}

 The calculation of the one-loop two-point and triangle graphs, with heavy non-Cartan  adjoint fields in the loops, drawn on Figs.~\ref{fig:03}, \ref{fig:05}, \ref{fig:06a}, \ref{fig:08}, leads to topological terms depending on the fields gauging the global $U(1)_A$ ($B, W$ from (\ref{axialbackground})) and the light, compared with $1/(NL)$, dynamical Cartan subalgebra fields ($\vec{A}, \vec{\phi}$). We denote these adjoint-induced topological terms by $L^{adj}$:
\begin{eqnarray}
\label{adjointBF1}
L^{adj} &=& {N \over \pi}\; B \wedge \vec{F} \cdot\left(\langle \vec{\phi} \rangle + \vec\phi' + {1 \over 2 N}\sum_{\beta\in\beta^+} \vec\beta\;  n(m_\beta,W) \right) - {A^a \wedge F^b \over 4 \pi} \sum\limits_{\beta^+} {\rm sign} (W) \; n(m_\beta, W) \; \beta_a \beta_b  \nonumber\\
 & &  +\frac{1}{4\pi} B\wedge dB \; \text{sign}(W) \left(\left(N^2-1\right)\frac23 \frac{L\left|W\right|}{2\pi} - (N-1)\left\lfloor \frac{L\left|W\right|}{2\pi}\right\rfloor - \sum_{\beta\in\beta^+} n(m_\beta,W) \right).\end{eqnarray}
  This Lagrangian is one of the main results of this paper.
 As explained in detail in the Appendix, only heavy non-Cartan fields contribute to the terms in the first line of $L^{adj}$. The second line, $\sim B \wedge d B$, receives also contributions from the heavy Kaluza-Klein modes of the Cartan subalgebra fields (as they are charged under $U(1)_A$).  These terms will be important when we discuss the matching of the cubic $U(1)_A$ anomaly.
We also remind the reader that, as per (\ref{defofphi}),  $\vec\phi'$ is the fluctuation of the field $\vec\phi$ around its vev and that $\langle \vec\phi \rangle = 0$ at the center symmetric point.\footnote{As described in the Appendix, the two parts of the  BF-term in (\ref{adjointBF1}) come from different diagrams: the part containing the vev $\langle\vec\phi\rangle$  is from the two-point function Fig.~\ref{fig:05}, and the part containing the fluctuation $\vec\phi'$ is from the triangle graph, Fig.~\ref{fig:06a}. That $\langle\vec\phi\rangle$  and  $\vec\phi'$  appear as shown is a check on the somewhat laborious calculation of Appendices \ref{appx:A.4} and \ref{appx:A.5}. }

The topological terms (\ref{adjointBF1})  in our Cartan-subalgebra EFT appear rather involved and we now explain the various factors. We begin with the mass parameter $m_\beta$, defined as \begin{eqnarray}\label{mbeta}
m_\beta &\equiv& {\langle \vec{A}_3\rangle \cdot \vec\beta}= {2 \pi \over L} {\vec\rho \cdot \vec\beta \over N} +  {2 \pi \over L}{\langle \vec\phi \rangle\cdot \vec\beta}~.
\end{eqnarray}
Its physical significance is that $|m_\beta|$,\footnote{\label{footnotewilson}For vanishing $U(1)_A$ Wilson line, $W=0$: at the center symmetric point, if the $U(1)$ Wilson line is taken to be in $\Z_{N}$ ($W = {B_3 } = {2 \pi \over L}{p\over N}$, with $p \in \Z$), there are massless non-Cartan adjoint fermions and the long distance dynamics needs to be reconsidered (recall that the derivation of (\ref{adjointBF1}) assumes that fields of mass $\ge 1/(NL)$ are integrated out, producing local terms in the  EFT of the lighter modes).} with $\beta = \beta^{AB}$,  equals the mass of the $AB$-th $W$-boson. For example for $\beta = \beta^{AB}$ and at the center symmetric point, $\langle \vec\phi \rangle=0$, we find $m_{\beta^{AB}} ={2 \pi \over L} {B - A \over N}$; as the minimal value of $B-A$ is unity, we obtain the already quoted expression for the lightest $W$-boson (and its adjoint fermion superpartner) mass, $m_W = {2 \pi \over LN}$.

Next, the functions $n(m_\beta, W)$ and $n'(m_\beta, W)$ appearing in (\ref{adjointBF1}) are
\begin{eqnarray}\label{nfunctions}
n(m_\beta,W) &\equiv & {1 \over 2} \sum_{k \in \Z}\left[1 - {\rm sign}( |m_\beta + {2 \pi \over L} k| - |W|)\right], \\
n'(m_\beta,W) &\equiv &{1\over 2} \sum_{k \in \Z} {\rm sign}(m_\beta + {2 \pi \over L} k)\left[1 - {\rm sign}( |m_\beta + {2 \pi \over L} k| - |W|)\right]~.\nonumber
\end{eqnarray} 
We stress right away that $n$ and $n'$ both vanish when $W=0$ and that, despite the way they are written, the sums involve only a finite number of terms.  We also note that the values of $W$ when $n$ and $n'$ jump correspond to points where non-Cartan adjoint fermions become massless and a breakdown of the EFT occurs (as, e.g. in footnote \ref{footnotewilson}).

The function $n(m_\beta, W)$  takes only nonnegative integer values. It counts the number of integers $k$ such that $|W| >  |m_\beta + {2 \pi \over L}k|$. Under a large gauge transformation, $W \rightarrow W+{2 \pi \over L}$, it is easy to see from (\ref{nfunctions}) that \begin{eqnarray} 
\label{nfunctions2}
{\rm sign}(q(W+{2 \pi \over L})) \;n(m_\beta,q(W+{2 \pi \over L})) &\equiv & {\rm sign}(qW) \;n(m_\beta,qW) + 2 q .
\end{eqnarray}
Here, we slightly generalized the definition of $n$ (and of $n'$ below) to be used in future cases where the fields have integer charge $|q|>1$ under the relevant $U(1)$. For further use, we also note that (\ref{nfunctions2}) holds if $\vec\beta$ in (\ref{mbeta},\ref{nfunctions}) is replaced by a  weight $\vec\lambda$ of an arbitrary representation.

The function $n'(m_\beta,W)$ counts the number of integers with the same property as $n(m_\beta,W)$, but now weighted by the sign of $m_\beta + {2 \pi \over L}k$. It is easy to see that, as opposed to $n(m_\beta, W)$, $n'$ is a periodic function of $W$: 
\begin{eqnarray} 
\label{nfunctions3}
n'(m_\beta,q(W+{2 \pi \over L})) &\equiv &  n'(m_\beta, qW) ~.
\end{eqnarray}

The properties of the various terms in $L^{adj}$ just listed are important to understand how anomalies are reflected in the effective theory on $\R^3 \times \S^1$. We will next show that the effective Lagrangian (\ref{adjointBF1}), with $m_\beta$ from (\ref{mbeta}),  and $n$, $n'$  from (\ref{nfunctions}),  captures the  anomalies of the $\R^4$ theory. We will also show that  (\ref{adjointBF1}) implies that the  $\Z_{2N}$ subgroup of the classical chiral $U(1)_A$ symmetry is anomaly free. Further, it shows that there is a mixed anomaly between the $\Z_{2N}$ symmetry and the one-form $\Z_N^{(1)}$ center symmetry of the $SU(N)$ adjoint theory.

Before looking at 't Hooft anomalies, we  note that all terms in (\ref{adjointBF1}) are invariant under large or small gauge transformations in the unbroken Cartan subgroup of $SU(N)$. This is not a surprise, given that we used a gauge invariant regulator in the calculation. Nonetheless, as an explicit   check, we can also  see the gauge invariance of our topological Lagrangian (\ref{adjointBF1}). The only term that warrants some discussion is the Chern-Simons term for the   Cartan gauge fields.  Consider a gauge transformation $\Delta_\lambda A^a = d \lambda^a$,  and taking (for some arbitrary $p,q = 1,...,N-1$) $\oint d \lambda^a = 2 \pi \alpha_p^a$ and $\oint F^a  = 2 \pi \alpha_q^a$, as per our earlier discussion (after eq.~(\ref{defofphi})). The action corresponding to (\ref{adjointBF1}), $\oint L^{adj}$ changes  as
\begin{eqnarray}\label{gaugevariation1}
\oint \Delta_\lambda L^{adj} &=&  -\oint {d \lambda^a \wedge F^b \over 4 \pi} \sum\limits_{\beta^+} {\rm sign} (W) \; n(m_\beta, W) \; \beta_a \beta_b  \\
&=&- 2 \pi \sum\limits_{\beta^+} {\rm sign} (W) \; n(m_\beta, W) \; {\vec\beta \cdot\vec\alpha_p\; \vec\beta \cdot \vec\alpha_q} \in 2 \pi\; \Z  ~,\nonumber
\end{eqnarray}
showing that $e^{i \oint L^{adj}}$ is gauge invariant.\footnote{We also note that this term is gauge invariant in the $SU(N)/\Z_N$ theory, i.e. upon replacing $\vec\alpha_{p,q}$ with $\vec w_{p,q}$ in the argument above; this is relevant for maintaining gauge invariance in the 4d theory with gauged $\Z_{N}^{(1)}$.}

\subsubsection{{}{$U(1)_A$  anomaly from EFT vs. $\R^4$ theory} }
\label{sec:3.1.1}
Here, we start with the $\R^4$ anomaly and reduce on   $\S^1$, with the heavy non-Cartan components of the gauge fields omitted from the variation. Explicitly, we take the   4d field strength restricted to the Cartan subalgebra to be
\begin{equation}\label{F4d}
F^{a}_{4d} = d A^a + 2 \pi  d \phi^a \wedge {d x^3 \over L}~,
\end{equation}
where $A^a$ and $\phi^a$ are the $\R^3$-dependent fields defined in Section~\ref{sec:2.1}.
The variation of the $\R^4$ theory path integral measure under general $U(1)_A$ gauge transformations is, in such backgrounds,
\begin{eqnarray} \label{4dU1anomaly} 
\Delta_{U(1)} S_{4d} &=& -2 N \delta_{ab} \int\limits \omega \; {F^a_{4d} \wedge F^b_{4d} \over 8 \pi^2}~.
\end{eqnarray}
Let us first discuss the anomaly under large $U(1)_A$ gauge transforms. 
The Chern-Simons term in $L^{adj}$ (\ref{adjointBF1}) is the only term not invariant under $2 \pi \over L$ shifts of $W$, i.e. large $U(1)_A$ gauge transformations.\footnote{An earlier calculation by one of us \cite{Poppitz:2008hr}, using $\zeta$-function, obtained a different result for the CS coefficient, periodic under $W \rightarrow W + {2 \pi \over L}$, i.e. missing the effect of the anomaly of large-$U(1)_A$ transforms. Similar subtleties and the importance of using a gauge invariant regulator to define all divergent graphs have been discussed in the literature \cite{Corvilain:2017luj,Corvilain:2020tfb}.} Thus, denoting by $\Delta_W$ the change under a ${2 \pi \over L}$ shift of $W$, we find, using (\ref{nfunctions2}) with $q=1$:
\begin{eqnarray}\label{Wvariation1}
\int\limits_{\R^3} \Delta_W L^{adj} &=& -\int\limits_{\R^3} {A^a \wedge F^b \over 4 \pi} \sum\limits_{\beta^+} 2 \beta_a \beta_b =  -2 N \delta_{ab} \int\limits_{\R^3} {A^a \wedge F^b\over 4 \pi} ~.
\end{eqnarray}
We now show that this variation of (\ref{adjointBF1}) matches the anomaly of the $\R^4$ theory under large $U(1)_A$ transformation, $\Delta_W B = d \omega$, with $\omega = {2 \pi x^3 \over L}$ ($\oint d\omega = 2 \pi$). For such large $U(1)_A$ gauge transforms, $\oint\limits_{\S^1} d \omega = 2\pi$,  we obtain from (\ref{4dU1anomaly}), integrating by parts and  using $\epsilon^{3012}=-1$,\begin{eqnarray}
\Delta_{W} S_{4d}
&=&   2 N \delta_{ab} \int\limits_{\R^3 \times \S^1} d \omega \wedge {A^a \wedge dA^b \over 8 \pi^2} = - 2N\delta_{ab} \int\limits_{\R^3}  {A^a \wedge F^b \over 4 \pi},
\nonumber 
\end{eqnarray}
precisely matching the EFT variation (\ref{Wvariation1}). 

Next, we consider variations of (\ref{adjointBF1}) under $U(1)_A$ transforms on $\R^3$, $\Delta_w B_\mu = \partial_\mu \omega$, with $\omega$ only a function of $x^\mu$. Since the ($\vec{F}, \vec{A}_4 \sim \vec\phi$) fields  of our EFT also only depend on $\R^3$, we can integrate over $\S^1$ to obtain for the $\R^3$-dependent $U(1)_A$ variation, with $\vec{F}_{4d}$ from (\ref{F4d}):
\begin{eqnarray} \label{R4U1adj}
\Delta_\omega S_{4d} &=& -2 N \int\limits_{\R^3 \times \S^1} \omega \; {\vec{F}_{4d}  \wedge \vec{F}_{4d} \over 8 \pi^2} = -{N \over  \pi}  \; \int\limits_{\R^3}   \omega\; \vec{F}  \wedge d \vec{\phi}
\end{eqnarray}
As indicated, this precisely matches the variation of the ${N \over  \pi} B \wedge \vec{F} \cdot \vec{\phi}$ term in the EFT, as we show below (see  $\Delta_\omega\int\limits_{\R^3} L^{adj}$ computed in (\ref{R3U1adj})).

{\subsubsection{{}{Anomaly-free discrete $\Z_{2N}^{(0)}$  and mixed $\Z_{2N}^{(0)}$-$\Z_{N}^{(1)}$ anomaly}}}
\label{sec:3.1.2}

For the $U(1)_A$ variation of the EFT action under  $\Delta_w B = d \omega$, integrating by parts, we find
\begin{eqnarray}
\label{R3U1adj} 
\Delta_\omega \oint L^{adj} &=& {N \over  \pi} \oint d \omega \wedge \vec{F} \cdot \left(\langle \vec\phi \rangle + \vec{\phi}'\right) + {1 \over \pi} \oint d \omega \wedge \vec{F} \cdot \sum\limits_{\beta^+} \vec\beta \; n'(m_\beta, W)\nonumber \\
&=& - {N \over  \pi} \oint \omega \vec{F} \wedge d  \vec{\phi}'~,
\end{eqnarray}
which is the same as (\ref{R4U1adj}). We can then take the limit of a constant transformation parameter $\omega$, to get
\begin{equation}
-{N \omega \over \pi} \oint \vec{F} \wedge d \vec{\phi}'~.
\end{equation}
To see that $U(1)_A$ has an anomaly free subgroup, we now imagine that $\R^3$ is further compactified on a large three-torus. There, as discussed after (\ref{defofphi}), for an $SU(N)$ bundle, we have $\oint \vec{F} = 2 \pi \vec\alpha_p$ and $\oint d \vec\phi' =   \vec\alpha_q$, therefore
\begin{equation}\Delta_\omega \oint L^{adj} =  -2 N \omega \vec\alpha_p \cdot \vec\alpha_q \in - 2 N \omega \Z~. \label{znanomaly}
\end{equation}
Clearly, 
then, we can see that if $\omega = {2 \pi 
\over 2 N} k$, with $k \in \Z$, the path integral is invariant under $U(1)$ transformations---this is the EFT reflection of the $U(1) \rightarrow \Z_{2N}$ breaking of the classical $U(1)$ due to the anomaly.

Finally, we can also see the mixed anomaly between the $\Z_{2N}$ and the one-form center symmetry. For an $SU(N)/\Z_N$ bundle background, as discussed in Section \ref{sec:2.2}, we replace the root vectors in (\ref{znanomaly}) with weight vectors, and have instead, taking $\omega = {2 \pi \over 2 N}$, for  the $\Z_{2N}$ variation of the EFT
\begin{equation} \label{chiralcenter1}
\Delta_{\omega \in \Z_{2N}}\oint L^{adj} =  - 2 N \omega \vec w_p \cdot \vec w_q  =  - 2 \pi \vec w_p \cdot \vec w_q \in  2 \pi (-{p q \over N} + \Z)~~,
\end{equation}
showing that the $\Z_{2N}$ anomaly-free chiral symmetry is anomalous in an $SU(N)/\Z_N$ background.

\subsubsection{{}{\(U(1)_A^3\) anomaly matching from the EFT}}
\label{sec:3.1.3}

The classical \(U(1)_A\) symmetry of the adjoint theory also has a \(U(1)_A^3\) chiral anomaly.\footnote{As already discussed, in the purely adjoint SYM theory, only  a $\Z_{2N}$ subgroup of   $U(1)_A$ is a symmetry. It only becomes an anomaly free $U(1)$ symmetry in the $A+F$ (or $A+R$) theory. Thus, in the adjoint theory we should really be talking about a cubic $\Z_{2N}$ anomaly; indeed this cubic anomaly is contained in the $\sim B \wedge dB$ terms, see  \cite{Cordova:2018acb}. We can also consider the $SU(N)$  and $U(1)_A$ as global backgrounds in a free-fermion theory (gauging subgroups of $U(N^2-1)$), in which case we can discuss cubic $U(1)_A$ matching, in addition to its various mixed anomalies.}
 This anomaly is matched in the EFT, as we shall now discuss, by the local  \(\sim B\wedge dB\) terms in (\ref{adjointBF1}). There are, however, important subtleties, related to the matching of the anomaly under small vs. large $U(1)_A$ gauge transformations that are discussed below.
In the 4d theory, the \(U(1)_A^3\) 't Hooft anomaly under a shift \(\Delta B = d\omega\) is given by 
\begin{equation}\label{U13_anom}
\Delta_\omega S_{4d}= -\left(N^2-1\right) \int_{\mathbb{R}^3\times S^1}  \omega  \frac{dB_{4d}\wedge dB_{4d}}{24\pi^2}~,~ ~ B_{4d} \equiv B + (W + B'_3) dx^3~, ~~\langle B'_3 \rangle = 0,
\end{equation}
where the expression for the 4d 1-form $B_{4d}$ includes the 3d 1-form $B$ and the scalar $B'_3$, as in (\ref{axialbackground}).
First, let us consider small $U(1)_A$ gauge transformations on \(\mathbb{R}^3\). In order to see the matching of the cubic $U(1)_A$ anomaly from $L^{adj}$, we need to introduce the dependence of the \(B \wedge dB\) terms on the scalar field \( B'_3 \). As for   $\langle \vec\phi \rangle$ and $\vec\phi'$ in (\ref{adjointBF1}),  the $x^\mu$-dependent fluctuation of $B_3$ should appear added to its vev, which is $\sim W$ as per (\ref{U13_anom}). Further, in $L^{adj}$, the continuous dependence of $W$ appears only in the term  $\sim$sign$(W)|W| = W$, and we conclude\footnote{The careful reader may notice that---as opposed to the case of $\phi' B \wedge F$ term in (\ref{adjointBF1}) calculated in Appendix \ref{appx:A.5}---we did not calculate the triangle graph contributing to the $B'_3 B \wedge B$ coupling, but included if from consistency. We note, however, that the triangle graph calculation for the $U(1)_A^3$ anomaly for a Weyl fermion on $\R^3\times \S^1$ appears in \cite{Corvilain:2017luj}, with an identical result.} that $B'_3$ appears in  $\int_{\R^3}L^{adj}$ as (for brevity, denoting $B_3 \equiv W+B_3'$ in the next few equations)
\begin{equation}
S_{BdB} = \frac{1}{4\pi} \int_{\mathbb{R}^3} B\wedge dB \; \left[\left(N^2-1\right)\frac23 \frac{L B_3}{2\pi} - \chi\left(\left|W\right|\right)\right]~.
\end{equation}
Here we denoted the integer-valued $W$-dependent terms in (\ref{adjointBF1}) as \(\chi\left(\left|W\right|\right)\) for some cleanliness in this discussion.\footnote{Integer-valued terms are not relevant to the matching of the anomaly under small $\R^3$-dependent transformations. It is easy to check that $\chi(|W|)$ vanishes at $W=0$ and is continuous around $W=0$.} Under the small \(U(1)_A\) transformation, this term in $\int_{\R^3}L^{adj}$ transforms as 
\begin{equation}  
\begin{split}
\Delta_\omega S_{BdB} =  & \frac{1}{4\pi} \int_{\mathbb{R}^3} d\omega \wedge dB \; \left[\left(N^2-1\right)\frac23 \frac{LB_3}{2\pi} - \chi\left(\left|W\right|\right)\right]\\& = -\left(N^2-1\right)  \frac{L}{2\pi} \int_{\mathbb{R}^3} \omega \frac{ dB'_3 \wedge dB }{6\pi} ~.
\end{split}
\end{equation}
Here the second line follows from the first by integration by parts. \par

To compare with the 4d cubic anomaly, from equation (\ref{U13_anom}), we separate the \(B'_3\) component and integrate around the \(S^1\) to find
\begin{equation}
\Delta_\omega S_{4d} = -\left(N^2-1\right)\frac{L}{2\pi} \int_{\mathbb{R}^3} \omega \left(\frac{dB'_3 \wedge dB}{12\pi} + \frac{dB \wedge dB'_3}{12\pi} \right) = -\left(N^2-1\right) \frac{L}{2\pi} \int_{\mathbb{R}^3} \omega \frac{dB'_3 \wedge dB}{6\pi}
\end{equation} 
Thus, the variation of the topological terms in the EFT under small $\R^3$-dependent $U(1)_A$ gauge transformation matches the cubic $U(1)_A$ anomaly of the 4d theory. We stress again that this cubic $U(1)_A$  anomaly matching is due entirely due to massive modes that are integrated out of the EFT. The cubic $U(1)_A$ anomaly (under small gauge transformations) is represented in the EFT by local background-dependent topological terms, the $B \wedge dB$ terms in $L^{adj}$. The massless Cartan subalgebra fermions (the zero Kaluza-Klein modes) do not contribute to the cubic $U(1)_A$ anomaly.

Now, we consider the large $U(1)_A$ gauge transformations \(\omega = \frac{2\pi k}{L} x^3\) with \(k\in\mathbb{Z}\). With an integration by parts and \(\varepsilon^{3012} = -1\), the 4d variation according to the \(U(1)_A^3\) anomaly is
\begin{equation}\label{largea4d}
\Delta_W S_{4d} = -\left(N^2-1\right) \; \int_{\mathbb{R}^3} k \frac{B \wedge dB}{12\pi} ~.
\end{equation}
As before, these large gauge transformations are represented in the EFT by shifting \(W\) by \(\frac{2\pi k}{L}\). In order to properly match this in the EFT, we need to integrate out the remaining massless Cartan (Kaluza-Klein zero modes) fermions. In calculating (\ref{adjointBF1}), we did not integrate out the zero modes of the fermions in the Cartan subalgebra, since they only have mass from the background holonomy, \(W\), which could be taken arbitrarily small. However, when considering shifts of \(W\) of order \(\frac{2\pi}{L}\), these fermions cannot remain low mass and therefore must be integrated out. Moreover, the large gauge transformation rearranges the Kaluza-Klein momentum modes of the fermion fields, so they must all play a role in matching the anomaly. 

Including these extra fermions, as shown in Appendix \ref{appx:A.6.1}, results in the \(B \wedge dB\) term in (\ref{adjointBF1}) being replaced by
\begin{equation}
\frac{\text{sign}(W)}{4\pi} \left[\left(N^2-1\right)\frac23 \frac{L\left|W\right|}{2\pi} - (N-1)\left(\frac12 + \left\lfloor \frac{L\left|W\right|}{2\pi}\right\rfloor\right) - \sum_{\beta\in\beta^+} n(m_\beta,W) \right]  B\wedge dB~,
\end{equation}
where the contribution of the Cartan KK zero modes is the $(N-1){1\over 2}$ factor in the second term. 
Notice that under \(\Delta_W W =  \frac{2\pi k}{L}\) we have
\begin{equation}
\Delta_W \text{sign}(W)\left(\frac12 + \left\lfloor \frac{L\left|W\right|}{2\pi}\right\rfloor\right) = k~,
\end{equation}
and that the \(\frac12\) term is necessary for this shift in the case that \(W\) changes sign; this is how we can see the importance of integrating out all the fermions. Equipped with this identity and eq.~(\ref{nfunctions}), we can see that the whole term shifts as
\begin{equation}
 \frac{1}{4\pi} \int_{\mathbb{R}^3} B \wedge dB \left[\left(N^2-1\right) \frac23 k - (N-1) k - \left(\frac{N^2-N}{2}\right) 2k\right] = -\left(N^2-1\right)\int_{\mathbb{R}^3} k \frac{B \wedge dB}{12\pi}  ~.
\end{equation}
This matches exactly with the variation of the 4d action (\ref{largea4d}).

 More comments on the matching of the $U(1)_A^3$ anomaly are in Section \ref{sec:3.2.1}, pertaining to a theory where $U(1)_A$ is a genuine anomaly free global symmetry.

{\flushleft \bf Summary of Section \ref{sec:3.1}:}
The main result of this Section are the topological terms shown in (\ref{adjointBF1}), whose calculation is presented in detail in Appendix \ref{appx:A}. These terms are  induced by integrating out the KK modes of both the non-Cartan and Cartan fields.
Here, we showed that
the anomalies of the global symmetries in 4d and in the EFT on the Coulomb branch, governing distances greater than $LN$ on $\R^3 \times \S^1$, match as expected. The matching of the 't Hooft anomalies of all symmetries (continuous, discrete, and 1-form), except those under large $U(1)_A$ transformations, is due to the local background-field dependent terms of our $L^{adj}$ of eq.~(\ref{adjointBF1}). 

On the other hand, the matching of anomalies under large $U(1)_A$ transformations can not be achieved by the local terms given in $L^{adj}$. The Cartan fields, charged under $U(1)_A$ must then also be integrated out. That this is so can be understood from the nature of such transformations, which mix fields in the  EFT with heavy KK modes. In general  matching anomalies of large $U(1)_A$ transformations should not be expected from a purely EFT perspective.\footnote{We showed that the mixed $U(1)_A$-gauge anomaly under large $U(1)_A$ gauge transformations is matched by   the local terms already present in (\ref{adjointBF1}). In the Cartan subalgebra EFT, this is simply because the Cartan modes have no $U(1)^{N-1}$ charges. Nonetheless, even in this case, the large $U(1)_A$ involves a rearrangement of the KK modes of the Cartan fermions.}

\subsection{{}{Topological terms from a    Dirac fermion and  $A+F$ theory anomalies}}
\label{sec:3.2}

Next, we consider the contribution of a massless Dirac fermion  to the topological terms in the $\R^3 \times \S^1$ EFT and the matching of various anomalies.
The calculation of the topological terms  in Appendix \ref{appx:B} is done for a general representation $R$ and for a generic point on the Coulomb branch. For brevity, we limit our discussion here to the fundamental representation and to the center symmetric point $\langle \vec\phi \rangle = 0$.\footnote{This is only made for brevity of the presentation. The fact that  $\langle \vec\phi \rangle = 0$ and $\mu=0$ was used to obtain the $\vec w_{N\over 2}$ term in (\ref{fundBF1})---as we shall see later, it is the ${N \over2}$-th monopole-instanton where the fundamental zero modes are localized. When $\langle\vec\phi \rangle \ne 0$, the fundamental zero mode can shift to other monopole instantons and $\vec w_{N/2}$ has to be replaced by the appropriate one, see Appendix \ref{appx:B.2} and \cite{Poppitz:2008hr}.} 

In addition to a classical $U(1)_A$ symmetry, the Dirac fermion also has a vector $U(1)_V$. The corresponding background (\ref{vectorbackground}) consists of the $\R^3$ 1-form, $V = V_\lambda dx^\lambda$, and a scalar, $\mu + V_3'$, where $\mu$ denotes the constant part and $V_3'$ the fluctuation of the $\S^1$-component of the $U(1)_V$ gauge field.
As our goal is to discuss the $A+F$ theory, we now take the $U(1)_A$ axial charge of the Dirac fermion to be $-N$, so that the anomaly cancels that of the  adjoint. The $U(1)_V$ charge of the Dirac fermion can be taken to be unity, without loss of generality.

Performing a similar calculation to the adjoint case, with  gauge invariant PV regulators fields (see Appendix \ref{appx:B}), 
we find the following topological terms:
\begin{eqnarray}\label{fundBF1}
L^{fund} &=& -{N \over  \pi} B \wedge \vec{F} \cdot \left( \vec\phi' + { \vec\rho \over N} -  \vec w_{N\over 2} + {1 \over 2}  \sum\limits_{A=1}^N \vec\nu^A n'(m_A, NW) \right) \nonumber \\
&&
 + {A^a \wedge F^b \over 4 \pi} \sum\limits_{A=1}^N {\rm sign} (W) n(m_A, NW) \nu_a^A \nu_b^A \\
&&- \frac{N^2}{4\pi} B\wedge dB \; \text{sign}(W) \left[N^2 \frac{4}{3}\frac{L\left|W\right|}{2\pi} - \sum_{A=1}^N n(m_A,NW)\right] \nonumber \\
&&- \frac{LN^2}{2\pi^2} \left(\mu + V'_3\right) \;B\wedge dV  -  \frac{N}{2\pi} B\wedge dV \sum_{A=1}^N \left(n'(m_A,NW) - 2\left\lfloor\frac{Lm_A}{2\pi}\right\rfloor - 1\right) \nonumber\\
&&+ \frac{1}{4\pi} V \wedge dV  \sum_{A=1}^N \text{sign}(W) n(m_A,NW) \nonumber \\
& &+\frac{1}{2\pi} A^a \wedge dV  \sum_{A=1}^N \text{sign}(W) n(m_A,NW) \nu^A_a~. \nonumber
\end{eqnarray}
We stress again that, for brevity, we set $\langle \vec\phi \rangle =0$, assumed that $N$ is even, and that $\mu$ is infinitesimal---all reflected in the presence of the $N/2$-th fundamental weight in the first line above. A general
expression for an arbitrary representation $R$ can be found in (\ref{abcd},\ref{massfundam},\ref{generalrepvector}).
Here, $m_A = \mu+ {2 \pi \over L} {\vec\rho \cdot \vec\nu^A \over N}$, i.e. it is defined the same way as $m_\beta$ in (\ref{nfunctions}), but with the adjoint representation weight $\beta$ replaced by the $A$-th weight of the fundamental, $\vec\nu_A$, and shifted by the $U(1)_V$ holonomy $\mu$ (see also (\ref{massfundam})). The holonomy $\mu$ is chosen such that there are no massless components of the fundamental fermions at the chosen point on the Coulomb branch (here, the center-symmetric one).  The functions $n(m_A, NW)$ and $n'(m_A, NW)$ are as defined in (\ref{nfunctions}) and obey (\ref{nfunctions2}) and (\ref{nfunctions3}). 
One can show, like we did in the adjoint case, that this local Lagrangian matches all anomalies of the $U(1)_{A,V}$ symmetries of the fundamental Dirac theory, inlcluding those under large gauge transforms (as all components of the Dirac fermion are massive). We leave this straightforward exercise for the reader.  We shall provide more detail on anomaly matching in the framework of the $A+F$ theory below.

Now, we add $L^{adj}$ to $L^{fund}$, (\ref{adjointBF1}) and (\ref{fundBF1}), respectively, to obtain the following topological terms for the theory with a massless fundamental and a massless adjoint:
\begin{eqnarray}\label{faBF1}
L^{f+a} &=& -{1 \over  \pi} B \wedge \vec{F} \cdot \left( \vec\rho -  N \vec w_{N\over 2} \right)  \\ &&+{A^a \wedge F^b \over 4 \pi} \left[\sum\limits_{A=1}^N {\rm sign} (W) n(m_A, NW) \nu_a^A \nu_b^A - \sum\limits_{\beta^+} {\rm sign} (W) n(m_\beta, W) \beta_a \beta_b \right] \nonumber  \\
&&- {1 \over 2 \pi} B \wedge \vec{F}\cdot  \left[ \sum\limits_{A=1}^N  N \vec\nu^A n'(m_A, NW) - \sum\limits_{\beta^+} \vec\beta n'(m_\beta, W) \right] \nonumber \\
&&-{1 \over 4\pi} B \wedge dB \; \text{sign}(W) \;(2N^4-N^2+1)\; \frac23\; \frac{L\left|W\right|}{2\pi} \nonumber \\
&&+ {1 \over 4 \pi} B \wedge dB \; \text{sign}(W) \left[N^2 \sum_{A=1}^N n(m_A,NW) - \sum_{\beta\in\beta^+}n(m_\beta,W) - (N-1) \left\lfloor \frac{L\left|W\right|}{2\pi} \right\rfloor \right] \nonumber  \\
&&- \frac{LN^2}{2\pi^2} \left(\mu + V'_3\right) \;B\wedge dV  -  \frac{N}{2\pi} B\wedge dV \sum_{A=1}^N \left(n'(m_A,NW) - 2\left\lfloor\frac{Lm_A}{2\pi}\right\rfloor - 1\right) \nonumber\\
&&+ \frac{1}{4\pi} V \wedge dV  \sum_{A=1}^N \text{sign}(W) n(m_A,NW) \nonumber \\
& &+\frac{1}{2\pi} A^a \wedge dV  \sum_{A=1}^N \text{sign}(W) n(m_A,NW) \nu^A_a~. \nonumber
\end{eqnarray}
Eq.~(\ref{faBF1}) reflects the symmetries of the $A+F$ theory. First, the effective lagrangian is gauge invariant, as we used a gauge invariant regulator in the computation. Next, we notice that the continuously varying terms proportional to $\vec\phi'$ cancel between the adjoint (\ref{adjointBF1}) and fundamental (\ref{fundBF1}) contributions. These continuously varying terms are the ones that show the anomalous nature of the classical $U(1)_A$ symmetry of the adjoint theory, recall (\ref{R3U1adj}). 
Likewise, we note that the term multiplying the $U(1)^{N-1}$ Chern-Simons term is now periodic in $W$ with period $1$, as follows from the shift properties of $n$
from (\ref{nfunctions2}): the shifts, under $W \rightarrow W+{2 \pi \over L}$ of the two terms multiplying $A \wedge F$ cancel, yet again reflecting the absence of an anomaly in the $U(1)_A$ symmetry.

\subsubsection{{}{Matching the \(U(1)_A^3\) anomaly}}
\label{sec:3.2.1}

The matching of the $U(1)_A^3$ anomaly under $\R^3$-dependent small gauge transformations is reflected in the term $-{2 N^4 - N^2 +1 \over 6 \pi} \;\frac{LB'_3}{2\pi} \;B \wedge dB$, where we used the notation from (\ref{U13_anom}, \ref{vectorbackground}). The matching follows from the already discussed $U(1)_A^3$ anomaly matching for the adjoint theory, in addition to the fact that coefficient of this term precisely equals the contribution of $N^2-1$ components of the adjoint, of charge $1$, and the $2N$ Weyl components of the fundamental, of charge $-N$. It is easy to see that the anomaly under large $U(1)_A$ transforms, $W \rightarrow W + {2 \pi \over L}$, only matches if the Kaluza-Klein zero-modes of the adjoint are included, as explained at the end of Section \ref{sec:3.1} (and that it matches for the fundamental).

We now pause and  remark on the locality of the term matching the cubic $U(1)_A$ anomaly in the Coulomb branch EFT. On $\R^4$, it is well-known that anomalies can not be written as the variation of local counterterms. Instead, an  ``$1\over k^2$'' IR singularity, associated with massless particles \cite{Dolgov:1971ri,Frishman:1980dq,Coleman:1982yg}, is necessarily present  in the three-point function of the anomalous currents, corresponding to a nonlocal term in the effective action. To contrast this with our $\R^3\times \S^1$ case, we recall from (\ref{axialbackground}) that $B'_3$ is the fluctuation of $B_3$, and write the local term matching the  $(U(1)_A)^3$ anomaly in our EFT, dropping the overall constant, as $ \int\limits_{\R^3} B_3 L B \wedge dB$. For the background (\ref{axialbackground}), this can be also rewritten as  $\int\limits_{\R^3\times \S^1} d^4 x\; B_3\; \epsilon^{\mu\nu\lambda} B_\mu \partial_\nu B_\lambda$, showing that this term is   not 4d Lorentz invariant, even at large $\S^1$---an attempt to antisymmetrize all 4d  indices would make it vanish. 
The local  and 3d Lorentz invariant term $B'_3  B \wedge d B $ matches the anomaly under the restricted set of $U(1)_A$ gauge transformations in $\R^3$, $\delta B_\mu \sim \partial_\mu \omega$. We note that this cubic anomaly matching by local terms is also discussed  in \cite{Corvilain:2017luj}.

\subsubsection{{}{Matching the \(U(1)_AU(1)_V^2\) anomaly}}
\label{sec:3.2.2}

 Dirac fermions also have a global \(U(1)_V\) symmetry which participates in a mixed 't Hooft anomaly with the classical \(U(1)_A\) symmetry we have been exploring. If we introduce a background gauge field, \(V_M\), for the \(U(1)_V\) symmetry and perform a $U(1)_A$ transformation of the form \(\Delta_\omega B = d\omega\), we know that the 4d action shifts by
\begin{equation}\label{AV2_anom}
\Delta_\omega S_{4d} = \frac{N^2}{4\pi^2} \int_{\mathbb{R}^3\times S^1} \omega dV_{4d} \wedge dV_{4d}, ~ ~~ V_{4d} = V + (\mu + V_3') dx^3 ~.
\end{equation}
Adding the $U(1)_V$ field led to more topological terms in the EFT, the last three lines in (\ref{faBF1}). The   $B\wedge dV$ and $V \wedge dV$ terms in (\ref{faBF1}) reflect  this mixed anomaly (small and large axial transformations, respectively). 

For the small gauge transformations on \(\mathbb{R}^3\),  we apply \(\Delta_\omega B = d\omega\). After an integration by parts, in the $(\mu + V_3') B \wedge dV$ term in (\ref{faBF1}), we obtain 
\begin{equation}
\frac{LN^2}{2\pi^2} \int_{\mathbb{R}^3} \omega dV'_3 \wedge dV ~,
\end{equation}
precisely the variation we  get from integrating the 4d anomaly from eq. (\ref{AV2_anom}) around the \(S^1\). 

The large $U(1)_A$ gauge transformations shift \(W\) by \(\frac{2\pi k}{L}\) for some integer \(k\). These shifts affect the last two terms of eq. (\ref{faBF1}), according to eq. (\ref{nfunctions2}), to give a total shift of 
\begin{equation}
\frac{1}{4\pi} V \wedge dV  \sum_{A=1}^N 2Nk + \frac{1}{2\pi} A^a \wedge dV  \sum_{A=1}^N 2Nk \nu^A_a~.
\end{equation}
The latter term vanishes due to the fact that \(\sum_{A=1}^N \nu^A_a = 0\), while the former term simplifies to 
\begin{equation}\label{av2eft}
\frac{N^2}{2\pi} k \; V \wedge dV~.
\end{equation}
The 4d anomaly in this case, after an integration by parts, gives a shift of 
\begin{equation}\label{av2r4}
-\frac{N^2}{4\pi^2} \int_{\mathbb{R}^3\times S^1} d\omega \wedge V \wedge dV = \frac{N^2}{2\pi} k \int_{\mathbb{R}^3} V \wedge dV
\end{equation}
Here we used the facts that \(\oint_{S^1} d\omega = 2\pi k\) and that \(\varepsilon^{3012}=-1\). Thus, the \(U(1)_AU(1)_V^2\) anomaly of the $\R^4$ theory (\ref{av2r4}) is fully matched by the variation   (\ref{av2eft}) of the new terms of eq. (\ref{faBF1}).

 \section{{}{Topological terms in the magnetic dual EFT}}
 \label{sec:4}
 
In Section \ref{sec:3}, we discussed the ``kinematics''
of anomaly matching in the electric theory, obtained after integrating out the heavy non-Cartan and Kaluza-Klein degrees of freedom on $\R^3 \times \S^1$. The main result of that Section are the topological terms, $L^{adj}$ of (\ref{adjointBF1}), and $L^{f+a}$ of eq.~(\ref{faBF1}), which should be added to the free IR Lagrangian $L^{IR, kinetic}$ of (\ref{LIRkinetic}). We showed that these local topological terms represent the
't Hooft anomalies of the $\R^4$ theory in the Coulomb branch EFT. 

The most interesting dynamical aspect of the circle-compactified theories, however, has not been yet touched upon: the  nonperturbative effects that determine the true IR dynamics. These effects  make the IR look rather different from the free theory (\ref{LIRkinetic}). The nonperturbative semiclassical effects on $\R^3 \times \S^1$ are  understood to various degrees of detail, depending on the theory under consideration, and we shall not review them here; see \cite{Dunne:2016nmc} for references. The main point of essence to us is that the description of the nonperturbative IR effects requires another EFT,  magnetically dual to (\ref{LIRkinetic}).  

In this Section, we shall study how the topological terms and corresponding anomalies are reflected in the magnetic dual EFTs on the Coulomb branch. As our focus shifts from kinematics to dynamics,   we shall dispose of some of the background fields we turned on to study anomalies. In particular, as noted in 
Footnote \ref{footnotedynamicsw}, turning on a Wilson line for $U(1)_A$ can drastically change the dynamics, in a manner that has not been yet understood in detail. Thus, from now on we set the expectation value of the $U(1)_A$ holonomy $W=0$. Since the functions $n$ and $n'$ now both vanish, this ``kills'' many terms in (\ref{adjointBF1}) and (\ref{faBF1}), making these expressions more manageable. We now discuss the magnetic dual and the corresponding coupling of the background fields in the SYM and $A+F$ theories.

\subsection{{}{Topological terms in the magnetic $A+F$ EFT, $U(1)_A$-breaking and anomaly}}

\label{sec:4.1}

It turns out that it is easier to begin with a discussion of the magnetic dual of the $A+F$ theory. From the many terms in (\ref{faBF1}), the only remaining ones are\footnote{Assuming even $N$, $\mu=0$, and center symmetric gauge holonomy, making the integer-valued contribution to the $B \wedge dV$ term in (\ref{faBF1}) vanish.}
\begin{eqnarray}
\label{faBF2}
L^{f+a} &&= \\
&& -{1 \over  \pi} B \wedge \vec{F} \cdot \left( \vec\rho -  N \vec w_{N\over 2} \right)   -\frac{LN^2}{2\pi^2}  \;V'_3  \;B\wedge dV -{2N^4-N^2+1 \over 6\pi} \;\frac{LB'_3}{2\pi} \; B \wedge dB . \nonumber \end{eqnarray}
As already noted, in contrast to  the adjoint theory, the coefficient of the $B \wedge F$ term in the $A+F$  theory is constant, rather than a field-dependent quantity $\sim \vec\phi'$. This reflects the fact that  $U(1)_A$ is anomaly free in the UV theory and hence it is preserved in the IR theory obtained after integrating out the heavy non-Cartan fields. We also remind the reader that $V_3'$ and $B'_3$ above are proportional to the $x^\mu$-dependent fluctuations of the $x^3$-components of the $U(1)_V$ and $U(1)_A$ background gauge fields, introduced in (\ref{vectorbackground}) and (\ref{axialbackground}).  

For the following discussion, 
the presence of the constant $B\wedge F$ term in (\ref{faBF2}) is of utmost importance. It implies that the $U(1)_A$ current in the EFT has a topological contribution, $\sim \epsilon_{\mu\nu\lambda} F^{\nu \lambda}$, in addition to  the contribution of the massless Cartan fermions $\vec\psi$ from (\ref{LIRkinetic}) which are  charged under $U(1)_A$. Next, we recall our discussion in Section \ref{sec:2.2}.
In the  IR-free $U(1)^{N-1}$ 3d theory of the Cartan photons, there are 0-form ``magnetic center'' $U(1)^{N-1}$ symmetries, whose conserved currents are $\vec j_\mu = \epsilon_{\mu\nu\lambda} \vec F^{\nu\lambda}$. Background gauge fields $\vec{a}$ for these symmetries thus couple to the electric gauge fields via CS terms of the form ${a} \wedge {F}$.
The heavy-fermion generated BF term from (\ref{faBF2}),  $-{1 \over  \pi} B \wedge \vec{F} \cdot \left( \vec\rho -  N \vec w_{N\over 2} \right)$, induces an embedding of the $U(1)_A$ symmetry of the UV theory into the magnetic center of the  IR theory. The vector  $\vec\rho -  N \vec w_{N\over 2} $ fixes the direction of embedding.\footnote{As mentioned above and discussed in detail in Appendix \ref{appx:B.2}, the direction of this embedding depends on the point on the Coulomb branch and other background holonomies being considered.}

After a duality transformation, the constant $BF$ term implies that the dual photons shift, as they normally do under magnetic $U(1)$ symmetries, under a $U(1)_A$ transformation with parameter $\omega$, with the direction of the shift specified by the constant BF coupling from (\ref{faBF1}). Let us see this in detail for the present case.
We have, from (\ref{LIRkinetic}), (\ref{faBF1}), the part of the EFT Lagrangian containing the gauge and $U(1)_A$ fields 
\begin{eqnarray}\label{el3}
L(\vec F, \vec\sigma, B) =  - {L \over 4 g^2}\; \vec F_{\mu\nu} \cdot \vec F^{\mu\nu} - {1 \over 2 \pi} \;\epsilon^{\lambda\mu\nu} B_\lambda \vec F_{\mu\nu} \cdot \left( \vec\rho -  N \vec w_{N\over 2} \right) + {1 \over 4 \pi}\; \epsilon^{\lambda\mu\nu} \vec \sigma \cdot \partial_\lambda  \vec F_{\mu\nu}.
\end{eqnarray}
We can think of the $\vec\sigma$ field as a Lagrange multiplier field enforcing the Bianchi identity for $F$, where the last term implies that the periodicity of $\vec\sigma$ is in the weight lattice of $SU(N)$, $\vec\sigma \equiv \vec\sigma + 2 \pi \vec w_k c^k$, $c^k \in \Z$. Conversely, integrating out $\vec F_{\mu\nu}$, we have for the saddle point solution that we denote by $\vec F_{\mu\nu}^{\; 0}$:
\begin{eqnarray}\label{fzero}
\vec F_{\mu\nu}^{\; 0} = -{g^2 \over 2 \pi L} \; \epsilon_{\mu\nu\lambda} \left[ \partial^\lambda \vec\sigma + 2 B^\lambda (\vec \rho - N \vec w_{N\over 2}) \right]~.
\end{eqnarray}
Substituting (\ref{fzero}) back in (\ref{el3}), the IR lagrangian after a duality transform is
\begin{eqnarray}\label{el44}
L(\vec\sigma,B) &=& {L \over 4 g^2} \; \vec F_{\mu\nu}^0 \cdot \vec F^{ \mu\nu \; 0} \\
&=& {g^2 \over 8 \pi^2 L} [\partial_\lambda \vec\sigma + 2 B_\lambda (\vec \rho - N \vec w_{N\over 2}) ] \cdot [\partial^\lambda \vec\sigma + 2 B^\lambda (\vec \rho - N \vec w_{N\over 2}) ]~.\nonumber
\end{eqnarray}
Most importantly, the $U(1)_A$ invariance of this Lagrangian implies a the shift of the dual photon field under the $U(1)_A$:
\begin{eqnarray} \label{sigmashift}
B_\mu &\rightarrow&B_\mu + \partial_\mu \omega~, \nonumber \\
\vec\sigma &\rightarrow& \vec\sigma - 2 \omega (\vec\rho - N \vec w_{N\over 2})~.
\end{eqnarray}
The relevance of the loop-induced BF term to determining the transformation properties of  the dual photon under the $U(1)$ is identical to how it has been argued to work in $\R^3$ theories, where the $U(1)$ symmetry also does not suffer from an anomaly  \cite{Aharony:1997bx}. To the best of our knowledge,  a calculation and a   description of the role of the one-loop BF terms due to the heavy fermions on $\R^3 \times \S^1$ has not yet been given in the literature.

The perturbative contributions to the  magnetic dual EFT lagrangian is thus given by (\ref{el44}), supplemented by the kinetic terms for the Cartan fermions $\vec\psi$ from (\ref{LIRkinetic}). Depending on whether we consider scales above or below $m_{\vec\phi} \sim g \sqrt{N} m_W$ (recall Section \ref{sec:2.2}), we can include the $\vec\phi$ fields' kinetic term from (\ref{LIRkinetic}) as well as their ``GPY'' potential, see \cite{Unsal:2008ch,Shifman:2008ja}. The background field dependent local  terms from (\ref{faBF2}), $\sim V_3'\;  V \wedge dV, B'_3\; B \wedge dB$,  that match  the various $U(1)_{V/A}$  anomalies should also be included.

As discussed above, the electric and magnetic perturbative lagrangians are equivalent and which one is used is a matter of choice. However,  the IR physics is not captured correctly unless 
one  adds the contribution of nonperturbative effects. To leading order in semiclassics they are due to the $N$ monopole instantons on Coulomb branch of the $SU(N)$ theory on $\R^3 \times \S^1$, see \cite{Davies:2000nw} and references therein. The description of these nonperturbative effects is only local in the magnetic dual language. 

There are $N$ monopole instantons labeled by the simple and affine roots of $SU(N)$, which we shall collectively label by $\vec\alpha_A$, $A=1,...,N$, where $\alpha_N = - (\alpha_1 +...+\alpha_{N-1})$ is the affine root. In the long distance theory, they create local 't Hooft vertices. Their effect can be summarized in the following nonperturbative Lagrangian:\footnote{The 't Hooft vertices are given up to  overall dimensionless and $g^2$-dependent constants inessential for us. We have also written them at the center-symmetric point, i.e. we have integrated out $\vec\phi$.  We note that the 't Hooft vertex involving the fundamental Dirac fermion $\theta, \tilde\theta$ should really be integrated out, since it involves heavy fields with mass  of order $1/L$; it is left in the EFT only to show consistency with the symmetry (\ref{sigmashift}). A detailed discussion in a related case is in \cite{Poppitz:2013zqa}.}
\begin{eqnarray}\label{fathooft}
L^{a+f}_{n.p.} &&= \\
&&  {e^{- {8 \pi^2 \over g^2 N}} \over L} {\rm Re} \left(\sum\limits_{A \ne {N\over 2}} e^{i \vec\alpha_{A } \cdot \sigma} (\vec\psi\cdot\vec\alpha_{A})^2 + 
 e^{i \vec\alpha_{N\over 2} \cdot \vec\sigma} \;(\vec\psi\cdot\vec\alpha_{N\over 2})^2 \;L^2 {\theta \tilde\theta  }\right) +  {e^{- {16 \pi^2 \over g^2 N}} \over L^3} \;  \hat V_{\rm bion}(\vec\sigma). \nonumber
\end{eqnarray}
The parameter of the nonperturbative semiclassical expansion is $e^{- {8 \pi^2 \over g^2 N}}$ and we have  not explicitly written the second order term, the potential for the dual photons generated by magnetic bions, the dimensionless function $\hat V_{\rm bion}(\vec\sigma)$; see Footnote \ref{bionfootnote} and \cite{Poppitz:2009tw}. The presence of fermions in the 't Hooft vertex is due to the fact that the monopole instantons have fermion zero modes. Every monopole instanton has two adjoint zero modes,  proportional to $\vec\psi\cdot\vec\alpha_A$, and only the ${N \over 2}$-th monopole-instanton has fundamental zero modes, one for each Weyl component, $\tilde\theta, \theta$, of the Dirac fermion \cite{Poppitz:2008hr}.
 
As per (\ref{Lferm}), $\vec\psi$ are the adjoint fermions' Cartan components (with 3d canonical dimension)  transforming as $\vec\psi \rightarrow e^{i \omega} \vec\psi$ under $U(1)_A$. Likewise, $\theta$ and $\tilde\theta$ are the two-component Weyl fermions comprising the 4d fundamental Dirac fermion (this four component fermion is denoted by $\theta_i^k$ in (\ref{Diraclagrangian})) of axial charge $-N$. Thus, under $U(1)_A$,  $\theta \rightarrow e^{-i N \omega} \theta$, and similar for $\tilde\theta$. Combining the $U(1)_A$ transformations of $\vec\psi$, $\theta$, $\tilde\theta$,  with the $U(1)_A$ shift of the dual photon (\ref{sigmashift}), $\vec\sigma \rightarrow \vec\sigma - 2 \omega (\vec\rho - N \vec w_{N\over 2})$, and using the identities $\vec\alpha_A \cdot \vec\rho = 1 - N \delta_{AN}$, $\vec\alpha_A \cdot \vec w_{N\over 2} = \delta_{A {N\over 2}} - \delta_{AN} $, it is straightforward to see that the   nonperturbative terms  in (\ref{fathooft}) with fermion zero modes are invariant. 

The magnetic bion induced potential also inherits the shift symmetry of $\vec\sigma$.  
 We shall not dwell on the dynamics behind magnetic bions that generate $V_{\rm bion}$, see \cite{Unsal:2007vu,Unsal:2007jx, Anber:2011de} for a discussion of bions and \cite{Poppitz:2009tw} for this mixed-representation theory.\footnote{\label{bionfootnote}Briefly, as opposed to QCD(adj), here there are only $N-2$ magnetic bions:  $\hat V_{\rm bion}$ is  of the form 
 $\sum_{A\ne\{{N\over 2}, {N\over 2} -1\}} {\rm Re}(e^{i (\vec\alpha_A - \vec\alpha_{A+1}) \cdot \vec\sigma})$, with $N+1\equiv 1$. Explicitly, one can see that this gives mass to $N-2$ of the dual photons, leaving one field massless, due to the shift symmetry.}  
Even without the details, it is clear that, barring a symmetry reason, all components of  the $\vec\sigma$ fields will obtain mass from $\hat V_{\rm bion}$, except for a single component along the $ \vec\rho -  N \vec w_{N\over 2} $ direction, which remains massless due to the $U(1)_A$ shift symmetry (\ref{sigmashift}). We shall simply call this field $\sigma$, defining it by $\vec\sigma\vert_{massless} = 2 (\vec\rho - N \vec w_{N\over 2}) \sigma$. Then,  the deep-IR (governing dynamics below any mass scale) theory is that of a  massless $\sigma$, $\sigma \rightarrow \sigma - \omega$ under $U(1)_A$,  coupled to the background fields for $U(1)_A$ and $U(1)_V$:
\begin{eqnarray}
\label{faBF3}
L^{f+a}_{{\rm deep \; IR}} &=&   {g^2 c \over 2 \pi^2 L} |d  \sigma +  B|^2   -\frac{LN^2}{2\pi^2}  \;V'_3  \;B\wedge dV -{2N^4-N^2+1 \over 6\pi} \; \frac{LB'_3}{2\pi} \; B \wedge dB + \ldots 
\end{eqnarray}
where $\ldots$ denote higher-dimensional derivative terms involving the Goldstone field $\sigma$ and we defined $c = |\vec\rho - N \vec w_{N\over 2}|^2$. 

The fact that the dual photon is a Goldstone boson has been known since \cite{Affleck:1982as}, in the context of $\R^3$-theories. Here, this phenomenon emerges in a 4d context in circle compactification.
 In the long-distance theory, the $U(1)_A$ breaking can be viewed as arising from magnetic bion effects, with $\hat V_{\rm bion}$ having a flat direction along the $\sigma$ direction. However, the Goldstone field here does not participate in the matching of the $U(1)_A^3$ anomaly (by coupling to a Wess-Zumino term, as in the chiral lagrangian of QCD, or, for the theory at hand, the chiral lagrangian in the $U(1)_A$-broken phase of the 4d $A+F$ theory).  Instead, the cubic $U(1)_A$  (as well as the mixed $U(1)_A U(1)_V^2$) anomaly is matched by the purely local term (\ref{faBF3})   depending on background fields only and  induced by integrating out the heavy fermions along the Coulomb branch of the circle compactified theory. 
 
 We also note that the last two terms in (\ref{faBF3}) can be cast into the form of  Wess-Zumino terms upon replacing $B \wedge dV \rightarrow d \sigma \wedge dV$ and $B \wedge dB \rightarrow d \sigma \wedge dB$, since the transformation of the background $B$ and dynamical $d \sigma$ are identical. These terms then would have the form of the Wess-Zumino coupling of a Goldstone boson in 4d, dimensionally reduced to 3d.
  However, it is the local, background-field-only dependent form (\ref{faBF3}) of these terms that arises upon integrating out the fundamental. We stress that integrating out the fundamental and adjoint fermion KK modes, which lead to these local terms in (\ref{faBF3}), is consistent in the EFT framework. The  dynamics leading to the formation of the magnetic bions generating the potential for $\vec\sigma$ does not involve KK mode exchanges---bions are due solely to the monopole-instantons with only adjoint zero modes, whose exchange is calculable within the EFT.

\subsection{{}{Topological terms in the magnetic dual SYM EFT}}
\label{sec:4.2}
 
  The main differences between SYM and the $A+F$ theory discussed above involve subtleties 
associated with the explicit breaking $U(1)_A \rightarrow \Z_{2N}^{(0)}$ by the anomaly. In addition, SYM has a 1-form $Z_{N}^{(1)}$ symmetry with a (new) mixed anomaly with the chiral symmetry, whose matching in the calculable dynamics we aim to understand in explicit detail. 

Proceeding as in Section \ref{sec:4.1}, we set the expectation value of the $U(1)_A$ holonomy $W=0$, and find that the topological terms generated by integrating the non-Cartan and Kaluza-Klein modes of the adjoint fermion (\ref{adjointBF1}) become \begin{eqnarray}
\label{adjointBF2}
L^{adj} &=& {N \over \pi}\; B \wedge \vec{F} \cdot\left(\langle \vec{\phi} \rangle + \vec\phi' \right) 
+ \frac{N^2 -1}{6\pi} \;  \frac{LB'_3}{2\pi} \; B\wedge dB~.
 \end{eqnarray}
We momentarily retained the $x^\mu$-dependent fluctuation $B'_3$ of the $U(1)_A$ background $B_3$ (defined in (\ref{U13_anom})), which enters the term capturing the cubic $\Z_{2N}^{(0)}$ anomaly. As this term  depends on the background fields only, we shall  ignore it in our subsequent discussion of dynamics; the matching of this cubic anomaly is then similar to the matching of $U(1)_A^3$ discussed at the end of  Section \ref{sec:4.1}.\footnote{With all the subtleties related to replacing the $U(1)_A$ gauge field by a $\Z_{2N}^{(0)}$ one. We only offer a brief comment, related to our introduction of a $\Z_{2N}$ continuum gauge field   by replacing $B$ with $B^{(1)}$ defined in (\ref{zngauge}). One might be then tempted to conclude that (\ref{zngauge}) implies that $ \oint d B^{(1)} = 0$ and the cubic $\Z_{2N}$ anomaly can not be seen from (\ref{adjointBF2}). However, with $b^{(0)}$ the phase of a charge-$2N$ Higgs field, as discussed in Section \ref{sec:4.2.1}, one can introduce (in addition to codimension-1 defects with the $b^{(0)}$ phase changing by $2\pi$ upon crossing them, ensuring $\oint B^{(1)} = {2 \pi \over 2N}$) also monodromy defects where  the Higgs vev vanishes and where $d B^{(1)} \ne 0$. See \cite{Wang:2014pma} for  details; for us, it is important to note that the $B^{(1)}$ field obeys $\oint B^{(1)}= {2 \pi \over 2N}$ and $\oint dB^{(1)} = 2 \pi \Z$ on appropriate cycles, i.e. it is ``almost'' flat, with its curvature integrating to $2 \pi$ on closed cycles. With such backgroundss,  the cubic $\Z_{2N}$ anomaly can be inferred from (\ref{adjointBF2}). We also note that ultimately, this requires an underlying regularization, e.g. lattice, or the introduction of a new scale associated with the Higgs vev. For a related discussion, see  \cite{Cordova:2018acb}, as well as the more recent work on global anomalies \cite{Hsieh:2018ifc,Garcia-Etxebarria:2018ajm}.}
 
To study the matching of anomalies, we want to introduce background fields for $\Z_{2N}^{(0)}$ and $\Z_{N}^{(1)}$ in the magnetic dual Lagrangian on the Coulomb branch of SYM. In order  for the presentation to be less crowded, we consider turning on these backgrounds one by one.

\subsubsection{{}{Turning on   background gauge fields for the $\Z_{2N}^{(0)}$ chiral symmetry}}
\label{sec:4.2.1}

To restrict the $U(1)_A$ field to a $\Z_{2N}^{(0)}$ field, we shall use the continuum formalism for $\Z_{2N}$ gauge fields \cite{Kapustin:2014gua}. Thus, we consider a pair of a 1-form, $B^{(1)} = B$, the field appearing in (\ref{adjointBF2}),  and 0-form, $b^{(0)}$, fields, obeying
\begin{eqnarray} \label{zngauge}
2N B^{(1)} = d b^{(0)}
\end{eqnarray}
with transformation laws $B^{(1)} \rightarrow d \omega^{(0)}, b^{(0)} \rightarrow 2N \omega^{(0)}$, where $\omega^{(0)}$ has quantized  periods, $\oint d \omega^{(0)} = 2 \pi \Z$ over closed 1-cycles (one can think of $\R^3$ compactified on, e.g. a large  three torus). The gauge fields also have normalized integrals over  closed one-cycles, $\oint B^{(1)} = {2\pi \Z \over 2N}$ and $\oint d b^{(0)} = 2 \pi \Z$.   

Intuitively, one can think of the 0-form part $b^{(0)}$ of the 1-form $\Z_{2N}$ gauge field ($B^{(1)}, b^{(0)}$)   in (\ref{zngauge})  as the phase of a charge-$2N$ (under the $U(1)$ gauge field $B^{(1)}$) field Higgsing $U(1) \rightarrow \Z_{2N}$. In the IR, below the Higgsing scale, this leads to the above continuum description of a $\Z_{2N}$ gauge theory. Thus, a $2 \pi$ shift of $b^{(0)}$ represents an unbroken $\Z_{2N}$ gauge transformation. On the other hand, the gauge invariant object $e^{i \oint\limits_C B^{(1)}} = e^{i {2 \pi \Z \over 2N}}$, with $C$ a closed loop,  represents a $\Z_{2N}$-gauge field Wilson line; it can be thought of as a product of $\Z_{2N}$-valued link fields along the closed loop, as naturally occur in the lattice formulation of the theory.

Further, we replace $B = B^{(1)} = {d b^{(0)} \over 2N}$  in (\ref{adjointBF2}), integrate by parts, and add the kinetic terms for the Cartan photons and $\vec\phi$ from (\ref{LIRkinetic}) to obtain, as we did near   (\ref{el3}):
\begin{eqnarray}\label{el456}
L(F,\sigma, B, \phi)&&=\\
&&  - {L \over 4 g^2}\; \vec F_{\mu\nu} \cdot \vec F^{\mu\nu} - {1 \over 4 \pi} b^{(0)} \; \epsilon^{\lambda\mu\nu}  \vec F_{\mu\nu} \cdot \partial_\lambda  \vec\phi  + {1 \over 4 \pi} \epsilon^{\lambda\mu\nu} \vec \sigma \cdot \partial_\lambda  \vec F_{\mu\nu} + {2 \pi^2 \over g^2 L}\; \partial_\mu \vec\phi \cdot \partial^\mu \vec\phi~.\nonumber
 \end{eqnarray}
 We notice that this lagrangian (or, rather $e^{i \oint L}$) is invariant under $\omega = 2\pi \Z$ shifts of $b^{(0)}$, $b^{(0)} \rightarrow b^{(0)}+ 2 \pi k$, $k \in \Z$, provided the $SU(N)$ quantization conditions for $F$ and $d\phi'$ (both periods are in the root lattice) are obeyed. The upshot is that introducing background gauge fields for the $\Z_{2N}^{(0)}$ discrete chiral symmetry in SYM amounts to the introduction of a background theta angle $b^{(0)}$ and that $\Z_{2N}^{(0)}$ transformations correspond to $2 \pi$ shifts of this theta angle, consistent with our interpretation of $b^{(0)}$ after (\ref{zngauge}).\footnote{\label{footnoteeasy}We note that (\ref{el456}) could be obtained more directly as follows. One can introduce a background $b^{(0)}$ derivatively coupled in the kinetic terms of the 4d fermion lagrangian. Then, redefining the fermions by a $e^{i {b^{(0)} \over 2 N}}$ phase gives rise to the theta term in (\ref{el456}) and  to an IR theory of $\Z_{2N}^{(0)}$-invariant fermions.}
 
 Proceeding with the duality, now from (\ref{el456}), we find, instead of (\ref{fzero}), the saddle point value $\vec F_{\mu\nu}^{\; 0}$:
\begin{eqnarray}\label{fzero1}
\vec F_{\mu\nu}^{\; 0} = -{g^2 \over 2 \pi L} \; \epsilon_{\mu\nu\lambda} \left[ \partial^\lambda \vec\sigma +   b^{(0)} \; \partial_\lambda \vec\phi \right]~, 
\end{eqnarray}
and a  bosonic kinetic term in the dual theory, instead of (\ref{el44}),
\begin{eqnarray}\label{el45}
L(\vec\sigma,B,\phi) &&= \\
&& {g^2 \over 8 \pi^2 L} [\partial_\lambda \vec\sigma +   b^{(0)}  \partial_\lambda \vec\phi  ] \cdot [\partial^\lambda \vec\sigma + b^{(0)} \partial_\lambda \vec\phi]~+ {2 \pi^2 \over g^2 L}\; \partial_\mu \vec\phi \cdot \partial^\mu \vec\phi~.\nonumber 
\end{eqnarray}
Since (\ref{el456}) is $\Z_{2N}^{(0)}$ invariant, we find that under $\Z_{2N}^{(0)}$ shifts of $b^{(0)}$, the dual photon shifts as
\begin{eqnarray}\label{sigmashiftdiscrete}
\Z_{2N}^{(0)}: b^{(0)} &\rightarrow& b^{(0)} + 2 \pi \nonumber, \\ \vec\sigma &\rightarrow&  \vec\sigma - 2 \pi \vec\phi ~.
\end{eqnarray}

Let us now discuss how 't Hooft vertices for monopole instantons are consistent with the $\Z_{2N}^{(0)}$ shifts of $b^{(0)}$ and $\vec\sigma$. For the subsequent discussion, it is convenient (in fact, necessary, due to the form the 't Hooft vertex, see (\ref{thooftSYM}) below) to include the center symmetric vev back into the definition of the scalar (\ref{defofphi}).  Thus, we define 
\begin{equation}\label{varphi}
\vec{\varphi} \equiv {\vec\rho \over N} + \vec\phi= {L \vec{A}_3 \over 2 \pi}.
\end{equation} Since $\vec\varphi$ and $\vec\phi$ differ by a constant shift, the effective kinetic Lagrangian (\ref{el45}) for $\vec\varphi$ is obtained by replacing $\vec\phi \rightarrow \vec\varphi$, i.e. the derivative term for $\vec\sigma$ now reads 
$\partial_\lambda \vec\sigma +   b^{(0)}  \partial_\lambda \vec\varphi$. It is also consistent to replace   the transformation of $\vec\sigma$ (\ref{sigmashiftdiscrete})  (this is, in fact, necessary, to ensure invariance of the 't Hooft vertex (\ref{thooftSYM}))   with 
\begin{eqnarray}\label{sigmashiftdiscrete1}
\Z_{2N}^{(0)}: b^{(0)} &\rightarrow& b^{(0)} + 2 \pi \nonumber, \\ \vec\sigma &\rightarrow&  \vec\sigma - 2\pi   \vec\varphi .
\end{eqnarray}
In the presence of a background theta angle, the bosonic factor in the 't Hooft vertex of the $A$-th monopole-instanton is proportional to
\begin{equation}
\label{thooftSYM}
e^{ - {8 \pi^2 \over g^2 } \delta_{AN}} e^{- {8 \pi^2 \over g^2} \vec\alpha_A \cdot \vec\varphi + i \vec\alpha_A \cdot(\vec\sigma + {b^{(0)} } \vec\varphi)} ~.
\end{equation}
This form was obtained  already in ref.~\cite{Davies:2000nw},\footnote{We stress that, when comparing with that reference, the reader should be mindful of  Footnote \ref{footnoteconvention}.} which kept an explicit $\theta$ dependence in the massless theory. A quick argument in favour of the combination $\vec\sigma + b^{(0)} \vec\varphi$ appearing in the exponent of the 't Hooft vertex is that it is needed to correctly reproduce long-range monopole-instanton interactions via the $\vec\sigma$ and $\vec\varphi$ propagators from   (\ref{el45}).   Clearly, the bosonic 't Hooft vertex factor (\ref{thooftSYM}) is invariant under the $\Z_{2N}^{(0)}$ shift of $b^{(0)}$ and $\vec\sigma$ of (\ref{sigmashiftdiscrete1}).\footnote{\label{footnotedym}In passing, we note that the factor (\ref{thooftSYM}) should be used  in the bosonic dYM at nonvanishing $\theta$ angle, upon replacing $b^{(0)} \rightarrow \theta$ and with an additional multiplicative factor $e^{i \theta \delta_{AN}}$ included, whenever the theory is studied at length scales above $m_{\vec\phi}$.  Explicitly, the dYM 't Hooft vertex of the $A$-th monopole instanton is proportional to
 $$
 e^{( - {8 \pi^2 \over g^2 } + i \theta)\delta_{AN}} e^{- {8 \pi^2 \over g^2} \vec\alpha_A \cdot \vec\varphi + i \vec\alpha_A \cdot(\vec\sigma + \theta  \vec\varphi)} 
 $$
 This ensures that the effective theory incorporating monopole-instanton vertices retains invariance under $2 \pi$ shifts of the $\theta$ angle, governed by eq.~(\ref{sigmashiftdiscrete1}), also at scales where fluctuations of $\vec\varphi$  are relevant. The usual form of the monopole-instanton vertex used in dYM, which is the same  for all $A=1,\ldots,N$ \cite{Unsal:2008ch,Thomas:2011ee}  is obtained from (\ref{thooftSYM}) after integrating out $\vec\varphi$, by (to leading order) substituting its vev $\vec\rho\over N$ and using $\vec\alpha_A \cdot \vec\rho = 1 - N \delta_{AN}$. We stress    that, as opposed to dYM, in SYM there is no energy range where one can integrate out $\vec\varphi$ but keep $\vec\sigma$, hence we are forced to consider the form (\ref{thooftSYM}). }

As for the fermionic part of the 't Hooft vertex, we offer two points of view. Recall that fermion zero modes in monopole-instanton backgrounds multiply the 't Hooft factors (\ref{thooftSYM}) by $(\vec\alpha_A \cdot \vec\psi)^2$, as in the terms with two adjoint zero modes in (\ref{fathooft}).   Clearly, to be consistent with (\ref{sigmashiftdiscrete1}), the fermion zero mode insertions have to separately be $\Z_{2N}^{(0)}$-invariant. If we view our effective theory along the lines of footnote \ref{footnoteeasy}, the fermions $\vec\psi$ are already chirally invariant. Alternatively, if we follow our integration of KK modes and non-Cartan fields in the original $\Z_{2N}^{(0)}$-variant fermion basis (as  in Appendix \ref{appx:A}), the fermion zero-mode insertions should  be  $e^{- i {b^{(0)} \over N}} (\vec\alpha_A \cdot \vec\psi)^2$ instead, with the exponential factors obtained after solving for the Dirac zero-modes in the combined monopole-instanton and $b^{(0)}$ background.

In the next Section, we turn on a background for the center symmetry. Then, one can use the 't Hooft vertex (\ref{thooftSYM}) and the remarks of this and the following Sections to write the complete Lagrangian for the SYM Coulomb branch EFT in the global symmetry backgrounds. 

\subsubsection{{Turning on  background gauge fields for the center and chiral symmetries}}
\label{sec:4.2.2}

We now proceed to turning on the center symmetry backgrounds. 
We follow the formalism of \cite{Kapustin:2014gua} and the recent work on gauging center symmetry in  dYM  \cite{Tanizaki:2019rbk}.
The 4d theory 1-form electric center symmetry, when reduced on $\S^1$, splits into a 3d 1-form $\Z_{N}^{(1)}$, eqns.~(\ref{globalcenter1}, \ref{globalcenter2}) and 0-form $\Z_N^{\S^1}$ symmetries, eqn.~(\ref{zeroformcenter2}), as per our discussion at the end of Section \ref{sec:2.2}. 

To gauge these symmetries, we introduce two pairs $(C^{(2)}, C^{(1)}$)---a 2-form field gauging $\Z_N^{(1)}$, and $(D^{(1)}, D^{(0)})$---a 1-form field gauging $\Z_N^{\S^1}$. These obey the relations and transformation laws
\begin{eqnarray}\label{gaugingcenter}
\Z_N^{(1)}: N C^{(2)} &=& d C^{(1)}, \;\delta C^{(2)} = d \lambda^{(1)}, \;\delta C^{(1)} = N \lambda^{(1)}, \\
\Z_N^{\S^1}: N D^{(1)} &=& d D^{(0)}, \;\delta D^{(1)} = d \lambda^{(0)}, \;\delta D^{(0)} = N \lambda^{(0)}, \nonumber
\end{eqnarray}
with $\lambda^{(1)}$ and $\lambda^{(0)}$ the 1-form and 0-form transformation parameters. Notice that the fields gauging the 0-form center $\Z_N^{\S^1}$ transform are similar to the ones gauging the chiral $\Z_{2N}^{(0)}$ of eq.~(\ref{zngauge}) and obey similar conditions when integrated over closed 1-cycles, $\oint D^{(1)} = {2 \pi \Z \over N}$, $\oint d D^{(0)} = 2 \pi \Z$. Likewise, when integrated over closed 2-cycles, the 2-form $\Z_N$ gauge field obeys  $\oint C^{(2)} = {2 \pi \Z \over N}$, $\oint d C^{(1)} = 2 \pi \Z$.
We now want to couple these fields to the point and line operators charged under the 1-form and 0-form center symmetries, respectively. 

The action of the global $\Z_N^{(1)}$ and $\Z_N^{\S^1}$ 1-form and 0-form center symmetries was discussed in Section \ref{sec:2.2}, in a manner that used an embedding of these symmetries into the $U(1)^{N-1}$ emergent center symmetry of the Coulomb-branch EFT, used earlier in \cite{Cordova:2019uob,Tanizaki:2019rbk}. Here, we shall revert to the description in terms of an embedding in $U(N)$. This has an advantage of a more transparent action of the center symmetries, especially when the full SYM lagrangian is considered. The gauging of the center symmetry in the UV $SU(N)$ theory  in the continuum is also   described  by an $U(N)$ embedding.
 
 To establish notation we embed $SU(N)$ into $U(N)$ whose Cartan generators we take to be the $N\times N$ matrices $\hat H^A$ with matrix elements $(\hat H^A)_{BC} = \delta_{AB} \delta_{BC}$.  In the Coulomb branch theory, we have $U(N) \rightarrow U(1)^N$, and
 the $N\times N$ matrix valued Cartan gauge field is then $\hat A = A_B \hat H^B$, where a sum over $B = 1 \ldots N$ is understood. On the other hand, the $SU(N)$ Cartan matrices are $H^a$, $a=1,\ldots, N-1$, with $(H^a)_{BC} = \nu^a_B \delta_{BC}$, obeying $\tr H^a H^b = \delta^{ab}$. We add an extra generator $H^0$, $(H^0)_{BC} = {1 \over \sqrt{N}} \delta_{BC}$, obeying $\tr H^0 H^a = 0$ and $\tr H^0 H^0 = 1$, and have the relation $A_B \hat H^B = \vec{A} \cdot \vec{H} + A_0 H^0$. 
More explicitly, from the above definitions,  the relation between the $U(1)^N$ Cartan fields and the physical $N-1$ fields that were used so far in the paper (and denoted by $A^a$ or $\vec{A}$): 
\begin{eqnarray}\label{relations1}
A_B &=& \sum\limits_{a=1}^{N-1} \nu_B^a A^a + {1 \over \sqrt{N}} A_0 = \vec\nu_B \cdot \vec A + {1 \over \sqrt{N}} A_0~, \nonumber \\
\vec A &=& \sum\limits_{B=1}^N A_B  \vec\nu_B~ , ~A_0 =  {1 \over \sqrt{N}} \sum\limits_{B=1}^N A_B~,
\end{eqnarray}
where deriving the last two equations used $\sum\limits_{B=1}^N   \nu_B^a \nu_B^b = \delta^{ab}$, and $\sum\limits_{B=1}^N  \vec \nu_B = 0$. 
The relations of $\vec F$, $F_0$ to $F_A$ are obtained from (\ref{relations1}) by replacing $A$ with $F$.

 The $\Z_N^{(1)}$ 1-form center symmetry acts on the $U(1)^N$ fields  
 \begin{eqnarray}
 \label{center6}
\Z_N^{(1)}:  \hat A \rightarrow \hat A + \lambda^{(1)}~,
\end{eqnarray}
where $\lambda^{(1)}$ is a 1-form with  quantized periods ($\oint d \lambda^{(1)} = 2\pi \Z$). As we are gauging the center symmetry, we do not require that $\lambda^{(1)}$ be closed, as opposed to our earlier discussion (\ref{globalcenter1}), where we also imposed (\ref{globalcenter2}).
The trace of the $k$-th power of the fundamental Wilson loop 
\begin{eqnarray}
\label{wilson6}
\tr W^k = \tr e^{i k \oint\limits_C A_B \hat H^B} =  \sum\limits_{B=1}^N e^{i k \oint\limits_C A_B}~, ~{\rm under} \;  \Z_N^{(1)}: \tr W^k  \rightarrow e^{i k \oint\limits_C \lambda^{(1)}} \tr W^k~,
\end{eqnarray}
is not invariant under $\Z_N^{(1)}$. To remedy this, we attach a surface operator (if $C$ is noncontractible, $S$ has to end on another Wilson loop)
\begin{eqnarray}\label{wilsoninvt6}
\tr W^k \; e^{-i k \int\limits_{S, \partial S = C} C^{(2)}}~.
\end{eqnarray}
The $C^{(2)}$ transformation law of (\ref{gaugingcenter}) along with (\ref{wilson6}) ensures that the operator (\ref{wilsoninvt6}) is invariant under the gauged center symmetry.
As we have enlarged the gauge group of the EFT to $U(1)^N$, we impose a gauge and 1-form center invariant constraint on the field strengths of the $U(1)^N$ gauge fields $A_B$, $F_B = d A_B$, which we write, for now, as a delta function in the path integral, 
\begin{equation}\label{constraint6}
\delta (\sum\limits_{B=1}^N F_B - N C^{(2)})~.
\end{equation}
This constraint descends from the corresponding constraint in the $U(N)$ UV theory to the EFT upon integrating out  the massive off-diagonal gauge bosons.\footnote{We  note that one often works with a local solution of the constraint (\ref{constraint6}), identifying $A_0 = {C^{(1)} \over \sqrt{N}}$, or $A_B = \vec\nu_B \cdot \vec{A} + {C^{(1)} \over N}$, as follows from (\ref{relations1}). Our goal is, however, to perform a duality transformation and integrate over the electric variables in the path integral. Thus, we do not make such an identification.}
 
{\flushleft{\bf Relation to global $\Z_N^{(1)}$ of Section \ref{sec:2.2}:} }{\small Before we continue, it  is instructive to consider the relation between the gauged $\Z_N^{(1)}$ transformations considered here and the global ones of Section~\ref{sec:2.2}. The global limit of (\ref{center6}) consists of taking a closed 1-form  parameter, $\lambda^{(1)} = \epsilon^{(1)}$, $d \epsilon^{(1)} = 0$,  with $\oint\limits_C \epsilon^{(1)} = {2 \pi \over N}$, as in (\ref{globalcenter1},\ref{globalcenter2}). Consider now the global $\Z_N^{(1)}$ transformation of (\ref{globalcenter1}), $\vec A \rightarrow \vec A - N \vec w_1 \epsilon^{(1)}$. Using (\ref{relations1}), we rewrite (\ref{globalcenter1}) as $A_B \rightarrow A_B + \epsilon^{(1)} - \delta_{B1} N \epsilon^{(1)}$. This is the same as (\ref{center6}) with $\lambda^{(1)}=\epsilon^{(1)}$, {\it except} for the  shift of $A_1$ by the closed form $-N \epsilon^{(1)}$. This shift of $A_1$  integrates to $2 \pi$ over noncontractible loops $C$ and does not affect any Wilson loops, hence it is not part of the emergent center symmetry on the Coulomb branch. Its effect is that of a large gauge transformation (along the noncontractible loop $C \in T^3$) in the $\hat H_1$ direction of $U(1)^N$.
We conclude that reducing the global limit of (\ref{center6})   to (\ref{globalcenter1}) requires an appropriate large gauge transformation in the unbroken group (a similar interpretation holds for the 0-form part of the global center symmetries, (\ref{zerocenter6}) vs. (\ref{zeroformcenter}), as discussed below). The advantage of (\ref{center6}) is that it is valid throughout the RG flow and does not require an embedding of $\Z_N^{(1)}$ in the emergent center symmetry, as in \cite{Cordova:2019uob,Tanizaki:2019rbk}.}
 
{\flushleft{Continuing}} with introducing gauge backgrounds for the 1-form center, the $\Z_N^{(1)}$-invariant kinetic terms of the $U(1)^N$ gauge bosons are
\begin{eqnarray}
\label{gaugebosonkinetic6}
L_{gauge} = - {L \over 4 g^2} \sum\limits_{B=1}^N (F_{B \; \mu\nu} - C_{\mu\nu}^{(2)}) (F_{B}^{\mu\nu} - C^{(2) \; \mu\nu}) ~.
\end{eqnarray}
The relations (\ref{relations1}) imply that this kinetic term is equivalent to the one for the $\vec{F}$ gauge field from (\ref{LIRkinetic}), while the kinetic term for $F_0$ is trivial in the theory with gauged center symmetry.

The discussion for the scalars and the gauging of $\Z_N^{\S^1}$  is almost identical. Instead of the $N-1$ scalars $\varphi$ of (\ref{varphi}) we now have the $N$ scalars $\varphi_A$, for which
\begin{eqnarray}\label{scalarrelations1}
\varphi_B &=& \sum\limits_{a=1}^{N-1} \nu_B^a \varphi_a + {1 \over \sqrt{N}} \varphi_0 = \vec\nu_B \cdot \vec\varphi + {1 \over \sqrt{N}} \varphi_0~, \nonumber \\
\vec\varphi &=& \sum\limits_{B=1}^N \varphi_B  \vec\nu_B~,~ \varphi_0 = {1 \over \sqrt{N}} \sum\limits_{B=1}^N \varphi_B.
\end{eqnarray}
The scalars  shift under $\Z_N^{\S^1}$,
\begin{equation}\label{zerocenter6}
\Z_N^{\S^1}: ~\varphi_B \rightarrow\varphi_B+ {1 \over 2 \pi} \lambda^{(0)}~,
\end{equation}
where, as opposed to (\ref{zeroformcenter}), $\lambda^{(0)}$ now is not constant.
Next, consider the trace of the $k$-th power of the $\S^1$ holonomy
\begin{eqnarray}
\label{polyakov6}
\tr \Omega^k = \sum\limits_{B=1}^N e^{i k 2 \pi  \varphi_B}~,  ~{\rm under} \; \Z_N^{\S^1}: \tr \Omega^k  \rightarrow e^{i k \lambda^{(0)}} \tr \Omega^k~,
\end{eqnarray}
and, similar to (\ref{wilson6}), is not invariant under the gauged $\Z_N^{\S^1}$. To remedy this, we attach a line  operator, with the line $\ell$ ending on the operator (taken to be at $x \in \R^3$):
\begin{eqnarray}\label{polyakovinvt6}
\tr \Omega^k \; e^{-i k \int\limits_{\ell, \partial \ell = x} D^{(1)}}~.
\end{eqnarray}
That the line operator (\ref{polyakovinvt6}) is invariant follows from the $D^{(1)}$ transformation (\ref{gaugingcenter}).
 The constraint in the path integral corresponding to (\ref{constraint6}) reads
\begin{equation}\label{scalarconstraint6}
\delta (\sum\limits_{B=1}^N d \varphi_B - {N \over 2 \pi} D^{(1)})~.
\end{equation}
and the kinetic term for the holonomy scalars is 
\begin{eqnarray}
\label{scalarkinetic6}
L_{scalar} =  {2 \pi \over g^2 L} \sum\limits_{B=1}^N (\partial_\mu \varphi_{B} - {1\over 2 \pi} D_{\mu}^{(1)}) (\partial^\mu \varphi_{B} - {1\over 2 \pi} D^{(1) \; \mu})  ~.
\end{eqnarray}

{\flushleft{\bf Relation to global $\Z_N^{\S^1}$ of Section \ref{sec:2.2}:} }{\small Let us comment on the relation between the global $\Z_N^{\S^1}$ transformation considered in Section~\ref{sec:2.2},  eqn.~(\ref{zeroformcenter}), and the gauged 0-form center transformation (\ref{zerocenter6}), similar to our discussion of $\Z_N^{(1)}$. In the $\varphi_B$ basis, recalling (\ref{varphi}) and using (\ref{scalarrelations1}), we can rewrite (\ref{zeroformcenter}) as $\varphi_B \rightarrow \varphi_B + { \epsilon^{(0)} \over 2 \pi} - \delta_{1B} {N \epsilon^{(0)} \over 2 \pi}$. This is the same as (\ref{zerocenter6}) with a constant parameter $\lambda^{(0)} = \epsilon^{(0)} = {2 \pi \over N}$, {\it except} for the integer shift of $\varphi_1$. Since the fields  have unit periodicity, $\varphi_B \equiv \varphi_B + 1$  (as is  evident from (\ref{polyakov6})), this shift corresponds to an $x^3$-dependent large gauge transformation in the  $\hat H_1$ direction of the unbroken $U(1)^N$. Thus, in the global limit, the  transformation (\ref{zerocenter6}) reduces to  (\ref{zeroformcenter})  after an appropriate large gauge transformation.

The reader may further wonder about the image of the transformations (\ref{zeroformcenter2}) (which differ from (\ref{zeroformcenter}) by a Weyl group transformation) in the $\varphi_B$, $A_B$ variables. They are given by $\varphi_B\rightarrow \varphi_{B-1 ({\rm mod}N)} + { \epsilon^{(0)} \over 2 \pi} - \delta_{1B} {N \epsilon^{(0)} \over 2 \pi}$ and $A_B \rightarrow A_{B-1 ({\rm mod}N)}$. In other words, the additional cyclic Weyl transformation $\cal{P}$ of (\ref{zeroformcenter4}) acts as a cyclic shift on $\varphi_B, A_B$.}

{\flushleft{The}} chiral-symmetry background dependent term from (\ref{el456}), obtained by integrating out the fermions is, using the relations (\ref{relations1}):
\begin{eqnarray}
\label{bzeroterm}
L_{top.} &=&  - {b^{(0)} \over 4 \pi} \vec F_{\mu\nu} \epsilon^{\mu\nu\lambda} \partial_\lambda \vec\phi \\
& =& - {b^{(0)} \over 4 \pi} \left[ \sum\limits_{B=1}^N F_{B \; \mu\nu} \epsilon^{\mu\nu\lambda}  \partial_\lambda \varphi_B  -{1 \over N} (\sum\limits_{B=1}^N F_{B \; \mu\nu}) \epsilon^{\mu\nu\lambda}(\sum\limits_{C=1}^N \partial_\lambda \varphi_C)   \right]~.\nonumber
\end{eqnarray}
This loop-induced term is invariant under both $\Z_N^{(1)}$ and $\Z_N^{\S^1}$ center-symmetries, as can be explicitly seen from the above expression. Moreover, the loop-induced term is the same as obtained earlier since the adjoint fermions do not couple to the overall \(U(1) \subset U(N)\) nor the 
$(C^{(2)}, C^{(1)})$ and $(D^{(1)}, D^{(0)})$ backgrounds. 

To see that $L_{top.}$, in the form shown in the second line in (\ref{bzeroterm}),  reproduces the mixed chiral-center anomaly upon a $2 \pi$ shift of $b^{(0)}$, one needs to solve the constraints (\ref{constraint6}) and (\ref{scalarconstraint6}) and use the solutions that correspond to the insertion of nontrivial 't Hooft fluxes. A simple example\footnote{There are $N$ independent solutions of the constraint (\ref{constraint6}) of minimal energy: only one of the $F_A$ is taken to be nonzero, equal to $NC^{(2)}$ (since $\oint F_A = 2 \pi \Z$, there can not be smaller fluxes in any of the $N$ $U(1)$ factors).  This gives rise to $N$ degenerate classical ground states of the finite-volume (e.g. compactified on a large $T^3$) theory in a  't Hooft flux background, say one with only $C^{(2)}_{12} \ne 0$  \cite{Unsal:2020yeh}. 
Monopole-instantons are tunnelling events between these degenerate ground states.  As this goes beyond our main theme, we do not pursue it here.} is to take a solution of the constraints with the only nonzero backgrounds $F_1 = N C^{(2)}$ and $d \varphi_1 = {N\over 2 \pi} D^{(1)}$. The relations (\ref{relations1})   show that this corresponds to the insertion of  a usual 't Hooft flux configuration embedded in the $\vec{A}$ Cartan part of $SU(N)$: $\vec{F} = \vec\nu_1 N C^{(2)}$, with an additional 't Hooft flux embedded in the $A_0$ field, ${F_0\over \sqrt{N}} = C^{(2)}$.  The $\Z_N$ phase in the mixed anomaly is now due to the $1/N$ term in (\ref{bzeroterm}).\footnote{Equivalently, on the surface of the constraint, we can rewrite (\ref{bzeroterm}) as
\begin{eqnarray}
\label{bzeroterm1}
L_{top.} 
& =& - {b^{(0)} \over 4 \pi} \left[ \sum\limits_{B=1}^N F_{B \; \mu\nu} \epsilon^{\mu\nu\lambda}  \partial_\lambda \varphi_B  - N C^{(2)}_{\mu\nu} \epsilon^{\mu\nu\lambda}  D^{(1)}_\lambda \right]~,\nonumber
\end{eqnarray}
but in using this form, it is understood that  $F_B$ and $\phi_B$ satisfy (\ref{constraint6}) and 
(\ref{scalarconstraint6}), respectively. The anomaly under $2\pi$ shifts of $b^{(0)}$ is due to  the last, center-background dependent term.}

As a final step before performing a duality transformation, we enforce the Bianchi identity for the $U(1)^N$ fields $F_A$ by introducing $N$ dual photon fields $\sigma_A$ and coupling them to $F_A$ in an $\Z_N^{(1)}$-invariant manner. We also add a Lagrange multiplier field $u_\mu$ to enforce the constraint (\ref{constraint6}):
\begin{eqnarray}\label{dsigmaF}
L_{lagr.} = - {1 \over 4 \pi} \sum\limits_{A=1}^N \partial_\lambda \sigma_A \epsilon^{\lambda\mu\nu} F_{A \; \mu\nu} + {1 \over 4\pi} u_\lambda \epsilon^{\mu\nu\lambda}  \sum\limits_{B=1}^N (F_{B \; \mu\nu} -  C^{(2)}_{\mu\nu}).
\end{eqnarray}
The $U(1)^N$ fields $F_A$ have quantized fluxes, $\oint F_A = 2 \pi \Z$, hence the fields $\sigma_A$ are $2\pi$ periodic; $\Z_N^{(1)}$ invariance of (\ref{dsigmaF}) follows from the quantization of the $\sigma$ and $\lambda^{(1)}$ periods. Thus, now we have the bosonic Lagrangian, combining (\ref{gaugebosonkinetic6}, \ref{scalarkinetic6}, \ref{bzeroterm}, \ref{dsigmaF})
\begin{eqnarray}\label{ltotal}
L_{bos} &=&\nonumber \\
&&- {L \over 4 g^2} \sum\limits_{B=1}^N (F_{B \; \mu\nu} - C_{\mu\nu}^{(2)}) (F_{B}^{\mu\nu} - C^{(2) \; \mu\nu}) + {2 \pi \over g^2 L} \sum\limits_{B=1}^N (\partial_\mu \varphi_{B} - {1\over 2 \pi} D_{\mu}^{(1)}) (\partial^\mu \varphi_{B} - {1\over 2 \pi} D^{(1) \; \mu})\nonumber \\
&&- {1 \over 4 \pi} \sum\limits_{A=1}^N \partial_\lambda \sigma_A \epsilon^{\lambda\mu\nu} F_{A \; \mu\nu} + {1 \over 4\pi} u_\lambda \epsilon^{\mu\nu\lambda}  \sum\limits_{B=1}^N (F_{B \; \mu\nu} -  C^{(2)}_{\mu\nu}) \\ &&- {b^{(0)} \over 4 \pi} \left[ \sum\limits_{B=1}^N F_{B \; \mu\nu} \epsilon^{\mu\nu\lambda}  \partial_\lambda \varphi_B  -{1 \over N} (\sum\limits_{B=1}^N F_{B \; \mu\nu}) \epsilon^{\mu\nu\lambda}(\sum\limits_{C=1}^N \partial_\lambda \varphi_C)   \right]~.\nonumber
\end{eqnarray}
Integrating out $F_B$, we find the solution of the saddle point equation:
\begin{eqnarray}\label{fsaddle}
F_B^{\mu\nu} = C^{(2)\; \mu\nu} - {g^2 \over 2 \pi L} \epsilon^{\mu\nu\lambda} \left( \partial_\lambda \sigma_B - u_\lambda + b^{(0)} (\partial_\lambda \varphi_B - {1 \over N} \sum\limits_{C=1}^N \partial_\lambda \varphi_C)\right).
\end{eqnarray}
Some algebra shows that with (\ref{fsaddle}), (\ref{ltotal}) becomes:
\begin{eqnarray}\label{ltotal2}
L_{bos}&=& \nonumber \\
&&{g^2 \over 8 \pi^2 L} \sum\limits_{B=1}^N \left(\partial_\lambda \sigma_B - u_\lambda + b^{(0)}(\partial_\lambda \varphi_B - {1 \over N} \sum\limits_{C=1}^N \partial_\lambda \varphi_C)\right) \left(\partial^\lambda \sigma_B - u^\lambda + b^{(0)}(\partial^\lambda \varphi_B - {1 \over N} \sum\limits_{C=1}^N \partial^\lambda \varphi_C)\right)\nonumber
\\
&&+ {2 \pi \over g^2 L} \sum\limits_{B=1}^N (\partial_\mu \varphi_{B} - {1\over 2 \pi} D_{\mu}^{(1)}) (\partial^\mu \varphi_{B} - {1\over 2 \pi} D^{(1) \; \mu})\\
&& -{1 \over 4 \pi} \sum\limits_{B=1}^N \partial_\lambda\sigma_B \epsilon^{\lambda\mu\nu} C_{\mu\nu}^{(2)}\nonumber~.
\end{eqnarray}
We next eliminate the Lagrange multiplier $u_\lambda$ from its equation of motion
\begin{equation}
\label{ulambda}
u_\lambda = {1 \over N} \sum\limits_{B=1}^{N} \partial_\lambda \sigma_B
\end{equation}
to finally obtain for the kinetic term of the magnetic dual theory:
\begin{eqnarray}\label{dualsym1}
L_{dual, SYM}&=&{g^2 \over 8 \pi^2 L} \sum\limits_{B=1}^N\bigg\vert(\partial_\lambda \sigma_B - {1 \over N} \sum\limits_{C=1}^{N} \partial_\lambda \sigma_C) + b^{(0)}(\partial_\lambda \varphi_B - {1 \over N} \sum\limits_{C=1}^N \partial_\lambda \varphi_C) \bigg\vert^2\nonumber
\\
&&+ {2 \pi \over g^2 L} \sum\limits_{B=1}^N \bigg\vert\partial_\mu \varphi_{B} - {1\over 2 \pi} D_{\mu}^{(1)}\bigg\vert^2\\
&& -{1 \over 4 \pi} \sum\limits_{B=1}^N \partial_\lambda\sigma_B \epsilon^{\lambda\mu\nu} C_{\mu\nu}^{(2)}+ {1\over 4 \pi}  \sum\limits_{B=1}^N v^\mu \; (\partial_\mu \varphi_B - {D^{(1)}_\mu \over 2 \pi} )\nonumber~,
\end{eqnarray}
where on the last line we introduced a Lagrange multiplier field  $v^\mu$ to enforce the constraint (\ref{scalarconstraint6}). The form of the kinetic term for the $N$ dual photons implies that their diagonal fluctuation in the $(\sigma_1,...\sigma_N) \sim (1,...1)$ direction has no kinetic term and decouples. 

We stress again that (\ref{dualsym1}) is invariant under the gauged 1-form center symmetry, which only acts on $C^{(2)}$ in the magnetic dual theory. In form notation the corresponding term in (\ref{dualsym1}) is $- {1\over 2 \pi} \sum\limits_{B=1}^N d\sigma_B \wedge C^{(2)}$, whose variation under $\Z_{N}^{(1)}$, 
 $- {1\over 2 \pi} \sum\limits_{B=1}^N d\sigma_B \wedge d \lambda^{(1)}$, does not affect the partition function, given that both $d \sigma_B$ and $d\lambda^{(1)}$ have $2 \pi$ quantized periods.
 
Further, it follows from (\ref{dualsym1})
that under a $\Z_{2N}^{(0)}$ discrete chiral transformation, a $2 \pi k$ shift of $b^{(0)}$, the dual photons shift as
\begin{eqnarray}
\label{dualphotonschiral}
\Z_{2N}^{(0)}: \sigma_A \rightarrow \sigma_A - 2 \pi k \varphi_A~, ~ b^{(0)} \rightarrow b^{(0)} + 2\pi k~,
\end{eqnarray}
rederiving the earlier result (\ref{sigmashiftdiscrete}) in this basis.
The Lagrangian (\ref{dualsym1}) is invariant, except for the term coupling $d \sigma$ to $C^{(2)}$. Its variation $\Delta L_{dual,SYM}$,  returning to form notation and using the constraint (\ref{scalarconstraint6}), is
\begin{eqnarray}
\label{dualsymanomaly}
\Z_{2N}^{(0)}: \Delta L_{dual, SYM} = - k  \sum\limits_{B=1}^N d \varphi_B C^{(2)} = - {2 \pi k\over N}\; {N D^{(1)}\over 2 \pi} \wedge {N C^{(2)} \over 2 \pi}~
\end{eqnarray} 
Naturally, the magnetic dual theory gives the same mixed chiral-center anomaly as that computed from the electric theory in (\ref{chiralcenter1}).

Finally, we note that with the $U(1)^N$ parametrization of the scalars, the 't Hooft vertex
(\ref{thooftSYM}), reproduced here $e^{ - {8 \pi^2 \over g^2 } \delta_{AN}} e^{- {8 \pi^2 \over g^2} \vec\alpha_A \cdot \vec\varphi + i \vec\alpha_A \cdot(\vec\sigma + {b^{(0)} } \vec\varphi)}$, is explicitly invariant under the gauged $\Z_N^{\S^1}$ 0-form center. The 't Hooft vertex only depends on $\vec\alpha_A \cdot \vec\varphi$. According to (\ref{scalarrelations1}), it  can be written as  $\vec\alpha_A \cdot \vec\phi = \sum\limits_{B=1}^N \varphi_B \vec\alpha_A \cdot \vec\nu_B$. Thus, under the shift (\ref{zerocenter6}), $\vec\alpha_A \cdot \vec\phi$ shifts by ${\lambda^{(0)} \over 2 \pi} \sum\limits_{B=1}^N \vec\alpha_A \cdot \vec\nu_B = 0$.
Likewise, the terms in the 't Hooft vertex depend on the dual photon fields only through $\vec\alpha_A \cdot \vec\sigma$; we can use the relations (\ref{scalarrelations1}) with $\varphi_A \rightarrow \sigma_A$ to rewrite them in terms of the $N$-dimensional basis. The invariance of the 't Hooft vertices under chiral transformations (\ref{dualsymanomaly}) then is as described in Section \ref{sec:4.2.1}.

The results of this paper should be useful 
 to studying the questions of the interplay of anomaly matching and  IR dynamics, mentioned in Section \ref{sec:1}, but  left for future work. 

{\flushleft{\bf Acknowledgements:}} We thank Mohamed Anber for discussions. This work is supported by an NSERC Discovery Grant.

\appendix
\section{Adjoint Weyl fermion calculation}
\label{appx:A}
This appendix outlines the calculations done for the adjoint Weyl fermions leading to (\ref{adjointBF1}).

\subsection{Choosing a regulator}
\label{appx:A.1}
The first step in the calculation is to determine exactly how we want to introduce the Pauli-Villars (PV) fermions in order to maintain gauge invariance. The fermion terms of the original 4D Lagrangian are of the form:
\begin{equation}\label{Lferm}
L_{ferm}(\psi,A,B) = \Tr\left[i \psi^\dag  \bar\sigma^M \partial_M \psi + \psi^\dag \bar\sigma^M [A_M,\psi] + \psi^\dag \bar\sigma^M B_M \psi \right]
\end{equation}
 Here, $A_M$ is the $SU(N)$ gauge field and $B_M$ the nondynamical background field gauging the classical $U(1)$ symmetry. Moreover, \(\sigma^M\) are the Pauli matrices with \(\sigma^0\) defined to be the \(2\times2\) identity matrix and \(\bar{\sigma}^M\) are the same matrices except with opposite sign for \(M=1,2,3\).\par
We use traces to specify the contraction of colour indices since, in the adjoint representation, we can treat the fermions as \(\mathfrak{su}(N)\) valued. Explicitly, we break it up as:
\[
\psi = \sum_{a = 1}^{N-1} \psi^a H^a + \sum_{1\leq A\neq B\leq N} \psi^{\beta^{AB}} E_{\beta^{AB}}
\]
Here the \(H^a\) are the usual Cartan generators (scaled so that \(\Tr[H^a,H^b] = \delta^{ab}\)) and \(E_{\beta^{AB}} \equiv E_{AB}\) are \(N \times N\) matrices with zeroes everywhere except for a one in the Ath row and Bth column. While these are technically complex linear combinations of elements of \(\mathfrak{su}(N)\), they have the nice property that 
\[
\left[\vec{H},E_{\beta^{AB}}\right] = \vec\beta^{AB} E_{\beta^{AB}}.
\]
Thus, they correspond exactly to the root vectors. It is also helpful to keep in mind that \(E^\dag_{\beta^{AB}} = E^T_{\beta^{AB}} = E_{\beta^{BA}} = E_{-\beta^{AB}}\) and that \(\Tr\left[E_{\beta^{AB}},E_{\beta^{CD}}\right] = \delta^{AD}\delta^{BC}\).\par

The usual method of introducing PV fermions for Weyl fermions is to introduce an uncharged righthand component for the PV fermions and a Dirac mass term. However, such a mass term will mix the charged left handed components and uncharged right handed components in a way that cannot be gauge invariant. Instead, owing to the reality of the adjoint representation, we introduce the PV masses as Majorana masses, eliminating the need for any right handed modes.

Using the trace formulae for the Cartan basis generators given above, the Lorentz and gauge invariant mass term for a two-component adjoint fermion can be written as
\[
\Tr\left[\psi\psi\right] = \sum_{a=1}^{N-1}\psi^a\psi^a + \sum_{1\leq A \neq B \leq N} \psi^{\beta^{AB}} \psi^{\beta^{BA}}.
\] 
Thus, the Lagrangian for each PV fermion, looks like
\begin{equation}\label{Lpv}
L_{PV}(\psi_i,A,B) = \Tr\left[i \psi_i^\dag  \bar\sigma^M \partial_M \psi_i + \psi_i^\dag \bar\sigma^M [A_M,\psi_i] + \psi_i^\dag \bar\sigma^M B_M \psi_i + \frac{M_i}{2}\left(\psi_i\psi_i + \psi^\dag_i\psi_i^\dag\right) \right]
\end{equation}
Notice that we include an index, \(i\), on our PV fermions. We do this because, in order to fully regulate the theory, we need more than one PV fermion. If we denote the statistics of the particles by a variable \(s_i\) with usual fermionic statistics corresponding to \(s_i = +1\) and bosonic statistics corresponding to \(s_i=-1\), then the conditions our PV fermions need to satisfy in order to fully regulate the theory are
\begin{equation}\label{PVmassrelation}
\sum_{i} s_i = -1 ~ {\rm and} ~
\sum_{i} s_i M^2_i = 0.
\end{equation}
To ensure finiteness of all diagrams in our theory, we introduce 3 PV fermions with the following characteristics: \(s_1 = s_2 = -1\), \(s_3=+1\), \(M_1 = M_2 = M\), and \(M_3 = \sqrt{2} M\) (these conditions ensure that the short-distance behaviour of the $x$-space massless propagator is sufficiently smoothed out by the PV contributions, for  explicit expressions and discussion see e.g.~\cite{Bogolyubov:1980nc}). Here \(M\) is the regulator scale, which will be taken to infinity at the end of all calculations. \par

For simplicity in our notation, we include the physical fermions with index \(i=0\); thus, \(\psi_0 \equiv \psi\) with \(s_0=+1\) and \(M_0 = 0\). Using this notation, our whole Lagrangian becomes:
\begin{equation}\label{Lfermreg}
\begin{split}
L_{ferm,reg}(\psi_0,\psi_1,\psi_2,& \psi_3,A,B) = \\ & \sum_{i=0}^{3} \Tr\left[i \psi_i^\dag  \bar\sigma^M \partial_M \psi_i + \psi_i^\dag \bar\sigma^M [A_M,\psi_i] + \psi_i^\dag \bar\sigma^M B_M \psi_i + \frac{M_i}{2}\left(\psi_i\psi_i + \psi^\dag_i\psi_i^\dag\right) \right].
\end{split}
\end{equation}

\subsection{Finding the propagator}\label{appx:A.2}
Equipped with a regulator, we can now look for the Feynman rules needed for our calculation. In order to eventually find the effective 3D theory, we need Feynman rules that correspond to 3D fields. The way to do this is to perform an integral around the compact \(x^3\) direction:
\[
L_{3D} = \int_0^L dx^3 L_{full}.
\] 
This integral is straightforward to evaluate once we split the fields into Kaluza-Klein (KK) modes as follows:
\[
\chi(x^\mu,x^3) = \sum_{z\in \mathbb{Z}} e^{-i\frac{2\pi z}{L} x^3} \chi_z(x^\mu)
\]
Notice that here we are using \(\chi\) to stand for any of the fields at play. For fermions, we also specify the notation such that
\[
\psi_i^\dag(x^\mu,x^3) = \sum_{z\in \mathbb{Z}} e^{i\frac{2\pi z}{L} x^3} \psi^\dag_{i,z}(x^\mu).
\]
Following through with the \(x^3\) integral, we are left with factors of the circumference $L$. We absorb these into the definition of the fermion fields and the coupling constant in order to give them the correct mass dimensions for a 3D theory. Specifically,
\begin{equation}
\begin{split}
\sqrt{L} \psi_i \rightarrow & \psi_i\\
g_4/\sqrt{L} \rightarrow & g_3\\
\end{split}
\end{equation} 
This gives the following breakdown of mass dimensions:
\begin{center}
\begin{tabular} {| c | c | c |}
\hline
Quantity/Field & 4D Mass Dimension & 3D Mass Dimension\\
\hline
\hline
Vector Fields ($A_\mu$ and $B_\mu$) & 1 & 1 \\
\hline
Fermions ($\psi_i$) & $\frac32$ & 1\\
\hline
Scalars  & 1 & 1 \\ 
\hline
g & 0 & 1/2\\
\hline
\end{tabular}
\end{center}
Also, in the resulting 3D Lagrangian, we find that all 3D fields with \(z\neq 0\) develop a Kaluza-Klein (KK) mass \(\frac{2\pi z}{L}\) and that all adjoint field modes that correspond to a root vector \(\beta^{AB}\) develop a mass \(m_{\beta^{AB}} = \expval{\vec{A}_3} \cdot \vec{\beta}^{AB}\) due to the nonzero holonomy vev. Note that this assumes that we have picked a gauge in which the vev of \(A_3\) is contained in the Cartan subalgebra; since this can always be done and it significantly simplifies the following work, we make this choice of gauge. Additionally, if we are in the vicinity of the center symmetric point, \(\expval{\vec{A}_3} = \frac{2\pi}{NL} \vec{\rho}\),  the KK and  holonomy vev masses cannot cancel out\footnote{In fact, these masses cannot cancel for any vev of \(LA_3\) which is in the interior of the Weyl chamber, see e.g. \cite{Anber:2015wha} for a precise definition.} and \(m_{\beta^{AB}} + {2 \pi z \over L}\) is on the order of \(\frac{2\pi}{NL}\). Hence, it is sensible to consider the IR theory at scales below \(\frac{2\pi}{NL}\) and integrate out all the massive W-bosons and fermion modes. This is the basis of the calculations we present in this paper. \par
\begin{figure}[t]
  \includegraphics[width=1 \textwidth]{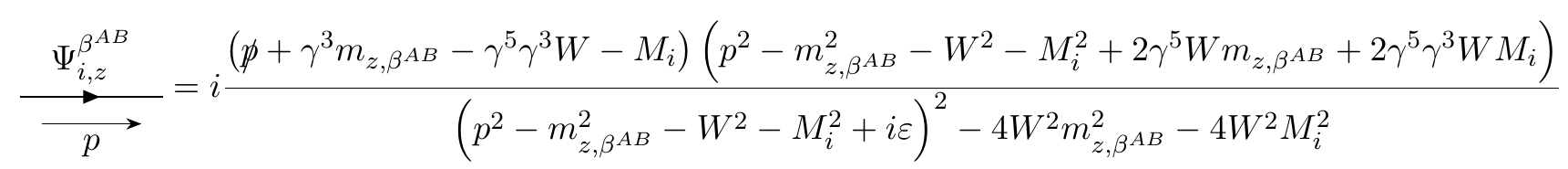}
  \caption{The propagator of the non-Cartan adjoint fermions, assembled into the four-component spinor (\ref{4components}), in the background of nontrivial gauge, $\langle \vec A_3 \rangle$, included in $m_{z, \beta^{AB}}$ of (\ref{mzbeta}), and global-$U(1)$, $W$, holonomies. When $W=0$, the propagator, shown on Figure \ref{fig:09a}, drastically simplifies.}
  \label{fig:01}
\end{figure}
\begin{figure}[h] 
\begin{subfigure}[h]{.3  \textwidth}
  \includegraphics[width= 1 \textwidth]{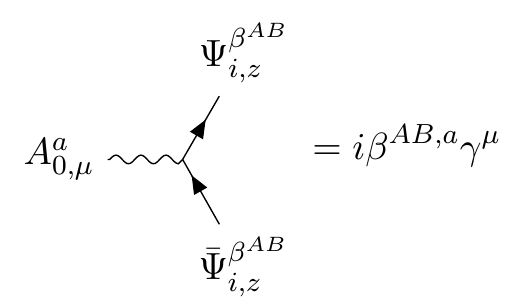}
\end{subfigure} 
\qquad
\begin{subfigure}[h]{.3 \textwidth}
  \includegraphics[width= 1\textwidth]{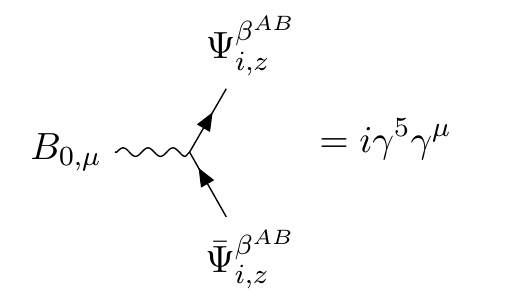}
\end{subfigure}
\qquad
\begin{subfigure}[h]{.3 \textwidth}
  \includegraphics[width= 1\textwidth]{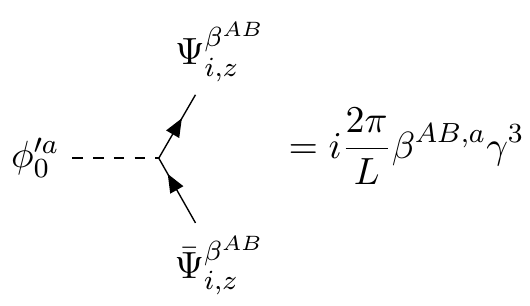}
\end{subfigure}
\caption{The interaction vertices between the non-Cartan adjoint fermions and the light gauge and global background fields.}
\label{fig:02}
\end{figure}

For our purposes, we do not need the entire \(L_{3D}\). We only care about having massless gauge bosons and scalars as external fields and charged massive fermions as internal fields. For propagators, we need the kinetic and mass terms of all the fermion fields corresponding to \(\mathfrak{su}(N)\) roots:
\[
\begin{split}
L = \sum_{i=0}^{3} \sum_{1\leq A<B\leq N} \sum_{z\in\mathbb{Z}} & \psi_{i,z}^{\beta^{AB}\dag} i \bar{\sigma}^\mu \partial_\mu \psi_{i,z}^{\beta^{AB}} + \psi_{i,-z}^{\beta^{BA}\dag} i \bar{\sigma}^\mu \partial_\mu \psi_{i,-z}^{\beta^{BA}} \\
& + \psi_{i,z}^{\beta^{AB}\dag} \bar{\sigma}^3 \left(m_{z,\beta^{AB}} + W\right) \psi_{i,z}^{\beta^{AB}} + \psi_{i,-z}^{\beta^{BA}\dag} \bar{\sigma}^3  \left(m_{-z,\beta^{BA}} +W\right)  \psi_{i,-z}^{\beta^{BA}} \\
& + M_i \left(\psi_{i,-z}^{\beta^{BA}}\psi_{i,z}^{\beta^{AB}} - \psi_{i,z}^{\beta^{AB}\dag}\psi_{i,-z}^{\beta^{BA}\dag}\right) 
\end{split} 
\] 
Here, the quantity \(m_{z,\beta^{AB}}\) is defined by
\begin{equation} \label{mzbeta}
 m_{z,\beta^{AB}} = \frac{2\pi z}{L} + m_{\beta^{AB}} =   \frac{2\pi z}{L} + \expval{\vec{A}_3} \cdot \vec{\beta}^{AB} = \frac{2\pi z}{L} + \frac{2\pi}{NL} \rho\cdot \vec{\beta}^{AB} + \frac{2\pi}{L} \expval{\vec{\phi}} \cdot \vec{\beta}^{AB} .
\end{equation} 
We also introduce the notation $W = B_3$, proportional to the $U(1)$-background holonomy. For vertices we need the following terms:
\[
\begin{split}
\sum_{i=0}^{3} \sum_{1\leq A<B\leq N} \sum_{z\in\mathbb{Z}}& \psi_{i,z}^{\beta^{AB}\dag} \bar{\sigma}^\mu \vec{\beta}^{AB} \cdot \vec{A}_{0,\mu} \psi_{i,z}^{\beta^{AB}} + \psi_{i,-z}^{\beta^{BA}\dag} \bar{\sigma}^\mu \vec{\beta}^{BA} \cdot \vec{A}_{0,\mu} \psi_{i,-z}^{\beta^{BA}} \\&+ \psi_{i,z}^{\beta^{AB}\dag} \bar{\sigma}^3 \frac{2\pi}{L}\vec{\beta}^{AB} \cdot \vec{\phi'} \psi_{i,z}^{\beta^{AB}} + \psi_{i,-z}^{\beta^{BA}\dag} \bar{\sigma}^\mu \vec{\beta}^{BA} \cdot \vec{\phi'} \psi_{i,-z}^{\beta^{BA}}\\&+\psi_{i,z}^{\beta^{AB}\dag} \bar{\sigma}^\mu B_\mu \psi_{i,z}^{\beta^{AB}} + \psi_{i,-z}^{\beta^{BA}\dag} \bar{\sigma}^\mu B_\mu \psi_{i,-z}^{\beta^{BA}} 
\end{split}
\]
Notice that for each \(z\in \mathbb{Z}\) and \(1\leq A<B\leq N\), \( \psi_{i,z}^{\beta^{AB}}\) and \(\psi_{i,-z}^{\beta^{BA}}\) are mixed by the Majorana mass term and not mixed with any other fermion modes. To clean up these expressions,\footnote{The calculation can also be done in the two-component formalism described in \cite{Dreiner:2008tw}. While some details are different---most importantly, the propagators are simpler, but there are more diagrams to consider if the 4-component Dirac representation (\ref{4components}) is not used---we have checked that, naturally, the final result is the same. Note that we use the two-component formalism in Section \ref{appx:A.6.1} when calculating the contribution of the KK zero-modes needed to match the anomaly under large gauge transformations.} we can pair these two 2-component fermions into a 4-component fermion:
\begin{equation}
\label{4components}
\Psi_{i,z}^{\beta^{AB}} = \begin{pmatrix} \psi_{i,z}^{\beta^{AB}} \\ \psi_{i,-z}^{\beta^{BA}\dag}  \end{pmatrix}.
\end{equation}
To find the vertices and propagators, we use the chiral representation of 4D  gamma matrices, $\gamma^5 = \left(\begin{array}{cc}1&0\cr 0&-1\end{array}\right)$,  $\gamma^0 = \left(\begin{array}{cc}0&1\cr 1&0\end{array}\right)$, $\gamma^i= \left(\begin{array}{cc}0&\sigma^i\cr -\sigma^i&0\end{array}\right)$, $i=1,2,3$, and \(\bar{\Psi} = \Psi^\dag \gamma^0\), so that  the kinetic terms of the fermions are
\[
\sum_{i=0}^{3} \sum_{1\leq A<B\leq N} \sum_{z\in\mathbb{Z}} \bar{\Psi}_{i,z}^{\beta^{AB}} \left(i\slashed{\partial} + \gamma^3 m_{z,\beta^{AB}} + \gamma^5\gamma^3 W + M_i \right)\Psi_{i,z}^{\beta^{AB}}~.
\]
The propagator  for the non-Cartan adjoint fields, assembled into the 4-component object (\ref{4components})  is shown in Figure \ref{fig:01}. 
The interaction vertices take the form
\[
\sum_{i=0}^{3} \sum_{1\leq A<B\leq N} \sum_{z\in\mathbb{Z}} \bar{\Psi}_{i,z}^{\beta^{AB}} \left(\vec{\beta}^{AB} \cdot \vec{\slashed{A}}_0 + \gamma^3 \vec{\beta}^{AB}\cdot\vec{\phi'} +   \gamma^5\slashed{B}_0 \right)\Psi_{i,z}^{\beta^{AB}}.
\]
and the vertex Feynman rules that follow are  shown on Figure \ref{fig:02}. 

\subsection{Chern-Simons term} 
\label{appx:A.3}

\begin{figure}[h]\centering
  \includegraphics[width=.4 \textwidth]{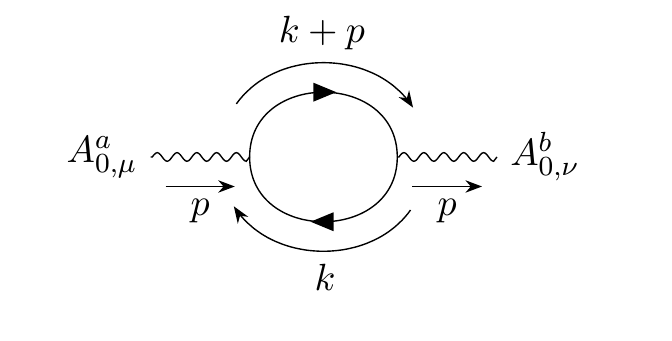}
  \caption{The diagram contributing to the Chern-Simons $A \wedge F$ term.}
  \label{fig:03}
\end{figure}

Equipped with the Feynman rules of Figures \ref{fig:01}, \ref{fig:02} we can now complete the calculations of the one loop topological terms in our IR theory. We start with the \(A\wedge F\) term, which involves the  diagram from Fig. \ref{fig:03},  
where we sum over all the \(\Psi_{i,z}^{\beta^{AB}}\) fields that can run in the internal loop. 
Fixing \(i\in\{0,1,2,3\}\), \(z\in\mathbb{Z}\), and \(\beta\in\{\beta^{AB}\in \mathbb{R}^{N-1} | 1\leq A<B\leq N\}\), the contribution from this individual diagram looks like:
\begin{equation}
\begin{split}
-\frac{s_i}{2} \beta^a\beta^b \int & \frac{d^3k}{(2\pi)^3}  \frac{1}{\left[\left(k^2 - m_{z,\beta}^2 - W^2 - M_i^2 + i\varepsilon \right)^2 - 4W^2m_{z,\beta}^2 - 4 W^2 M_i^2\right]}\\
 \times & \frac{1}{\left[\left((k+p)^2 - m_{z,\beta}^2 - W^2 - M_i^2 + i\varepsilon \right)^2 - 4W^2m_{z,\beta}^2 - 4 W^2 M_i^2\right]}\\
\times &\Tr\left[\left(\slashed{k} + \gamma^3 m_{z,\beta} - \gamma^5\gamma^3 W - M_i\right)\left(k^2 - m_{z,\beta}^2 - W^2 - M_i^2 + 2\gamma^5Wm_{z,\beta} + 2\gamma^5\gamma^3WM_i\right)\gamma^\nu\right.\\
&\, \left.\left(\slashed{k} + \slashed{p} + \gamma^3 m_{z,\beta} - \gamma^5\gamma^3 W - M_i\right)\left((k+p)^2 - m_{z,\beta}^2 - W^2 - M_i^2 + 2\gamma^5Wm_{z,\beta} + 2\gamma^5\gamma^3WM_i\right)\gamma^\nu\right]
\end{split}
\end{equation}
Note the overall factor of \(\frac{1}{2}\) comes from a symmetry factor and the overall minus follows from the fermion loop.\par
Since we only care about the topological terms, we can avoid all terms of the trace, except for those that give a Levi-Civita tensor. After working out this part of the trace, we are left with an overall factor \(p_\sigma \varepsilon^{\sigma\mu\nu}\). Here we can again simplify our calculation by recalling that we are only interested in the terms that are leading order in \(p\), since \(p\) is a small quantity in the IR. Therefore, since we have an overall linear factor of \(p\), we can safely set \(p=0\) everywhere else in the integrand. This leaves us with the following expression:
\begin{equation}
2i s_i \beta^a\beta^b W p_\sigma \varepsilon^{\sigma\mu\nu} \int \frac{d^3k}{(2\pi)^3} \frac{a(k)^2 + 4\left(m_{z,\beta}^2+M_i^2\right)a(k) + 4\left(m_{z,\beta}^2+M_i^2\right)W^2}{\left[\left(a(k)+i\varepsilon\right)^2 - 4W^2m_{z,\beta}^2 - 4 W^2 M_i^2\right]^2}
\end{equation} 
with \(a(k) = \left(k^2 - m_{z,\beta}^2 - W^2 - M_i^2\right)\). Using the standard technique of replacing our \(k\) with a Euclidean \(k_E\) and changing to spherical coordinates gives us:
\begin{equation}
-\frac{1}{\pi^2}s_i\beta^a\beta^b W p_\sigma \varepsilon^{\sigma\mu\nu} \int_0^{\infty} dk_E k_E^2 \frac{b(k_E)^2- 4(m_{z,\beta}^2+M_i^2)b(k_E) + 4(m_{z,\beta}^2+M_i^2)W^2}{\left[b(k_E)^2-4(m_{z,\beta}^2+M_i^2)W^2\right]^2}
\end{equation}
with \(b(k_E) = k_E^2 + m_{z,\beta}^2 + M_i^2 + W^2\).The integral has an exact antiderivative given by:
\[
\begin{split}
\frac14& \left[\frac{2k_E(k_E^2 -m_{z,\beta}^2 -M_i^2 + W^2)}{k_E^4 + 2k_E^2(m_{z,\beta}^2+M_i^2+W^2) + (m_{z,\beta}^2+M_i^2-W^2)^2}\right.\\
& - \frac{1}{W} \text{sign}\left(\sqrt{m_{z,\beta}^2+M_i^2} - W\right) \arctan{\left(\frac{k_E}{\left|\sqrt{m_{z,\beta}^2+M_i^2}-W\right|}\right)} \\
&\left.+ \frac{1}{W}  \text{sign}\left(\sqrt{m_{z,\beta}^2+M_i^2} + W\right)\arctan{\left(\frac{k_E}{\left|\sqrt{m_{z,\beta}^2+M_i^2}+W\right|}\right)}\right]
\end{split}
\]
The first term vanishes for \(k_E \rightarrow 0\) and \(k_E \rightarrow \infty\), so it does not contribute to the integral. For \(k_E\rightarrow 0\), both \(\arctan\) functions also vanish. Thus the only contribution is from the \(\arctan\) terms in the limit \(k_E\rightarrow \infty\). This limit gives us 
\begin{equation}
\frac{\pi}{8W} \left[\text{sign}\left(\sqrt{m_{z,\beta}^2 + M_i^2} + W\right) - \text{sign}\left(\sqrt{m_{z,\beta}^2 + M_i^2} - W\right)\right]
\end{equation}
To further simplify this expression, we note that it is even in \(W\), hence we can replace \(W\) with \(\left|W\right|\). This gives
\[
\frac{\pi}{8\left|W\right|} \left[1 - \text{sign}\left(\sqrt{m_{z,\beta}^2 + M_i^2} - \left|W\right|\right)\right]
\]
Thus, the Feynman diagram gives
\begin{equation}
-\frac{1}{4\pi}s_i\beta^a\beta^b \text{sign}\left(W\right) \frac12\left[1 - \text{sign}\left(\sqrt{m_{z,\beta}^2 + M_i^2} - \left|W\right|\right)\right]p_\sigma \varepsilon^{\sigma\mu\nu}
\end{equation}
For \(\sqrt{m_{z,\beta}^2 + M_i^2} < \left|W\right|\), this vanishes. This will always happen for the PV fermions, since we will take the \(M \rightarrow \infty\) limit.  Moreover, if \(\left|z\right|\) is sufficiently large, it will vanish. Thus, only the physical fermions with \(z\) close to zero can contribute, and the sum over infinite KK modes is really a finite sum over a small number of terms. We write this as
\begin{equation}
-\frac{1}{4\pi} \text{sign}\left(W\right) \sum_{\beta\in\beta^+} n(m_\beta,W) \beta^a\beta^b p_\sigma \varepsilon^{\sigma\mu\nu} 
\end{equation}
Where \(n(m_\beta,W)\) is the function from equation (\ref{nfunctions}), which we reproduce here for convenience:
\begin{eqnarray}\label{n1}
n(m_\beta,W) &\equiv & {1 \over 2} \sum_{k \in \Z}\left[1 - {\rm sign}( |m_\beta + {2 \pi \over L} k| - |W|)\right],
\end{eqnarray} 
 stressing again that the seemingly infinite sum only involves a finite number of nonzero terms.
 This precisely gives the \(A\wedge F\) term in equation (\ref{adjointBF1}):
\begin{equation}\label{AF_result}
-\frac{1}{4\pi} \text{sign}\left(W\right) \sum_{\beta\in\beta^+} n(m_\beta,W) \beta^a\beta^b A^a \wedge F^b.
\end{equation}

\subsection{The BF term}
\label{appx:A.4}

\begin{figure}[t]\centering
  \includegraphics[width= .4\textwidth]{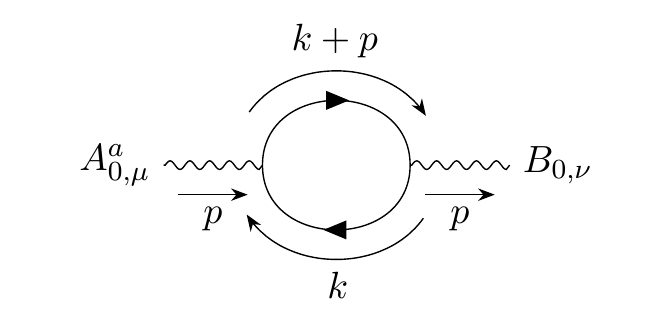}
  \caption{The diagram contributing to the $B \wedge F$ term.}
  \label{fig:05}
\end{figure}

Here we look at the contribution to the \(B \wedge F\) term of equation (\ref{adjointBF1}) coming from the diagram shown on Figure \ref{fig:05}.
This gives the expression
\begin{equation}
\begin{split}
-s_i \beta^a \int & \frac{d^3k}{(2\pi)^3} \frac{1}{\left[\left(k^2 - m_{z,\beta}^2 - W^2 - M_i^2 + i\varepsilon \right)^2 - 4W^2m_{z,\beta}^2 - 4 W^2 M_i^2\right]}\\
&\times\frac{1}{\left[\left((k+p)^2 - m_{z,\beta}^2 - W^2 - M_i^2 + i\varepsilon \right)^2 - 4W^2m_{z,\beta}^2 - 4 W^2 M_i^2\right]}\\
&\times\Tr\left[\left(\slashed{k} + \gamma^3 m_{z,\beta} - \gamma^5\gamma^3 W - M_i\right)\left(k^2 - m_{z,\beta}^2 - W^2 - M_i^2 + 2\gamma^5Wm_{z,\beta} + 2\gamma^5\gamma^3WM_i\right)\gamma^5\gamma^\nu\right.\\
&\, \left.\left(\slashed{k} + \slashed{p} + \gamma^3 m_{z,\beta} - \gamma^5\gamma^3 W - M_i\right)\left((k+p)^2 - m_{z,\beta}^2 - W^2 - M_i^2 + 2\gamma^5Wm_{z,\beta} + 2\gamma^5\gamma^3WM_i\right)\gamma^\nu\right]
\end{split}
\end{equation}
The main difference is the extra \(\gamma^5\) in the trace. As before, we look for the Levi-Civita terms and keep everything to leading order in \(p\), leaving
\begin{equation}
\begin{split}
4is_i m_{z,\beta} \beta^a p_\sigma \varepsilon^{\sigma\mu\nu}&  \int \frac{d^3k}{(2\pi)^3} \frac{a(k)^2 + 4W^2a(k) + 4\left(m_{z,\beta}^2+M_i^2\right)W^2}{\left[(a(k)+i\varepsilon)^2 -  4(m_{z,\beta}^2 + M_i^2)W^2\right]^2}\\
= & -\frac{2}{\pi} s_i m_{z,\beta} \beta^a p_\sigma \varepsilon^{\sigma\mu\nu} \int_0^{\infty} dk_E k_E^2 \frac{b(k_E)^2- 4W^2b(k_E) + 4(m_{z,\beta}^2+M_i^2)W^2}{\left[b(k_E)^2-4(m_{z,\beta}^2+M_i^2)W^2\right]^2}
\end{split}
\end{equation} 
The integral here is nearly identical to the one from the \(A \wedge F\) term except with the replacement \( \left(m_{z,\beta}^2 + M_i^2\right) \leftrightarrow W^2\). Hence, the integral gives 
\begin{equation}
\frac{\pi}{8\sqrt{m_{z,\beta}^2 + M_i^2} } \left[\text{sign}\left(W + \sqrt{m_{z,\beta}^2 + M_i^2}\right) - \text{sign}\left(W - \sqrt{m_{z,\beta}^2 + M_i^2} \right)\right]
\end{equation}
This is again even with respect to \(W\), so we replace \(W\) with \(\left|W\right|\). Then the Feynman diagram gives
\begin{equation}
-\frac{1}{2\pi} s_i \frac{m_{z,\beta}}{\sqrt{m_{z,\beta}^2+M_i^2}} \beta^a p_\sigma \varepsilon^{\sigma\mu\nu} \times \frac12 \left[ 1 - \text{sign}\left(\left|W\right| - \sqrt{m_{z,\beta}^2 + M_i^2} \right)\right]
\end{equation} 
This is now completely opposite to the \(A \wedge F\) case, as this contribution only vanishes for the finitely many physical fermion modes with \(\sqrt{m_{z,\beta}^2 + M_i^2} < \left|W\right|\) and  never vanishes for the regulators.
Summing over all modes now gives
\begin{equation}
\begin{split}
-\frac{1}{2\pi} \sum_{\beta\in\beta^+} & \beta^a \left[ \lim_{M\rightarrow\infty}  \sum_{z\in\mathbb{Z}}\left( \text{sign}\left(m_{z,\beta}\right) - 2\frac{m_{z,\beta}}{\sqrt{m_{z,\beta}^2+M^2}} +\frac{m_{z,\beta}}{\sqrt{m_{z,\beta}^2+2M^2}}\right)\right.\\
&\left. - \sum_{z\in\mathbb{Z}} \text{sign}\left(m_{z,\beta}\right) \frac12 \left(1 - \text{sign}\left( \left|m_{z,\beta}\right|-\left|W\right|\right) \right) \right] p_\sigma \varepsilon^{\sigma\mu\nu}
\end{split}
\end{equation} 
Notice that we have included \(\text{sign}\left(m_{z,\beta}\right)\) for every physical fermion in the first infinite sum, then used the second infinite sum to subtract off the finitely many that should vanish. This makes the first infinite sum easier to evaluate; one can show that:\footnote{This was used in \cite{Corvilain:2017luj}. For completeness, we give a detailed proof in Appendix \ref{appx:D}. We note that the expression for the terms not involving $W$ is identical to that obtained by $\zeta$ function regularization (as the only nonzero holonomy background involved is one of the gauge group, periodicity properties are ensured by gauge invariance; as noted earlier, periodicity in the holonomy is automatic in the $\zeta$ function regulator).} 
\begin{equation}\label{identity1}
 \lim_{M\rightarrow\infty}  \sum_{z\in\mathbb{Z}}\left( \text{sign}\left(m_{z,\beta}\right) - 2\frac{m_{z,\beta}}{\sqrt{m_{z,\beta}^2+M^2}} +\frac{m_{z,\beta}}{\sqrt{m_{z,\beta}^2+2M^2}}\right) = 1 + 2\left\lfloor\frac{Lm_\beta}{2\pi}\right\rfloor - 2\frac{Lm_\beta}{2\pi}.
\end{equation}

Thus, overall for the Feynman diagram we have
\begin{equation}\label{BF_result_unpolished}
-\frac{1}{2\pi} \sum_{\beta\in\beta^+} \left(1 + 2\left\lfloor\frac{Lm_\beta}{2\pi}\right\rfloor - 2\frac{Lm_\beta}{2\pi} - n'\left(m_\beta,W\right)\right)\beta^a p_\sigma \varepsilon^{\sigma\mu\nu}
\end{equation}
where $n'$ was defined earler in (\ref{nfunctions}): 
\begin{eqnarray}\label{nfunctions31}
n'(m_\beta,W) &\equiv &{1\over 2} \sum_{k \in \Z} {\rm sign}(m_\beta + {2 \pi \over L} k)\left[1 - {\rm sign}( |m_\beta + {2 \pi \over L} k| - |W|)\right]~.
\end{eqnarray} 
If we assume that we are in the Weyl chamber (i.e. close to the center symmetric point), the term with the floor function vanishes. Recalling the identities \(\sum_{\beta\in\beta^+} \beta^a = 2\vec{\rho}\) and \(\sum_{\beta\in\beta^+} \beta^a\beta^b  = N\delta^{ab}\), we see that the constant term gives \(2\rho^a\) and the continuous term linear in \(m_\beta\) gives \(-2\rho^a - 2\expval{\phi^a}\). Hence, we get
\begin{equation}
\frac{1}{2\pi} \left(2 \expval{\phi^a} + \sum_{\beta\in\beta^+} n'\left(m_\beta,W\right)\beta^a\right) p_\sigma \varepsilon^{\sigma\mu\nu}.
\end{equation}
This is in agreement with equation (\ref{adjointBF1}), since it suggests the following IR term:
\begin{equation}
\frac{1}{2\pi} \left(2 \expval{\phi^a} + \sum_{\beta\in\beta^+} n'\left(m_\beta,W\right)\beta^a\right) B\wedge F^a.
\end{equation}

\subsection{Triangle diagram contributions} 
\label{appx:A.5}

\begin{figure}[h] 
\begin{subfigure}[t]{.48  \textwidth}
  \includegraphics[width= 1 \textwidth]{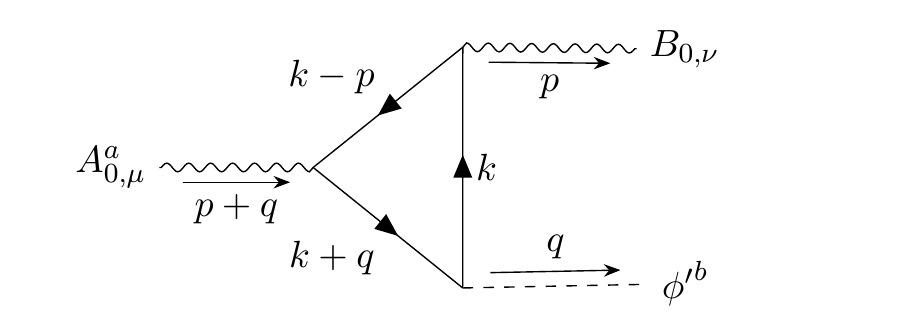}
\end{subfigure} 
\qquad
\begin{subfigure}[t]{.48 \textwidth}
  \includegraphics[width= 1\textwidth]{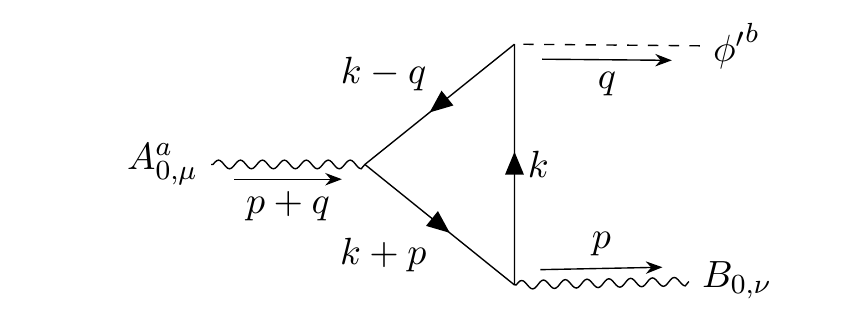}
\end{subfigure}
\caption{The triangle diagrams contributing to the $B \wedge dA \; \phi'$ term.}
\label{fig:06a}
\end{figure}

In this section, we will finish off the determination of equation (\ref{adjointBF1}) by finding the \(\phi'\) dependence in the \(B \wedge F\) term. This obtained from the  diagrams shown on Figure \ref{fig:06a}.

\begin{figure}[h]\centering
  \includegraphics[width= .5 \textwidth]{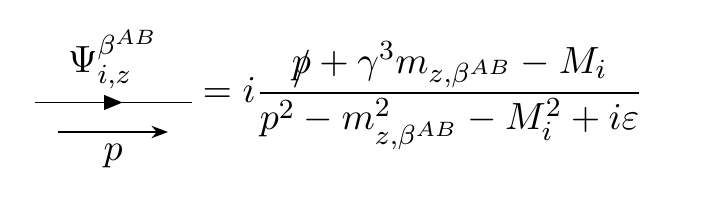}
  \caption{The propagator in the $W=0$ background.}
  \label{fig:09a}
\end{figure}

Since we are only concerned with \(\phi'\) dependence, we drop \(W\) from these calculations. This leaves us with a much simpler propagator, shown in Figure \ref{fig:09a}. 
With this propagator, we find that the first diagram on Fig.~\ref{fig:06a} gives us
\begin{equation}
\begin{split}
\frac{2\pi}{L} s_i \beta^a\beta^b \int \frac{d^3k}{(2\pi)^3} & \frac{1}{\left[k^2 - m_{z,\beta}^2 - M_i^2\right]\left[(k+q)^2 - m_{z,\beta}^2 - M_i^2\right]\left[(k-p)^2 - m_{z,\beta}^2 - M_i^2\right]} \\
& \times \Tr\left[\gamma^5\gamma^\nu \left(\slashed{k} + \gamma^3 m_{z,\beta} - M_i\right)\gamma^3 \left(\slashed{k} + \slashed{q} + \gamma^3 m_{z,\beta} - M_i\right)\gamma^\mu\left(\slashed{k} - \slashed{p} + \gamma^3 m_{z,\beta} - M_i\right)\right]
\end{split}
\end{equation}
we evaluate the trace, keeping only the terms that develop a Levi-Civita tensor, then use Feynman parameters to combine the denominator. This leads to an expression that is not entirely pleasant, which can be written more simply as\footnote{For brevity, we omit the $\mu\nu$ indices from the definitions of $b_1, b_2, b_3$, and of similar definitions further in this Section.}
\begin{equation}
\frac{16\pi i}{L} s_i \beta^a\beta^b \int_0^1dx dy dz \delta(x+ y+z-1) \int \frac{d^3k}{(2\pi)^3} \frac{M_i^2 b_1 + b_2 + k^2 b_3}{\left(a-M_i^2\right)^3} 
\end{equation}
where \(a\), \(b_1\), \(b_2\), and \(b_3\) are defined as follows:
\begin{equation}
\begin{split}
a = & k^2 + 2k\cdot(yq-xp) + xp^2 + yq^2 - m_{z,\beta}^2\\
b_1 = & (k-p-q)_\sigma \varepsilon^{\sigma\mu\nu} \\
b_2 = &  m_{z,\beta}^2(k+p+q)_\sigma \varepsilon^{\sigma\mu\nu} - k^\nu k_\delta (p+q)_\lambda \varepsilon^{\delta\mu\lambda} - k^\nu q_\delta p_\lambda\varepsilon^{\delta\mu\lambda} \\
& + (k-p)^\mu k_\sigma q_\delta \varepsilon^{\delta\nu\sigma} - (k+q)^\mu k_\sigma p_\delta \varepsilon^{\delta\nu\sigma} + (p\cdot q + k\cdot p - k\cdot q) k_\sigma \varepsilon^{\sigma\mu\nu}\\
b_3 = & -k_\sigma \varepsilon^{\sigma\mu\nu} 
\end{split}
\end{equation}
Unlike the previous diagrams, this is divergent. To obtain a finite expression, we perform the sum over the \(i\) index, adding the physical and the three PV fermion contributions. This leads to the following expression for the first diagram on Fig.~\ref{fig:06a}:
\begin{equation}
\begin{split}
8i\beta^a\beta^b \int_0^1dx dy dz& \delta(x+ y+z-1) \int \frac{d^3k}{(2\pi)^3}  \frac{1}{a^3\left(a-M^2\right)^3\left(a-2M^2\right)^3}\\
\times & \left[\left(6M^4a^5 - 18M^6a^4 + 14M^8a^3\right)b_1 \right.\\
& \left.+ \left(12M^4 a^4-48M^6a^3+66M^8a^2-36M^{10}a+8M^{12}\right)\left(b_2 + k^2b_3\right)\right]
\end{split}
\end{equation}
The second diagram   on Fig.~\ref{fig:06a} is nearly identical and gives the same expression with the following replacements:
\begin{equation}
\begin{split}
a \rightarrow c = & k^2 + 2k(xp-yq) + xp^2 + yq^2 - m_{z,\beta}^2\\
b_1 \rightarrow d_1 = & - (k+p+q)_\sigma \varepsilon^{\sigma\mu\nu}\\
b_2 \rightarrow d_2 = & m_{z,\beta}^2(p+q-k)_\sigma \varepsilon^{\sigma\mu\nu} + (k+p)^\nu k_\lambda q_\delta \varepsilon^{\mu\lambda\delta} - k\cdot q (k+q)_\sigma \varepsilon^{\sigma\mu\nu}\\
& - k^\mu \left(p_\lambda k_\delta - k_\lambda q_\delta - p_\lambda q_\delta \right)\varepsilon^{\delta\nu\lambda} + (k-q)^\mu k_\delta p_\lambda\varepsilon^{\delta\nu\lambda}\\
b_3 \rightarrow d_3 = & -(k+p)_\sigma \varepsilon^{\sigma\mu\nu}
\end{split}
\end{equation}
The expressions obtained finally give convergent integrals, so one can shift the integration variable. Using a shift \(k\rightarrow k' = k + yq-xp\) for the first diagram and a shift \(k\rightarrow k'' = k+xp-yq\) for the second proves useful. Under these shifts and using the facts that odd functions of \(k'\) or \(k''\) will vanish and that under an otherwise Lorentz invariant integral \(k'_\sigma k'_\lambda = \frac{1}{3}k^2 g_{\sigma\lambda}\), we can simplify the \(a\), \(b_i\), \(c\), and \(d_i\) variables to
\begin{equation}
\begin{split}
a = & {k'}^2 - (xp-yq)^2 + xp^2 + yq^2 - m_{z,\beta}^2\\
b_1 = & -\left[(1-x)p_\sigma + (1+y) q_\sigma\right]\varepsilon^{\sigma\mu\nu}\\
b_2 = & m_{z,\beta} \left[(1+x)p_\sigma + (1-y) q_\sigma\right]\varepsilon^{\sigma\mu\nu} + {k'}^2\left[p_\sigma - \frac13 q_\sigma\right]\varepsilon^{\sigma\mu\nu} 
\end{split}
\end{equation}
\begin{equation}
\begin{split}
k^2 b_3 = & {k'}^2 \left[\frac53 y q_\sigma - \frac53 x p_\sigma \right]\varepsilon^{\sigma\mu\nu}\\
c = & {k''}^2 - (xp-yq)^2 + xp^2 + yq^2 - m_{z,\beta}^2 \nonumber
\end{split}
\end{equation}
\begin{equation}
\begin{split}
d_1 = & -\left[(1-x)p_\sigma + (1+y)q_\sigma \right]\varepsilon^{\sigma\mu\nu}\\
d_2 = & m_{z,\beta} \left[(1+x)p_\sigma + (1-y) q_\sigma\right]\varepsilon^{\sigma\mu\nu} - \frac13 {k''}^2q_\sigma \varepsilon^{\sigma\mu\nu}\\
k^2 d_3 = & {k''}^2\left[\frac53 yq_\sigma - \left(\frac53 x - 1\right)p_\sigma\right]\varepsilon^{\sigma\mu\nu}
\end{split}
\end{equation}
Notice that the expression for \(a\) and \(c\) are identical apart from the different symbol for the integration variable. Hence, we can now combine the expressions from the two diagrams. This leads to
\begin{equation} \label{Int_setting}
\begin{split}
\frac{16\pi i}{L} \beta^a\beta^b & \int_0^1 dx dy dz \delta(x+y+z-1)  \int \frac{d^3k'}{(2\pi)^2} \frac{1}{a^3\left(a-M^2\right)^3\left(a-2M^2\right)^3}\\
& \times \left[\left(6M^4a^5- 18M^6a^4 + 14M^8a^3\right)(b_1+d_1)\right. \\
& \, \left.+ \left(12M^4a^4-48M^6a^3+66M^8a^2 - 36M^{10} a + 8M^{12} \right)\left(b_2+d_2+k^2b_3+k^2d_3\right)\right]\\
= - \frac{16\pi i}{L} \beta^a\beta^b  & p_\sigma \varepsilon^{\sigma\mu\nu}\left[\frac23 I_1 - \frac43  I_2 - \frac49 I_3\right] 
-\frac{16\pi i}{L}  \beta^a\beta^b  q_\sigma \varepsilon^{\sigma\mu\nu} \left[\frac43  I_1 -\frac23 I_2 - \frac29 I_3\right]~,
\end{split}
\end{equation}
where we have simplified the integral into bite-sized pieces as follows
\begin{equation}
\begin{split}
I_1 = & \int \frac{d^3k'}{(2\pi)^3} \frac{1}{a^3\left(a-M^2\right)^3\left(a-2M^2\right)^3} \left(6M^4a^5- 18M^6a^4 + 14M^8a^3\right)\\
= &\frac{i}{16\pi} M^4 \left(\frac{1}{\left(M^2+m_{z,\beta}^2\right)^{3/2} \left(2M^2+m_{z,\beta}^2\right)} + \frac{1}{\left(M^2+m_{z,\beta}^2\right)\left(2M^2+m_{z,\beta}^2\right)^{3/2}} \right.\\
& \, \left.- \frac{1}{\left(M^2+m_{z,\beta}^2\right)^{3/2} \left(2M^2+m_{z,\beta}^2\right)+\left(M^2+m_{z,\beta}^2\right)\left(2M^2+m_{z,\beta}^2\right)^{3/2}}\right)~, \end{split}
\end{equation}
\begin{equation} \begin{split}
I_2 = & m_{z,\beta}^2 \int \frac{d^3k'}{(2\pi)^3} \frac{1}{a^3\left(a-M^2\right)^3\left(a-2M^2\right)^3} \left(12M^4a^4-48M^6a^3+66M^8a^2 - 36M^10 a + 8M^12 \right)\\
= & \frac{-i}{32\pi} \left(\frac{1}{\left|m_{z,\beta}\right|} - \frac{2m_{z,\beta}^2}{\left(M^2+m_{z,\beta}^2\right)^{3/2}} + \frac{m_{z,\beta}^2}{\left(2M^2+m_{z,\beta}^2\right)^{3/2}}\right)~,
\end{split}
\end{equation}
and
\begin{equation}\begin{split}
I_3 = & \int \frac{d^3k'}{(2\pi)^3} \frac{{k'}^2}{a^3\left(a-M^2\right)^3\left(a-2M^2\right)^3} \left(12M^4a^4-48M^6a^3+66M^8a^2 - 36M^10 a + 8M^12 \right)\\
= & \frac{3i}{32\pi} \left(\frac{1}{\left|m_{z,\beta}\right|} - \frac{2}{\sqrt{m_{z,\beta}^2+M^2}} + \frac{1}{\sqrt{m_{z,\beta}^2 + 2M^2}}\right)~.
\end{split}
\end{equation}
These expressions must be summed over all \(z\in\mathbb{Z}\) and then taken to the limit \(M\rightarrow\infty\). Luckily, this is fairly straightforward, since we can use the definition of Riemann integration to replace \(\lim_{M\rightarrow\infty} \frac1M \sum_{z\in\mathbb{Z}, z>0}\) with \(\int_{0}^{\infty} dz \lim_{M\rightarrow\infty}\). Pulling out a factor of \(M\) then, we can evaluate these sums as integrals, apart from terms that leave a harmonic sum:\footnote{Technically, here we should replace \(M\) with a dimensionless scaled version, $ML \over 2\pi$, but the idea works out the same. We skip some of the details of obtaining the harmonic sum as well as the further expressions below. We only note that the harmonic sum arises after splitting the sum over $z\in \Z$ into terms with $|z|\le \lfloor\frac{LM}{2\pi}\rfloor$ (giving the harmonic sum) and ones with $|z|>\lfloor\frac{LM}{2\pi}\rfloor$ (the latter, after a shift of the summation variable give rise to convergent Riemannian integrals in the $ML \rightarrow \infty$ limit).}
\[
\lim_{M\rightarrow\infty} \sum_{z = -\left\lfloor\frac{LM}{2\pi}\right\rfloor}^{\left\lfloor\frac{LM}{2\pi}\right\rfloor} \frac{1}{\left|z \pm \mu\right|}~.
\]
However, we can use the identity \(\sum_{z=1}^{N} \frac{1}{z} \rightarrow \log\left(N\right) + \gamma_E\) to convert these sums into divergent logarithms that will eventually cancel (one has to add and subtract a $1/|z|$ sum to the harmonic sum above). Using these techniques, we get
\begin{equation}
\begin{split}
I_1 \rightarrow & \frac{iL}{32 \pi^2}\\
I_2 \rightarrow & \frac{-iL}{64\pi^2} \left(\log\left(\frac{L^2M^2}{32\pi^2}\right) - F(\mu) - F(1-\mu) + 2\right)\\
I_3 \rightarrow & \frac{3iL}{64\pi^2} \left(\log\left(\frac{L^2M^2}{32\pi^2}\right) - F(\mu) - F(1-\mu)\right)
\end{split}
\end{equation}
Here \(F(x)\) is the digamma function, defined by \(F(x) = \Gamma'(x) / \Gamma(x)\), and \(\mu \equiv \frac{Lm_\beta}{2\pi} - \left\lfloor \frac{Lm_\beta}{2\pi}\right\rfloor\). Plugging these back into equation (\ref{Int_setting}), we get
\begin{equation}\label{tri_result_unpolished}
\frac{1}{\pi} \beta^a\beta^b (p_\sigma + q_\sigma)\varepsilon^{\sigma\mu\nu}  
\end{equation}
Summing over the positive roots gives
\begin{equation}
\frac{N}{\pi} \delta^{ab} (p_\sigma + q_\sigma)\varepsilon^{\sigma\mu\nu} 
\end{equation} 
Finally, this justifies the inclusion of the \(\phi'\) term in equation (\ref{adjointBF1}) since it perfectly matches an IR term of the form
\begin{equation}
\frac{1}{\pi} \vec{\phi'} \cdot\vec{F}\wedge B ~.
\end{equation} 
We note that the calculation of this section is a check on our calculations of the BF term contribution proportional to the vev of the previous section, as the appearance of $\vec\phi'$ in the combination $\vec\phi' + \langle \vec\phi\rangle$ is expected.

\subsection{The chiral Chern-Simons term}
\label{appx:A.6}

In this subsection, we find the \(B\wedge dB\) term of the EFT from equation \ref{adjointBF1}. This calculation is quite similar to the \(A \wedge F\) calculation, but with one major difference: the \(B\) field couples to all fermions the same, meaning we also must include the massive Cartan fermions.\par
\begin{figure}[t]\centering
  \includegraphics[width= .4 \textwidth]{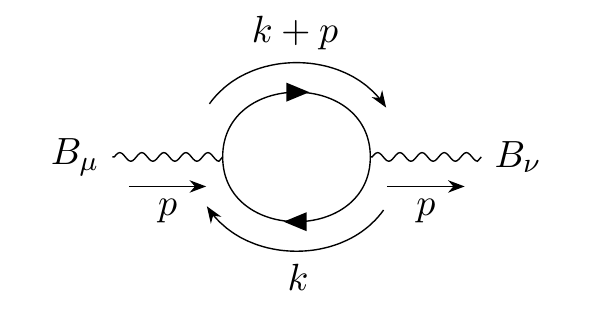}
  \caption{The diagram generating the $B \wedge dB$ term.}
  \label{fig:08}
\end{figure}

Before we start that, however, we check the contribution from the non-Cartan fermions. This is based on the diagram shown in Figure \ref{fig:08}, 
which gives rise to to the expression 
\begin{equation}
\begin{split}
-\frac{s_i}{2}  \int & \frac{d^3k}{(2\pi)^3}  \frac{1}{\left[\left(k^2 - m_{z,\beta}^2 - W^2 - M_i^2 + i\varepsilon \right)^2 - 4W^2m_{z,\beta}^2 - 4 W^2 M_i^2\right]}\\
 \times & \frac{1}{\left[\left((k+p)^2 - m_{z,\beta}^2 - W^2 - M_i^2 + i\varepsilon \right)^2 - 4W^2m_{z,\beta}^2 - 4 W^2 M_i^2\right]}\\
\times &\Tr\left[\left(\slashed{k} + \gamma^3 m_{z,\beta} - \gamma^5\gamma^3 W - M_i\right)\left(k^2 - m_{z,\beta}^2 - W^2 - M_i^2 + 2\gamma^5Wm_{z,\beta} + 2\gamma^5\gamma^3WM_i\right)\gamma^5\gamma^\nu\right.\\
&\, \left.\left(\slashed{k} + \slashed{p} + \gamma^3 m_{z,\beta} - \gamma^5\gamma^3 W - M_i\right)\left((k+p)^2 - m_{z,\beta}^2 - W^2 - M_i^2 + 2\gamma^5Wm_{z,\beta} + 2\gamma^5\gamma^3WM_i\right)\gamma^5\gamma^\nu\right]
\end{split}
\end{equation}
Thus the only differences between this and the \(A\wedge F\) term calculation are the lack of \(\beta\) out front and the inclusion of two extra \(\gamma^5\) from the vertices. The \(\gamma^5\) do change the results of taking the traces, but the procedure is nearly identical to before: keep only terms that give a Levi-Civita and consider only the leading order term in \(p\). From this, we are left with one term that looks like \(k_\sigma k\cdot p\), so we use the fact that in a Lorentz in variant 3d-integral, we can make the replacement \(k_\sigma k_\rho \rightarrow \frac13 g_{\sigma\rho} k^2\) to obtain the integral:
\begin{equation} 
\begin{split}
2is_iW p_\sigma \varepsilon^{\sigma\mu\nu} & \int \frac{d^3k}{(2\pi)^3} \frac{a(k)^2 + 4m_{z,\beta}a(k) + 4\left(m_{z,\beta}^2 - M_i^2\right)W^2 + \frac83 M_i^2k^2}{\left[\left(a(k) + i\varepsilon\right)^2 - 4\left(m_{z,\beta}^2+M_i^2\right)W^2\right]^2}
\end{split}
\end{equation}
here we are again using \(a(k) \equiv k^2 - m_{z,\beta}^2 - M_i^2 -W^2\). We Wick rotate our integration variable and change to spherical coordinates to find
\begin{equation}
-\frac{s_i}{\pi^2} W p_\sigma \varepsilon \int dk_E k_E^2 \frac{b(k_E)^2 - 4m_{z,\beta}^2b(k_E) + 4\left(m_{z,\beta}^2 - M_i^2\right)W^2 - \frac83 M_i^2k_E^2}{\left[b(k_E)^2 - 4\left(m_{z,\beta}^2+M_i^2\right)W^2\right]^2}
\end{equation}
We can again find an exact antiderivative:
\begin{equation}
\label{bbint_exact}
\begin{split}
\frac{1}{12W^3\left(m_{z,\beta}^2 + M_i^2\right)} & \left[\frac{f(k_E,W)}{k_E^4 + 2k_E^2\left(m_{z,\beta}^2+M_i^2+W^2\right) + \left(m_{z,\beta}^2+M_i^2-W^2\right)^2}\right.\\
 & + \frac{g(W)}{\left|\sqrt{m_{z,\beta}^2+M_i^2}-W\right|\sqrt{m_{z,\beta}^2+M_i^2}} \arctan\left(\frac{k_E}{\left|\sqrt{m_{z,\beta}^2+M_i^2}-W\right|}\right) \\
 & - \left.\frac{g(-W)}{\left|\sqrt{m_{z,\beta}^2+M_i^2}+W\right|\sqrt{m_{z,\beta}^2+M_i^2}}  \arctan \left(\frac{k_E}{\left|\sqrt{m_{z,\beta}^2+M_i^2}+W\right|}\right) \right]
\end{split}
\end{equation}
with
\begin{align}
\begin{aligned}
f(k_E,W) \equiv & -2k_E^3W\left(3m_{z,\beta}^2W^2 + m_{z,\beta}^2M_i^2 + W^2M_i^2 + M_i^4\right)\\
& \, + 2k_EW\left(m_{z,\beta}^2+M_i^2-W^2\right)\left(3m_{z,\beta}^2W^2 - m_{z,\beta}^2M_i^2 + W^2M_i^2 - M_i^4\right)
\end{aligned}\\
\begin{aligned}
g(W) \equiv &  -3\mzb^4W^2 + \mzb^4M_i^4 + 3\mzb^2W^3\sqmzmi - 6 \mzb^2W^2M_i^2 \\
& \, - \mzb^2 W M_i^2 \sqmzmi +2\mzb^2 M_i^4 -2W^4M_i^2+5W^3 M_i^2\sqmzmi\\
& \, -3W^2M_i^4 - W M_i^4 \sqmzmi + M_i^6
\end{aligned}
\end{align}
This formula is considerably uglier than before, but it is nice in the necessary limits; it vanishes under \(k_E \rightarrow 0\), so we need only consider the \(k_E \rightarrow \infty\) limit. For the case \(\sqmzmi > \left|W\right|\), which occurs for the PV fermions and physical fermions with sufficiently high \(z\), we get 
\begin{equation}
\frac{\pi}{6} \frac{M_i^2}{\left(\mzb^2 + M_i^2 \right)^{3/2}}
\end{equation} 
This vanishes for the physical \(i=0\) fermions. For the PV fermions we get
\begin{equation}
\lim_{M_i \rightarrow \infty} \sum_{z\in\mathbb{Z}} \frac{\pi}{6} \frac{M_i^2}{\left(\mzb^2 + M_i^2 \right)^{3/2}} = \frac{\pi}{6} \frac{L}{2\pi} \int_{-\infty}^{\infty} \frac{dx}{\left(x^2 + 1\right)^{3/2}} = \frac{L}{6} 
\end{equation}
Hence for the non-Cartan PV fermions we get a total contribution of 
\begin{equation}
\frac{LW}{6\pi^2} \left(\frac{N^2-N}{2}\right) p_\sigma \varepsilon^{\sigma\mu\nu} 
\end{equation}
where \(\frac{N^2-N}{2}\) is just the number of positive roots, over which we have to sum. \par 
For the physical fermions, we have seen that the \(k_E \rightarrow \infty\) limit of equation (\ref{bbint_exact}) vanishes if \(\sqmzmi = \left|\mzb\right| > \left|W\right|\), so we now only need to look at the case where \(\left|\mzb\right| < \left|W\right|\). In this case (and remembering that \(M_0 = 0\)), the limit gives
\begin{equation}
\frac{\pi}{4\left|W\right|}
\end{equation}
This leads to the non-Cartan physical contribution taking the familiar form:
\begin{equation}
-\frac{1}{4\pi} \text{sign}\left(W\right)\sum_{\beta\in\beta^+} n\left(m_\beta,W\right) p_\sigma\varepsilon^{\sigma\mu\nu}
\end{equation}
Now we just need to find the contribution of the Cartan fermions. Fortunately, the calculation is rather similar. \par
Since the \(B_\mu\) field couples the same way to all the adjoint fermion modes, the vertices do not change. Thus the only change is in the propagators. Selecting the relevant term from the 3d Lagrangian, we find
\begin{equation}
\begin{split}
\sum_{i=0}^3 \sum_{z=1}^{\infty} \sum_{a=1}^{N-1} & \psi_{i,z}^{a\dag} \bar{\sigma}^\mu \partial_\mu \psi_{i,z}^{a} + \psi_{i,-z}^{a\dag} \bar{\sigma}^\mu \partial_\mu \psi_{i,-z}^{a} \\
& + \psi_{i,z}^{a\dag} \bar{\sigma}^3 \left(\frac{2\pi z}{L} + W\right) \psi_{i,z}^{a} + \psi_{i,-z}^{a\dag} \bar{\sigma}^3  \left(\frac{-2\pi z}{L} +W\right)  \psi_{i,-z}^{a} \\
& + M_i \left(\psi_{i,-z}^{a}\psi_{i,z}^{a} - \psi_{i,z}^{a\dag}\psi_{i,-z}^{a\dag}\right) 
\end{split}
\end{equation}
Notice that we are only summing over positive \(z\). This is because, unlike with the non-Cartan fermions, all the negative \(z\) terms are already accounted for due to the pairing of \(\psi^a_{i,z}\) and \(\psi^a_{i,-z}\) terms. For now, we purposefully leave out the \(z=0\) mode, because the physical \(z=0\) mode is massless and reamins in the theory\footnote{This breaks down when \(W\) gets too large, so we deal with integrating out the \(z=0\) at the end of the section.}. 
Similarly to our treatment of the non-Cartan fermions, we combine the paired Cartan modes into 4-component fermions:
\begin{equation}
\Psi^a_{i,z} = \begin{pmatrix} \psi^a_{i,z} \\ \psi^{a\dag}_{i,-z} \end{pmatrix}
\end{equation}
Then the Lagrangian terms can be rewritten as 
\begin{equation}
\sum_{i=0}^3 \sum_{z=1}^{\infty} \sum_{a=1}^{N-1} \bar{\Psi}_{i,z}^a \left(i\slashed{\partial} + \gamma^3 \frac{2\pi z}{L} + \gamma^5\gamma^3 W + M_i\right) \Psi^a_{i,z} 
\end{equation}
Hence the only details that change between the Cartan and non-Cartan calculations is the replacement \(\mzb \rightarrow \frac{2\pi z}{L} \) and the difference in the sums. Hence from the PV fermions we get the following contribution from the integral 
\begin{equation}
\begin{split}
 \frac{\pi}{6} \lim_{M_i\rightarrow\infty} \sum_{z=1}^{\infty} \frac{M_i^2}{\left(\right(\frac{2\pi z}{L}\left)^2+M_i^2\right)^{3/2}} = \frac{\pi}{6} \frac{L}{2\pi} \int_0^{\infty} \frac{dx}{\left(x^2+1\right)^{3/2}}  = \frac{L}{12} 
\end{split} 
\end{equation}
This means we get exactly half the contribution for each Cartan mode of \(\Psi\) as we did for each non-Cartan mode. And since there are only \(N-1\) Cartan modes, we find the Cartan PV fermion contribution to be
\begin{equation}
\frac{LW}{6\pi^2} \frac{1}{2} \left(N-1\right) p_\sigma \varepsilon^{\sigma\mu\nu} 
\end{equation}
The physical Cartan modes are very similar to the physical non-Cartan modes in that for each mode with \(\left|W\right|> \frac{2\pi z}{L}\), we get a contribution
\begin{equation}
-\frac{1}{4\pi} \text{sign}\left(W\right) p_\sigma\varepsilon^{\sigma\mu\nu} 
\end{equation}
Recalling that we only consider \(z\geq1\) and that there are \(N-1\) Cartan generators, we can count the contributing modes to get: 
\begin{equation}
-\frac{1}{4\pi} \text{sign}\left(W\right) (N-1)\left\lfloor \frac{L\left|W\right|}{2\pi}\right\rfloor p_\sigma\varepsilon^{\sigma\mu\nu} 
\end{equation}
Hence, combining all contributions and factoring out a \(\text{sign}(W)\), gives the expression
\begin{equation}
\frac{\text{sign}(W)}{4\pi} \left[\left(N^2-1\right)\frac23 \frac{L\left|W\right|}{2\pi} - (N-1)\left\lfloor \frac{L\left|W\right|}{2\pi}\right\rfloor - \sum_{\beta\in\beta^+} n(m_\beta,W) \right] p_\sigma\varepsilon^{\sigma\mu\nu} 
\end{equation}
This corresponds to the IR term
\begin{equation}
\frac{\text{sign}(W)}{4\pi} \left[\left(N^2-1\right)\frac23 \frac{L\left|W\right|}{2\pi} - (N-1)\left\lfloor \frac{L\left|W\right|}{2\pi}\right\rfloor - \sum_{\beta\in\beta^+} n(m_\beta,W) \right]  B\wedge dB
\end{equation}

\subsubsection{Including the Kaluza-Klein zero-modes}
\label{appx:A.6.1}

In the case that \(W\) is of the order \(\approx\frac{2\pi}{L}\), it makes sense to integrate out the remaining \(z=0\) Cartan modes. This is for example necessary when dealing with large chiral gauge transformations that shift \(W \rightarrow W + \frac{2\pi}{L} \mathbb{Z}\); indeed, as demonstrated in the main text, such shifts only produce the correct anomaly when all the fermions have been integrated out. Here we outline the calculation. \par
\begin{figure}[h] 
\begin{subfigure}[t]{.48  \textwidth}
  \includegraphics[width= 1 \textwidth]{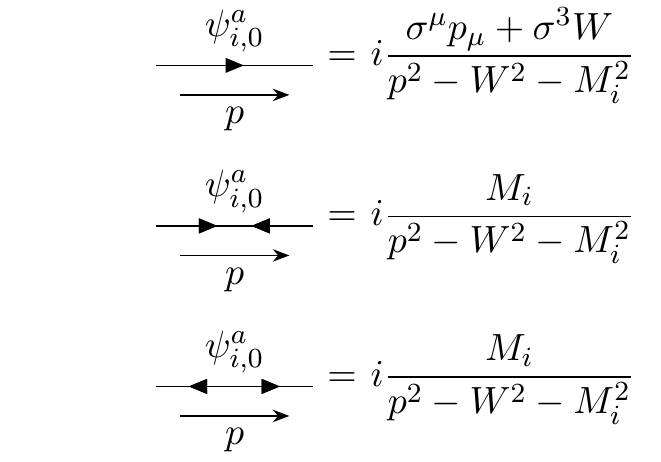}
\end{subfigure} 
\qquad
\begin{subfigure}[t]{.48 \textwidth}
  \includegraphics[width= 1\textwidth]{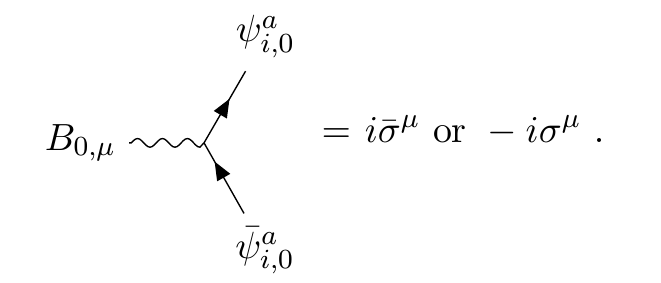}
\end{subfigure}
\caption{The Feynman rules for the physical Cartan fermions and PV regulators.}
\label{fig:09}
\end{figure}
\begin{figure}[h] 
\begin{subfigure}[t]{.44  \textwidth}
  \includegraphics[width= 1 \textwidth]{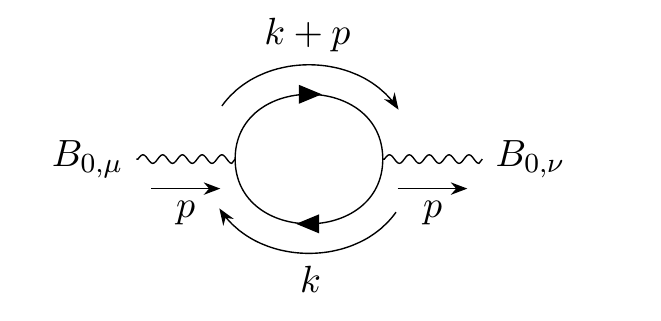}
\end{subfigure} 
\qquad
\begin{subfigure}[t]{.44 \textwidth}
  \includegraphics[width= 1\textwidth]{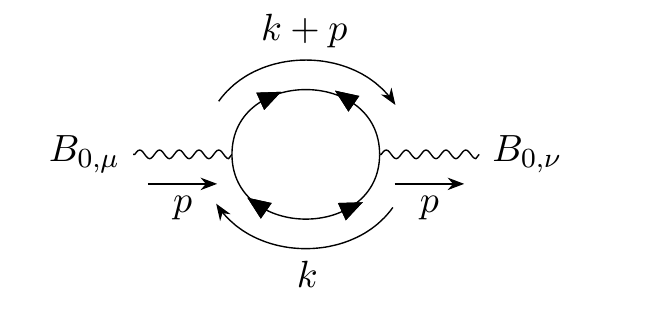}
\end{subfigure}
\caption{The Cartan fermion contributions to the $B \wedge dB$ term.}
\label{fig:10}
\end{figure}

As before, we start with the relevant terms in the Lagrangian:
\begin{equation}
\begin{split}
\sum_{a=1}^{N-1} \sum_{i=0}^3 & \psi^{a\dag}_{i,0} \left(i\bar{\sigma}^\mu \partial_\mu + \bar\sigma^3 W + \bar\sigma^\mu B_\mu \right) \psi^a_{i,0} \\
& + \frac{M_i}{2} \left(\psi^a_{i,0}\psi^a_{i,0} + \psi^{a\dag}_{i,0}\psi^{a\dag}_{i,0} \right)
\end{split}
\end{equation}
Using standard two-component notation \cite{Dreiner:2008tw}, the Feynman rules are shown in Figure \ref{fig:09}. Using these, there are two possible diagrams to consider, shown on Figure \ref{fig:10}. 

The latter of these (the ones with the mass insertions) involves \(\Tr\left[\bar\sigma^\mu \sigma^\nu \right] = 2g^{\mu\nu}\) and thus cannot contribute to our topological terms. The former gives
\begin{equation}
-\frac{s_i}{2} \int \frac{d^3k}{(2\pi)^3} \frac{1}{\left[k^2 - W^2 - M_i^2\right]\left[(k+p)^2 - W^2 - M_i^2\right]} \Tr\left[\bar\sigma^\mu\left(\sigma^\lambda k_\lambda + \sigma^3 W\right)\bar\sigma^\nu\left(\sigma^\rho (k+p)_\rho + \sigma^3 W \right) \right] ~.
\end{equation}  
Using the identity, \(\Tr\left[\bar\sigma^a \sigma^b \bar\sigma^c \sigma^d\right] = 2\left(g^{ab}g^{cd} - g^{ac}g^{bd} + g^{ad}g^{bc} -i\varepsilon^{abcd} \right)\), we see that we need exactly one of the four sigma matrices in the trace to be \(\sigma^3\) in order to get a Levi-Civita. This selects the terms
\begin{equation}
-\frac{s_i}{2} \int \frac{d^3k}{(2\pi)^3} \frac{1}{\left[k^2 - W^2 - M_i^2\right]\left[(k+p)^2 - W^2 - M_i^2\right]} \left(W k_\lambda\Tr\left[\bar\sigma^\mu\sigma^\lambda \bar\sigma^\nu \sigma^3\right] + W (k+p)_\lambda \Tr\left[\bar\sigma^\mu\sigma^3 \bar\sigma^\nu \sigma^\lambda\right]\right) ~.
\end{equation} 
Evaluating the traces and ignoring everything except the Levi-Civita terms, we find
\begin{equation}
is_i W p_\lambda \varepsilon^{\lambda\mu\nu} \int \frac{d^3k}{(2\pi)^3} \frac{1}{\left[k^2 - W^2 - M_i^2\right]\left[(k+p)^2 - W^2 - M_i^2\right]}
\end{equation}
Evaluating this integral gives 
\begin{equation}
-\frac{s_i}{8\pi} \text{sign}(W) \frac{1}{\sqrt{1 + \frac{M_i^2}{W^2}}} p_\lambda\varepsilon^{\lambda\mu\nu}
\end{equation}
Looking at the \(M_i \rightarrow \infty\) limit, we see that the PV fermions give no contribution. On the other hand, combining all the non-vanishing contributions from the physical fermions gives
\begin{equation}
 -\frac12(N-1) \frac{1}{4\pi} \text{sign}(W) p_\lambda \varepsilon^{\lambda\mu\nu} 
\end{equation}
Hence, including these contributions results in the \(B\wedge dB\) term
\begin{equation}
\frac{\text{sign}(W)}{4\pi} \left[\left(N^2-1\right)\frac23 \frac{L\left|W\right|}{2\pi} - (N-1)\left(\frac12 + \left\lfloor \frac{L\left|W\right|}{2\pi}\right\rfloor\right) - \sum_{\beta\in\beta^+} n(m_\beta,W) \right]  B\wedge dB
\end{equation}

\section{Dirac fermion calculations}
\label{appx:B}
In this section, we detail the calculations of the Dirac fermion contributions leading to equation (\ref{fundBF1}).
\subsection{Topological terms for general vectorlike representations} 

\label{appx:B.1}
In this subsection we deal with general Dirac fermions denoted by $\theta$, charged under an unspecified representation, \(R\). In this section, the generators of \(R\) are denoted by \(T^a_{jk}\) and the weights of \(R\) are labelled with \(\lambda_j^a\) which denotes the \(jj^{\text{th}}\) component of the \(a^{\text{th}}\) Cartan generator in this representation. We assume that the action of the gauged \(U(1)\) from the adjoint Weyl case on the Dirac fermions is a charge \(q\) axial transformation. We also add similar PV fermions to before (and reuse our index notation \(i=0,1,2,3\)), but now with a simple Dirac mass. This all leads to the 4D Lagrangian:
\begin{equation}\label{Diraclagrangian}
L_{vec} = \sum_{i=0}^3 \bar{\theta^j_i} \left(\delta_{jk} \gamma^M i\partial_M + \gamma^M A_M^a T^a_{jk} + \delta_{jk}  \gamma^5 \gamma^M qB_M - \delta_{jk} M_i\right) \theta^k_i
\end{equation}
As before, the mass \(M_i\) is the PV mass. $\theta^k_i$ are four-component Dirac fermions, where $k$ is the representation index (it runs from 1 to dim $R$) and   \(i=0\) denotes the physical fermions with \(M_0 = 0\), while $i=1,2,3$ denotes the PV fermions with masses chosen as before. Integrating around the compactified dimension and extracting the terms that are relevant for our calculation, gives us
\begin{equation} 
\begin{split}
\sum_{i=0}^3 \sum_{z\in\mathbb{Z}}\sum\limits_{j=1}^{{\rm dim} R} \bar{\theta}^j_{z,i} & \left(i \gamma^\mu\partial_\mu + \gamma^3 m_{z,j} + \gamma^5\gamma^3 qW + M_i\right.\\
&\left. + \gamma^\mu A^a_{0,\mu}\lambda^a_j + \gamma^5\gamma^\mu qB_{\mu} + \gamma^3 \phi^{' a} \lambda^a_j \right) \theta^j_{z,i} 
\end{split}
\end{equation}
where \(m_{z,j} \equiv \frac{2\pi z}{L} + \frac{2\pi}{NL} \rho^a \lambda^a_j + \frac{2\pi}{L} \expval{\phi^{a}} \lambda^a_j\), recalling our definition (\ref{defofphi}) of $A_3^a$.

 In certain cases, even at the center symmetric point, this mass may vanish\footnote{To see this, consider the case of, e.g.   fundamental Dirac fermions and odd $N$ at the center symmetric point.} If that occurs, for any mode, we simply add a background holonomy for the vectorlike \(U(1)\) symmetry. This holonomy contributes an extra non-zero term \(\mu\) to \(m_{z,j}\), which we can tune to ensure it does not vanish for any mode. \par
Notice that the only differences between these terms and the ones for the \(\Psi\) fermions constructed from the adjoint Weyl fermions are the inclusion of \(q\), the sign of \(M_i\), and the change from positive roots to weights of \(R\). Hence, the Feynman rules and all the calculations that follow take a very similar form. In fact, we can merely take the results of the previous section (namely equations (\ref{AF_result}, \ref{BF_result_unpolished}, \ref{tri_result_unpolished})) and make these replacements. This gives
\begin{equation}
\begin{split}
L^{Dirac} = \frac{q}{2\pi} B\wedge F^a &\sum_{j} \left(2\frac{Lm_{0,j}}{2\pi} - 2\left\lfloor\frac{Lm_{0,j}}{2\pi}\right\rfloor - 1+n'(m_{0,j},qW)\right) \lambda^a_j \\&- \frac{1}{2\pi} A^a\wedge F^b \sum_{j} \text{sign}(qW) n(m_{0,j},qW) \lambda_j^a\lambda_j^b \\
& + \frac{q}{\pi} B\wedge F^a {\phi'}^b \sum_j \lambda_j^a\lambda_j^b
\end{split}
\end{equation}
Using the identities \(\sum_j \lambda_j^a \lambda_j^b = C(R)\delta^{ab}\) and \(\sum_j \lambda_j^a=0\), this equation can be further simplified to
\begin{equation}
\begin{split}\label{abcd}
L^{Dirac} = \frac{qC(R)}{\pi} B\wedge F^a  & \left(\frac{\rho^a}{N} + \expval{\phi^a} + {\phi'}^a \right) - \frac{q}{\pi} B\wedge F^a \sum_j \left\lfloor \frac{Lm_{0,j}}{2\pi} \right\rfloor \lambda_j^a\\
&- \frac{1}{4\pi} A^a\wedge F^b \sum_{j} \text{sign}(qW) n(m_{0,j},qW) \lambda_j^a\lambda_j^b \\
& + \frac{q}{2\pi} B\wedge F^a \sum_{j}  n'(m_{0,j},qW) \lambda^a_j \\
& + \frac{q^2}{4\pi} B\wedge dB\; \text{sign}(qW) \sum_{j} \left[\frac{4}{3}\frac{L\left|qW\right|}{2\pi} - n(m_{0,j},qW)\right]~.
\end{split}
\end{equation}
We remind the reader that the functions $n$ and $n'$ are defined in (\ref{nfunctions2}), or (\ref{n1},\ref{nfunctions3}), and that 
\begin{equation}
\label{massfundam}
m_{0,j} \equiv \mu + \frac{2\pi}{NL} \rho^a \lambda^a_j + \frac{2\pi}{L} \expval{\phi^{a}} \lambda^a_j~,
\end{equation}
where $\mu$ is the  $M=3$ component of the $U(1)_V$ background gauge field (i.e., its holonomy along $\S^1$) that may be turned on to avoid having massless fermions (depending on the representation $R$ and the value of $\langle \phi^a \rangle$ one is interested in). The weights of $R$ are denoted by $\lambda_i^a$, $i = 1,...$ dim$R$,  and $C(R)$ is  the quadratic Casimir.
Eq. (\ref{abcd}) is our general result for the topological terms due to  Dirac fermions in general representations $R$. It is easy to see that the anomaly of the chiral-$U(1)_q$ small and large transformations for these fermions follows from the above formulae.

Unfortunately, in a general representation, even assuming we are a the center symmetric point, we cannot simplify the term with all the floor functions. However, it is possible when considering specifically the fundamental representation, which is the focus of the next subsection. 

\subsection{Fundamental fermions}
\label{appx:B.2}
In this subsection, we show how, in the case that the Dirac fermions are in the fundamental representation, the results from the previous subsection give us equation (\ref{fundBF1}). To obtain the same equation we also now pick \(q = -N\), which is the $U(1)_A$ anomaly-free chiral charge of the fundamentals in the mixed $A+F$ theory. With these specifics, we find
\begin{equation}
\begin{split}\label{fund1}
\frac{-N}{\pi} B\wedge F^a  & \left(\frac{\rho^a}{N} + \expval{\phi^a} + {\phi'}^a \right) + \frac{N}{\pi} B\wedge F^a \sum_{A = 1}^N \left\lfloor \frac{Lm_{A}}{2\pi} \right\rfloor \nu_a^A \\
& + \frac{1}{4\pi} A^a\wedge F^b \sum_{A = 1}^N \text{sign}(W) n(m_A,NW) \nu_a^A \nu_b^A \\
& - \frac{N}{2\pi} B\wedge F^a \sum_{A = 1}^N  n'(m_A,NW) \nu_a^A \\
& - \frac{N^2}{4\pi} B\wedge dB \;\text{sign}(W) \left[N^2 \frac{4}{3}\frac{L\left|W\right|}{2\pi} - \sum_{A=1}^N n(m_A,NW)\right]~,
\end{split}
\end{equation}
where we use $\nu^A_a$, $A=1,...,N$, to denote the weights of the fundamental representation.
Now, all that remains is to deal with the floor functions. For simplicity, we shall assume that  we are at the center symmetric point (i.e. \(\expval{\phi} = 0\)) (however, note that the presence of the floor functions guarantees that this contribution must vary discontinuously within the Weyl chamber as stated in the main text). \par
Given \(\expval{\phi} = 0\), the second $B \wedge F$ term  in (\ref{fund1}) becomes 
\begin{equation}
 \frac{N}{\pi} B\wedge F^a \sum_{A = 1}^N \left\lfloor \frac{1}{N} \vec{\rho} \cdot \vec{\nu}^A + \frac{L\mu}{2\pi} \right\rfloor \nu_a^A 
\end{equation}
For what follows, we assume that \(\mu\) is tuned to be very small and positive,\footnote{If $N$ is even, we can take $\mu=0$ right away.} so that the value of the floor function is determined by the first term. Then we make use of the identity \(\vec{\rho} \cdot \vec{\nu}^A = \frac{N+1}{2} - A\) to get
\begin{equation}
 \frac{N}{\pi} B\wedge F^a \sum_{A = 1}^N \left\lfloor \frac{N+1-2A}{2N} \right\rfloor \nu_a^A 
\end{equation} 
Notice that for \(1\leq A \leq \frac{N+1}{2}\) the floor evaluates to 0, whereas for \(A > \frac{N+1}{2}\) it evaluates to -1. Thus, defining 
\begin{equation}
A_* = \begin{cases} \frac{N}{2} & N\text{ even} \\ \frac{N+1}{2} & N\text{ odd} \end{cases}
\end{equation}
we can rewrite this term as 
\begin{equation}
- \frac{N}{\pi} B\wedge F^a \sum_{A = A_* + 1}^N  \nu_a^A
\end{equation}
But recall that \(\sum_{A = 1}^N \vec{\nu}^A=0\) or equivalently \(- \sum_{A = A_* + 1}^N  \vec{\nu}^A = \sum_{A=1}^{A_*} \vec{\nu}^A = \vec{w}_{A_*}\). Hence, this term contributes a single fundamental weight:
\begin{equation}
 \frac{N}{\pi} B\wedge \vec{F} \cdot \vec{w}_{A_*}
\end{equation}
Including this with the rest, we recover equation (\ref{adjointBF1}) recalling that \(A_{*} = \frac{N}{2}\) when \(N\) is even (for even $N$, at the center symmetric point, there is no need to turn on $\mu$):
\begin{equation}
 \begin{split}
\frac{-N}{\pi} & B\wedge F^a   \left(\frac{\rho^a}{N} - w^a_{A_*}+ {\phi'}^a \right)  \\
& + \frac{1}{4\pi} A^a\wedge F^b \sum_{A = 1}^N \text{sign}(W) n(m_A,qW) \nu_a^A \nu_b^A \\
& - \frac{N}{2\pi} B\wedge F^a \sum_{A = 1}^N  n'(m_A,qW) \nu_a^A \\
& - \frac{N^2}{4\pi} B\wedge dB\; \text{sign}(W) \left[N^2 \frac{2}{3}\frac{L\left|W\right|}{2\pi} - \sum_{A=1}^N n(m_A,NW)\right]~.
\end{split}
\end{equation}

\subsection{Including background \(U(1)_V\)} 
\label{appx:B.3}
\begin{figure}[h] 
\begin{subfigure}[t]{.44  \textwidth}
  \includegraphics[width= 1 \textwidth]{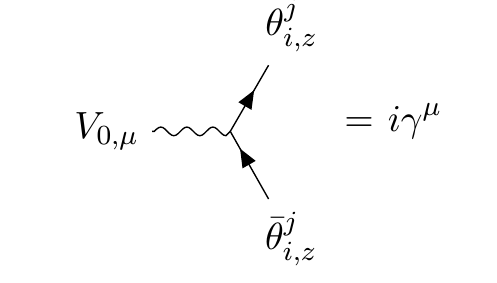}
\end{subfigure} 
\qquad
\begin{subfigure}[t]{.44 \textwidth}
  \includegraphics[width= 1\textwidth]{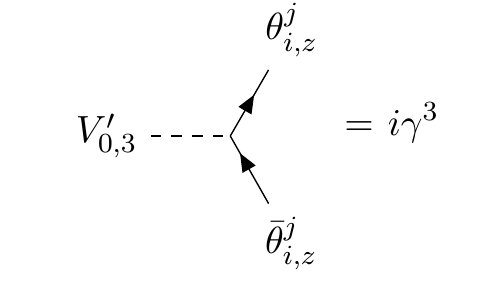}
\end{subfigure}
\caption{The vertices coupling the Dirac fermion to the $U(1)_V$ background.}
\label{fig:11}
\end{figure}

For the Dirac fermions, it is possible to introduce a background field for the \(U(1)_V\) vector symmetry. We will denote this field by \(V_M\). We split the \(V_3\) component up into \(V_3 = \mu + V'_3\) with \(\mu\) being the expectation value. This adds the following term to the 4d Lagrangian
\begin{equation}
\sum_{i=0}^3 \bar{\theta}^j_i \delta_{jk} \gamma^M V_M \theta^k_i~.
\end{equation}
In the 3d EFT term, this term becomes
\begin{equation}
\sum_{i=0}^3 \sum_{z\in\mathbb{Z}} \bar{\theta}^j_{i,z} \left(\gamma^3 \mu + \gamma^3 V'_{0,3} + \gamma^\mu V_{0,\mu} \right) \theta^k_{i,z} ~.
\end{equation}
This term does not change the form of the propagator; it simply adds a term \(\mu\) to \(m_{z,j}\). It also introduces the two new vertices shown in Figure \ref{fig:11}.

The first vertex is identical to the \(A\bar{\theta}\theta\) vertex with the replacement \(\beta^a \rightarrow 1\) and the second vertex is identical to the \(\phi'\bar{\theta}\theta\) vertex this the replacement \(\frac{2\pi}{L} \beta^a \rightarrow 1\). Therefore, we can merely adapt the outcomes of our calculations involving the \(A\) and \(\phi'\) terms. This leads to the following topological terms for a general Dirac fermion:
\begin{equation}
\begin{split}
 \frac{q}{2\pi} B\wedge dV &\sum_{j} \left(2\frac{Lm_{0,j}}{2\pi} - 2\left\lfloor\frac{Lm_{0,j}}{2\pi}\right\rfloor - 1+n'(m_{0,j},qW)\right) \\
&- \frac{1}{4\pi} V \wedge dV  \sum_{j} \text{sign}(qW) n(m_{0,j},qW) \\
& -\frac{1}{2\pi} A^a \wedge dV  \sum_{j} \text{sign}(qW) n(m_{0,j},qW) \lambda_j^a\\
& + \frac{Lq}{2\pi^2} B\wedge dV V'_3 \sum_j 1\\
& + \frac{q}{\pi} B\wedge dV {\phi'}^a \sum_j \lambda_j^a ~.
\end{split}
\end{equation}
The last term vanishes due to the identity \(\sum_j \lambda^a_j = 0\). This identity also allows us to simplify the first term to
\begin{equation}
\frac{q}{\pi} B\wedge dV \; \text{dim}(R) \frac{L\mu}{2\pi} +  \frac{q}{2\pi} B\wedge dV \sum_{j} \left(n'(m_{0,j},qW) - 2\left\lfloor\frac{Lm_{0,j}}{2\pi}\right\rfloor - 1\right)~.
\end{equation}
This leaves us with
\begin{equation}
\begin{split}\label{generalrepvector}
\frac{q}{\pi} B\wedge dV & \; \text{dim}(R) \frac{L\mu}{2\pi} \\
& +  \frac{q}{2\pi} B\wedge dV \sum_{j} \left(n'(m_{0,j},qW) - 2\left\lfloor\frac{Lm_{0,j}}{2\pi}\right\rfloor - 1\right) \\
& + \frac{Lq}{2\pi^2} \text{dim}(R) B\wedge dV V'_3 \\
&- \frac{1}{4\pi} V \wedge dV  \sum_{j} \text{sign}(qW) n(m_{0,j},qW) \\
& -\frac{1}{2\pi} A^a \wedge dV  \sum_{j} \text{sign}(qW) n(m_{0,j},qW) \lambda_j^a~.
\end{split}
\end{equation}
In the specific case discussed in the text, with fundamental fermions and \(q=-N\), we thus find
\begin{equation}
\begin{split}
- \frac{LN^2}{2\pi^2} &\left(\mu + V'_3\right) \;B\wedge dV \\
& -  \frac{N}{2\pi} B\wedge dV \sum_{A=1}^N \left(n'(m_A,NW) - 2\left\lfloor\frac{Lm_A}{2\pi}\right\rfloor - 1\right) \\
&+ \frac{1}{4\pi} V \wedge dV  \sum_{A=1}^N \text{sign}(W) n(m_A,NW) \\
& +\frac{1}{2\pi} A^a \wedge dV  \sum_{A=1}^N \text{sign}(W) n(m_A,NW) \lambda_j^a~.
\end{split}
\end{equation}

\section{On Chern-Simons terms in the unbroken phase: holonomies vs. boundary conditions} 
\label{appx:C}

Here, we point out a subtlety regarding the induced Chern-Simons term in the unbroken phase, involving the relation between background Wilson lines and boundary conditions. We discuss this subtlety because it directly follows from our calculations. The remark  may be relevant if one wants to study topological phases of the compactified theory, a topic which is outside of our main interest here.

To begin we note that, formally, our calculation of the Chern-Simons (CS) term in the adjoint theory holds also at vanishing holonomy, $\langle A_3^a \rangle =0$, i.e. in the phase with unbroken gauge $SU(N)$, provided $L W \ne 2 \pi \Z$. This is because in such a background, all components of the fermions are  massive and the fermion mass gap is of order $1/L$, hence the fermions can be integrated out.
 Then, we should keep all gauge bosons\footnote{\label{fnote1}Having $W \ne 0$ breaks supersymmetry (of the single-adjoint theory) and a one-loop potential (trustable when $L$ is small)  for the holonomy is introduced \cite{Gross:1980br}. From this potential (whose form is to be found in e.g. \cite{Anber:2017pak}) it is easy to see that  $LW/(2\pi)=1/2$ forces a vanishing expectation value for $A_3^a$, while an infinitesimal deviation of $LW/(2\pi)$ from $1/2$ produces a small vev for $A_3^a$. However, the gauge boson masses remains  small, allowing us to   keep all gauge bosons in the external legs. An alternative case  to consider, for SYM, is that of an infinitesimal $W$,  tuned to to be small enough to ignore the potential on the Coulomb-branch.} in the external legs and the sum over $a,b$ in the CS term in (\ref{adjointBF1}) will be extended over all $SU(N)$ generators.
 
 The long-distance 3D YM theory  has a dimensionful gauge coupling, $g_3^2 \sim {1 \over L} g^2({1 \over L})$, which sets the energy scale where the 3D dynamics is expected to become strong. In order to decouple the fermions of mass $1/L$ at weak coupling,  we need to have $g^2({1 \over L}) \ll 1$, or $\Lambda L \ll 1$, i.e. we need to be in a typical small-$L$ regime.
With all these caveats, we obtain from (\ref{AF_result}), recalling that $\sum\limits_{\beta+} \beta^a \beta^b = N \delta^{ab}$ and extending the sum over all generators,  the CS term: 
\begin{eqnarray}\label{lcs1}
L^{CS} &=& -{N\; {\rm sign}(W) \; n(0, W) \over 2 \pi} \;  \tr (A  \wedge F + \ldots) ~.
\end{eqnarray}
Admittedly, we have not computed the nonlinear terms in this paper, but expect that the usual form of the CS term  follows from $SU(N)$ gauge invariance.
Now, from the definition of $n(0,W)$ from (\ref{nfunctions}) or (\ref{n1}) we see that
 \begin{eqnarray}\label{lcs2}
n(0,W) &=& {1\over 2} \sum\limits_{k \in \Z} \left[ 1 - {\rm sign}({2 \pi \over L}|k|-|W|)\right] = 1, ~{\rm for} \; 0< |W| < 1,
\end{eqnarray}
and that the periodicity property (\ref{nfunctions2}) then implies that $n(0,W) = 3$, for $1<|W|<2$, etc. Thus, we find that the (small-$L$) adjoint-fermion theory, in the background of the $U(1)_A$ Wilson line $W$,  flows to  an  $SU(N)$ CS theory, whose Lagrangian we can write as
\begin{eqnarray}
L^{CS} &=&- {N p  \over 2 \pi}\; {\rm sign}(W)   \tr (A \wedge F + \ldots)~ = - {4 N p \; {\rm sign}(W) \over 8 \pi}   \tr (A \wedge F + \ldots),
\end{eqnarray}
showing that the CS level is $-4 N p\; {\rm sign}(W)$, with $p = 1 + 2  \left \lfloor{|W|}\right \rfloor$, where $\left \lfloor{|W|}\right \rfloor$ is the largest integer smaller than $|W|$.
We stress that the above result for the CS term holds for the theory  on $\R^3 \times \S^1$, defined with a physical fermion and PV fields, all taken in the background of a $U(1)$ Wilson line $W$. The PV masses explicitly break the  $U(1)_A$ and the  PV fields  obey the same periodic boundary conditions as the physical fermion.

 It is tempting to restrict to $\Z_{2N}$-valued Wilson lines, $B_3 L = L W = {2 \pi q \over 2 N}$, and make an identification of these with a boundary condition, where fermions are periodic up to a  $\Z_{2N}$ phase. In this Appendix, we shall make several comments on this identification. 
 
 First, we notice that unless one considers antiperiodic boundary conditions $\psi(L) = - \psi(0)$, one can not consistently impose $\Z_{2N}$ periodicity on the PV fields---periodicity up to $\Z_{2N}$ is not consistent with a translationally invariant (along $\S^1$, see below) PV Majorana mass terms. These terms  only allow periodic or antiperiodic boundary conditions. 

To elaborate this further, consider a $\Z_2$-valued Wilson line, $WL = B_3 L = \pi$. The CS term calculated above is not zero in this background---in fact, the CS level is $4 N$, formally the same for all values of $W$ between $0$ and $1$ (however, as discussed in footnote \ref{fnote1}, for $W$ significantly different from $1/2$, the calculation assuming unbroken $SU(N)$ at small $L$ is  not  meaningful). On the other hand, it is well known that no CS is generated in the thermal theory, where  both regulators and physical fields are taken antiperiodic.

 The  point we wish to make is that a nonzero Wilson line and a $\Z_{2N}$-twisted boundary condition are not equivalent  for the adjoint theory with an anomalous $U(1)$. To see this, imagine that $WL = B_3 L = {2 \pi q \over 2N}$ is a $\Z_{2N}$ valued Wilson line and  fermions and regulators are periodic. Classically, to remove this Wilson line, we redefine 
\begin{equation}\label{lambda1}
\psi(x^3) = \psi'(x^3) e^{i  {2 \pi   q\over 2 N} {x^3\over L} }, ~ \psi(L)=\psi(0), ~{\rm hence} ~ \psi'(L) = e^{- i {2 \pi   q\over 2 N}} \psi'(0)~.
\end{equation}
Thus, if $\psi$ is periodic around $\S^1$, the new field $\psi'$ is only periodic up to a $\Z_{2N}$ phase. Hence, periodicity up to $\Z_{2N}$ and a nonzero $L W = {2 \pi q \over 2 N}$ are classically equivalent, as is clear from considering the physical (massless) adjoint-fermion lagrangian.
However, the
transformation (\ref{lambda1}) is a local $x^3$-dependent $U(1)$ phase rotation, not an anomaly-free $\Z_{2N}$ transformation (it is  a  $U(1)$ transformation periodic up to a $\Z_{2N}$ phase). The $x^3$-dependent phase gets ``stuck'' in the PV mass terms, leading to a non-translationally invariant regulator.

For simplicity, let us further restrict to the case where $\psi'$ is antiperiodic, i.e. take $q=N$.  The transformation (\ref{lambda1}) with $q=N$ changes the partition function. That this is so is clear from our observation that the following two calculations yield different results for the CS level:
 
{\it i}.) a background Wilson line $WL={\pi}$, with periodic fermions and regulators, and PV Majorana mass $M_{PV} \int dx^3 \psi \psi$, as in (\ref{Lpv}), yields a nonzero CS level, as per (\ref{lcs1},\ref{lcs2}).

 {\it ii}.) a vanishing Wilson line $W=0$ and  antiperiodic fermions and regulators, with a translationally invariant PV Majorana mass $M_{PV} \int dx^3 \psi' \psi'$, where $\psi'$ are antiperiodic regulators, yields a vanishing CS term
 
We note that {\it i.}) can be made equivalent to {\it ii.}) by a field redefintion (as in (\ref{lambda1})). However, the Majorana mass becomes $x^3$-dependent, i.e. takes the form $M_{PV}\int dx^3 e^{- i {2 \pi x^3 \over L}} \psi'
  \psi'$, with antiperiodic regulators $\psi'$ (this generalizes to arbitrary values of $W$, where the fermions can be thought as having arbitrary periodicity property, and the PV mass term accommodates this twisting at the cost of being non translationally invariant).
One can argue that an $x^3$-dependent regulator mass breaks translational invariance, so if we are to study thermal partition function and want to have time translations in the regulated theory, we have to proceed with option {\it ii.}).  Our point is that a nonzero background Wilson line, $L W=\pi$,  with periodic fermions is not equivalent to the antiperiodic-fermion and vanishing Wilson line case, provided regulators respect translational invariance along $\S^1$.

\section{A useful infinite sum}\label{appx:D}
Above, see (\ref{identity1}), we used the fact that for $Lm_\beta/2\pi \notin \mathbb{Z}$
\begin{equation}
\lim_{M\rightarrow \infty} \sum_{z\in\mathbb{Z}} \frac{m_{z,\beta}}{\sqrt{m_{z,\beta}^2}} - \frac{2m_{z,\beta}}{\sqrt{m_{z,\beta}^2 + M^2}} + \frac{m_{z,\beta}}{\sqrt{m_{z,\beta}^2 + 2M^2}} = 1+2\left\lfloor\frac{Lm_\beta}{2\pi}\right\rfloor - 2\frac{Lm_\beta}{2\pi}
\end{equation}
Now we will prove this relation. Firstly, we define new dimensionless quantities $\mu \equiv Lm_\beta/2\pi$ and $A = LM/2\pi$. Then we can rewrite the sum as 
\begin{equation} 
 \sum_{z\in\mathbb{Z}} \frac{z + \mu}{\sqrt{(z+\mu)^2}} - \frac{2(z+\mu)}{\sqrt{(z+\mu)^2+A^2}} + \frac{z+\mu}{\sqrt{(z+\mu)^2 + 2A^2}} 
 \end{equation}
For cleanliness we make the definition:
\begin{equation}
C_{\mu,A}(z) \equiv \frac{z + \mu}{\sqrt{(z+\mu)^2}} - \frac{2(z+\mu)}{\sqrt{(z+\mu)^2+A^2}} + \frac{z+\mu}{\sqrt{(z+\mu)^2 + 2A^2}} 
\end{equation}
Now, we can bound the behavior of $C_{\mu,A}$ as $z\rightarrow \pm \infty$ to prove that it converges. Consider $z >  \text{max}\{2A - \mu,1-\mu\}$ and $A>0$, then we have
\begin{equation}
\begin{split}
C_{\mu,A}(z) = & 1 - \frac{2(z+\mu)}{\sqrt{(z+\mu)^2+A^2}} + \frac{z+\mu}{\sqrt{(z+\mu)^2 + 2A^2}} <  2\left(1 - \frac{z+\mu}{\sqrt{(z+\mu)^2+A^2}}\right)\\
= & \frac{2(z+\mu)}{\sqrt(z+\mu)^2+A^2} \left(\sqrt{1 + \frac{A^2}{(z+\mu)^2}}-1\right)
<  2\left(\sqrt{1 + \frac{A^2}{(z+\mu)^2}}-1\right)\\
= & \frac{A^2}{(z+\mu)^2} + \mathcal{O}\left(\frac{A^4}{(z+\mu)^4}\right)
\end{split}
\end{equation}
Similarly, 
\begin{equation}
\begin{split}
-C_{\mu,A}(z) = & -1 + \frac{2(z+\mu)}{\sqrt{(z+\mu)^2+A^2}} - \frac{z+\mu}{\sqrt{(z+\mu)^2 + 2A^2}}
<  -1 + 2 - \frac{z+\mu}{\sqrt{(z+\mu)^2 + 2A^2}}\\
= & 1 - \frac{z+\mu}{\sqrt{(z+\mu)^2 + 2A^2}}
=  \frac{(z+\mu)}{\sqrt(z+\mu)^2+2A^2} \left(\sqrt{1 + \frac{2A^2}{(z+\mu)^2}}-1\right)\\
< & \sqrt{1 + \frac{2A^2}{(z+\mu)^2}}-1
=  \frac{A^2}{(z+\mu)^2} + \mathcal{O}\left(\frac{A^4}{(z+\mu)^4}\right)
\end{split}
\end{equation}
Hence, 
\begin{equation}
\left|C_{\mu,A}(z)\right| <  \frac{A^2}{(z+\mu)^2} + \mathcal{O}\left(\frac{A^4}{(z+\mu)^4}\right) \Rightarrow \lim_{z\rightarrow\infty} \left|C_{\mu,A}(z)\right|z^{3/2} = 0
\end{equation}
Therefore, since $\sum_{z=1}^\infty z^{-3/2} = \zeta\left(3/2\right) \approx 2.61$ we have necessarily that $\sum_{z=1}^\infty C_{\mu,A}(z)$ converges absolutely. In fact, we have that $\sum_{z=n}^\infty C_{\mu,A}(z)$ converges absolutely for any $n\in\mathbb{Z}$. Now consider, $z > \text{max}\{2A - \mu,1-\mu\}$ and $A>0$, then similarly to the above we have
\begin{equation}
\begin{split}
C_{\mu,A}(-z) = &  -1 + \frac{2(z-\mu)}{\sqrt{(z-\mu)^2+A^2}} - \frac{z-\mu}{\sqrt{(z-\mu)^2 + 2A^2}}\\
< & 1 - \frac{z-\mu}{\sqrt{(z-\mu)^2 + 2A^2}} \\ 
< & \frac{A^2}{(z-\mu)^2} + \mathcal{O}\left(\frac{A^4}{(z+\mu)^4}\right)
\end{split}
\end{equation}
and
\begin{equation}
\begin{split}
-C_{\mu,A}(-z) = & 1 - \frac{2(z-\mu)}{\sqrt{(z-\mu)^2+A^2}} + \frac{z-\mu}{\sqrt{(z-\mu)^2 + 2A^2}}\\
< & 2\left(1 - \frac{(z-\mu)}{\sqrt{(z-\mu)^2+A^2}}\right)
<  \frac{A^2}{(z-\mu)^2} + \mathcal{O}\left(\frac{A^4}{(z+\mu)^4}\right)
\end{split}
\end{equation}
Hence, we have 
\begin{equation}
\left|C_{\mu,A}(-z)\right| <  \frac{A^2}{(z-\mu)^2} + \mathcal{O}\left(\frac{A^4}{(z-\mu)^4}\right) \Rightarrow \lim_{z\rightarrow\infty} \left|C_{\mu,A}(-z)\right|z^{3/2} = 0
\end{equation}
Therefore, as above we have $\sum_{z=-n}^\infty C_{\mu,A}(-z) = \sum_{z=-\infty}^n C_{\mu,A}(z)$ converges absolutely. Combining these two results we see that the total sum $\sum_{z\in\mathbb{Z}}C_{\mu,A}(z)$ converges for any non-zero value of $A$. \par
Notice that $z$ and $\mu$ always appear together as $z+\mu$ within $C_{\mu,A}(z)$ and since we are summing over all $z\in\mathbb{Z}$, we can shift $z$ or equivalently $\mu$ by any integer.  Hence, using $\tilde{\mu} \equiv \mu - \lfloor\mu\rfloor \in (0,1)$,\footnote{In general, $\tilde{\mu}\in [0,1)$, but we are using our assumption that $\mu\notin\mathbb{Z}$ which is guaranteed by $\langle A_3^a\rangle$ being in the interior of the Weyl chamber.} we have
\begin{equation}
\sum_{z\in\mathbb{Z}}C_{\mu,A}(z) = \sum_{z\in\mathbb{Z}}C_{\tilde\mu,A}(z)
\end{equation}
Now, we want to find 
\begin{equation}
\lim_{A\rightarrow\infty} \sum_{z\in\mathbb{Z}}C_{\tilde\mu,A}(z).
\end{equation}
We do not know \emph{a priori} that this is a well defined limit, but we have some tricks to help us. Notice that our argument for the convergence of each limit of the infinite sum can also be used verbatim to show that $\int_1^\infty C_{\tilde\mu,A}(z) dz$ and $\int_{-\infty}^{-1} C_{\tilde\mu,A}(z) dz$ are convergent. Hence, 
\begin{equation}
\frac{1}{2} C_{\tilde\mu,A}(1) + \sum_{z=2}^\infty C_{\mu,A}(z) - \int_1^\infty C_{\tilde\mu,A}(z) dz
\end{equation}
and 
\begin{equation}
\frac{1}{2} C_{\tilde\mu,A}(-1) + \sum_{z=-\infty}^2 C_{\tilde\mu,A}(z) - \int_{-\infty}^{-1} C_{\tilde\mu,A}(z) dz
\end{equation}
are finite. By the Euler-Maclaurin formula,\footnote{As a reminder, in physicists' notation it reads ${1 \over 2} F(0) + F(1) + F(2) + ... - \int\limits_0^\infty F(n) =  -{1 \over 2!} B_2 F'(0) - {1 \over 4!} B_4 F^{'''}(0) + ...$, where $B_{2k}$ denote Bernoulli numbers.} these particular expressions are equal to a linear function of the derivatives of $C_{\tilde\mu,A}$. Notice though that for $z>1$ or $z<-1$, the first term of $C_{\tilde\mu,A}(z)$ is a constant (1 or -1, respectively), so the only terms that survive are those that are dependent on $A$. These terms and all derivatives of these terms clearly vanish as $A\rightarrow\infty$. Hence, we can say that 
\begin{equation}
\lim_{A\rightarrow\infty} \left( \frac{1}{2} C_{\tilde\mu,A}(1) + \sum_{z=2}^\infty C_{\mu,A}(z) - \int_1^\infty C_{\tilde\mu,A}(z) dz\right) = 0
\end{equation}
and
\begin{equation}
\lim_{A\rightarrow\infty} \left( \frac{1}{2} C_{\tilde\mu,A}(-1) + \sum_{z=-\infty}^2 C_{\tilde\mu,A}(z) - \int_{-\infty}^{-1} C_{\tilde\mu,A}(z) dz\right) = 0
\end{equation}
From here we can calculate:
\begin{equation}
\begin{split}
1 = & \lim_{A\rightarrow\infty} \left(C_{\tilde\mu,A}(0) + \frac{1}{2}C_{\tilde\mu,A}(1) + \frac{1}{2}C_{\tilde\mu,A}(-1)\right)\\
= & \lim_{A\rightarrow\infty} \left(C_{\tilde\mu,A}(0) +   \sum_{z=1}^\infty C_{\mu,A}(z) + \sum_{z=-\infty}^1 C_{\tilde\mu,A}(z) - \int_1^\infty C_{\tilde\mu,A}(z) dz - \int_{-\infty}^{-1} C_{\tilde\mu,A}(z) dz \right)\\
= &  \lim_{A\rightarrow\infty} \left( \sum_{z=-\infty}^\infty C_{\mu,A}(z) - \int_1^\infty C_{\tilde\mu,A}(z) dz - \int_{-\infty}^{-1} C_{\tilde\mu,A}(z) dz \right)
\end{split}
\end{equation}
Now notice that 
\begin{equation}
\begin{split}
& \int_1^\infty C_{\tilde\mu,A}(z) dz + \int_{-\infty}^{-1} C_{\tilde\mu,A}(z) dz \\
=&  \int_1^\infty \left(1 - \frac{2(z+\tilde\mu)}{\sqrt{(z+\tilde\mu)^2+A^2}} + \frac{z+\tilde\mu}{\sqrt{(z+\tilde\mu)^2 + 2A^2}} \right)dz  - \int_1^\infty \left(1 - \frac{2(z-\tilde\mu)}{\sqrt{(z-\tilde\mu)^2+A^2}} + \frac{z-\tilde\mu}{\sqrt{(z-\tilde\mu)^2 + 2A^2}}\right) dz
\\ 
&= \int_{1+\tilde\mu}^\infty \left(1 - \frac{2t}{\sqrt{t^2+A^2}} + \frac{t}{\sqrt{t^2+2A^2}}\right)  -  \int_{1-\tilde\mu}^\infty \left(1 - \frac{2t}{\sqrt{t^2+A^2}} + \frac{t}{\sqrt{t^2+2A^2}}\right)\\
  &= -  \int_{1-\tilde\mu}^{1 + \tilde\mu} \left(1 - \frac{2t}{\sqrt{t^2+A^2}} + \frac{t}{\sqrt{t^2+2A^2}}\right)
\end{split}
\end{equation}
Hence,
\begin{equation}
\lim_{A\rightarrow\infty}  \int_1^\infty C_{\tilde\mu,A}(z) dz + \int_{-\infty}^{-1} C_{\tilde\mu,A}(z) dz = -  \int_{1-\tilde\mu}^{1 + \tilde\mu} dz = -2\tilde\mu
\end{equation}
Therefore, finally we have that 
\begin{equation}
\begin{split}
1-2\tilde\mu = & \lim_{A\rightarrow\infty}   \left( \sum_{z=-\infty}^\infty C_{\mu,A}(z) - \int_1^\infty C_{\tilde\mu,A}(z) dz - \int_{-\infty}^{-1} C_{\tilde\mu,A}(z) dz \right)\\
& + \lim_{A\rightarrow\infty} \left( \int_1^\infty C_{\tilde\mu,A}(z) dz + \int_{-\infty}^{-1} C_{\tilde\mu,A}(z) dz\right)\\
= & \lim_{A\rightarrow\infty} \sum_{z=-\infty}^\infty C_{\mu,A}(z)
\end{split}
\end{equation}
Finally, the observation that $1-2\tilde\mu = 1 + 2\left\lfloor\frac{Lm_\beta}{2\pi}\right\rfloor - 2\frac{Lm_\beta}{2\pi}$, finishes the proof.

  \bibliography{topterms1.bib}
  
  \bibliographystyle{JHEP}
\end{document}